\documentclass[twocolumn,showpacs,preprintnumbers,amsmath,amssymb]{revtex4}

\usepackage{graphicx}
\usepackage{dcolumn}

\usepackage{amsmath,amssymb}

\usepackage[mathscr]{euscript}

\usepackage{calc}

\usepackage{nicefrac}

\usepackage{color}

\usepackage{mathptmx, courier, pifont}
\usepackage[scaled=0.92]{helvet}
\usepackage[T1]{fontenc}
\usepackage{textcomp} 

\usepackage{bm}

\newcommand{\singlefig}[6]{%
\begin{figure}[tpb]\vspace{#3}%
\includegraphics*[scale=#5]{#2}%
\caption{\label{fig:#1} #6}%
\vspace{#4}%
\end{figure}}

\newcommand{\doublefig}[6]{%
\begin{figure*}[tpb] \vspace{#3}%
\includegraphics*[scale=#5]{#2}%
\caption{\label{fig:#1} #6}
\vspace{#4}
\end{figure*}}

\newtheorem{conjecture}{Conjecture}

\newcommand{\tsub}[1]{_{\mbox{\scriptsize#1}}}
\newcommand{\tsup}[1]{^{\mbox{\scriptsize#1}}}
\newcommand{\comm}[2]{\left[#1 , #2\right]}
\newcommand{\bra}[1]{\langle#1|}
\newcommand{\ket}[1]{|#1\rangle}
\newcommand{\ev}[1]{\langle#1\rangle}
\newcommand{\mel}[3]{\bra{#1}#2\ket{#3}}

\newcommand{\tightdot}{\hspace{-0.1em}\cdot\hspace{-0.1em}}

\newcommand{\spin}{\mathscr{S}}   

\newcommand{\shellfill}{f}

\newcommand{\llm}{{m_k}}

\newcommand{\kdegen}{\Omega_k}

\newcommand{\boldspin}{\pmb\spin}

\newcommand{\Piop}[2]{\Pi_{#1#2}}

\newcommand{\phantomdagger}{^{\vphantom{\dagger}}}

\newcommand{\sothree}{{\rm SO(3)}}

\newcommand{\sufour}{{\rm SU(4)}}
\newcommand{\sutwo}{{\rm SU(2)}}
\newcommand{\sofive}{{\rm SO(5)}}
\newcommand{\sosix}{{\rm SO(6)}}
\newcommand{\soseven}{{\rm SO(7)}}
\newcommand{\soeight}{{\rm SO(8)}}
\newcommand{\uone}{{\rm U(1)}}
\newcommand{\ufour}{{\rm U(4)}}

\newcommand{\casimir}[1]{C_{\scriptscriptstyle#1}}

\newcommand{\suiso}{\sutwo_{\tau}}
\newcommand{\uonecharge}{\uone_{{\scriptstyle\rm c}}}
\newcommand{\suspin}{\sutwo_{\sigma}}

\newcommand{\supsetne}{\mathbin{\rotatebox[origin=c]{45}{$\supset$}}}
\newcommand{\supsetse}{\mathbin{\rotatebox[origin=c]{-45}{$\supset$}}}

\newcommand{\eq}[1]{Eq.~(\ref{#1})}

\newcommand{\eqnoeq}[1]{(\ref{#1})}
\newcommand{\fig}[1]{Fig.~\ref{fig:#1}}
\newcommand{\tableref}[1]{Table~\ref{tb:#1}}

\newcommand{\units}[1]{\mbox{\ #1}}

\newcommand{\onematrix}[2]{\begin{pmatrix}#1\\#2\end{pmatrix}}
\newcommand{\twomatrix}[4]{\begin{pmatrix}#1&#2\\#3&#4\end{pmatrix}}

\renewcommand{\vec}{\bm}

\newcommand{\halfthin}{\kern 0.0834em}
\newcommand{\neghalfthin}{\kern -0.0834em}

\newcommand{\quarterthin}{\kern 0.0417em}
\newcommand{\negquarterthin}{\kern - 0.0417em}

\newcommand{\pardiv}[2]{\frac{\partial {\negquarterthin #1}}
{\partial {\negquarterthin #2}}}

\newcommand{\clebsch}[6]{C_{#1 #2 #3 #4}^{#5#6}}

\newcommand{\Adag}[2]{A^\dagger_{#1#2}}
\newcommand{\AdagPower}[2]{(\Adag{#1}{#2})^{N_{#1#2}}}

\newcommand{\Adagcoupled}[4]{A_{\scriptscriptstyle 
#1#2}^{\raisebox{0.2em}{$\scriptstyle\dagger$} {\scriptscriptstyle 
#3 #4}}}

\newcommand{\hws}{\ket{\rm HW}}

\newcommand{\ttfrac}[2]{\scriptscriptstyle\frac{#1}{#2}}

\newcommand{\sixj}[6]{\left\{ 
\begin{matrix} #1&#2&#3\\#4&#5&#6 \end{matrix}\right\}}

\newlength{\figup}
\newlength{\tabup}
\newlength{\tabtitlesep}
\newlength{\afterTableLineOne}
\newcommand{\tableLineOne}{\hline\rule{0pt}{\afterTableLineOne}}

\newlength{\afterTableLineTwo}
\newcommand{\tableLineTwo}{\\ \hline\rule{0pt}{\afterTableLineTwo}}

\newlength{\beforeTableLineThree}
\newcommand{\tableLineThree}{\\[\beforeTableLineThree] \hline}

\newlength{\tablerowMore}
\newlength{\tableLineShift}
\newcommand{\runningheads}[2]{\markboth{\hfill #1\hfill}{\hfill #2\hfill}}

\begin{document}

\title{SO(8) Fermion Dynamical Symmetry and Strongly-Correlated Quantum Hall\\ 
States in Monolayer Graphene}

\author{Lian-Ao Wu$^{(1)}$}
\email{lianaowu@gmail.com}
\author{Matthew Murphy$^{(2)}$}
\email{mhatt511@gmail.com}
\author{Mike Guidry$^{(2)}$}
\email{guidry@utk.edu}

\affiliation{
$^{(1)}$IKERBASQUE, Basque Foundation for Science, 48011 Bilbao, Spain,
and Department of Theoretical Physics and History of Science,
Basque Country University (EHU/UPV), Post Office Box 644, 48080 Bilbao, Spain
\\
$^{(2)}$Department of Physics and Astronomy, University of
Tennessee, Knoxville, Tennessee 37996, USA
}

\date{\today}

\begin{abstract}

A formalism is presented for treating strongly-correlated graphene quantum Hall 
states in terms of an SO(8) fermion dynamical symmetry that includes pairing as 
well as particle--hole generators.  The graphene SO(8) algebra is isomorphic to 
an SO(8) algebra that has found broad application in nuclear physics, albeit 
with physically very different generators, and exhibits a strong formal 
similarity to SU(4) symmetries that have been proposed to describe 
high-temperature superconductors.  The well-known SU(4) symmetry of quantum Hall 
ferromagnetism for single-layer graphene is recovered as one subgroup of SO(8), 
but the dynamical symmetry structure associated with the full set of SO(8) 
subgroup chains extends quantum Hall ferromagnetism and allows analytical 
many-body solutions for a rich set of 
collective states exhibiting spontaneously-broken symmetry that may be important 
for the low-energy physics of graphene in strong magnetic fields. The SO(8) 
symmetry permits a natural definition of generalized coherent states that 
correspond to symmetry-constrained Hartree--Fock--Bogoliubov solutions, or 
equivalently a microscopically-derived Ginzburg--Landau formalism, 
exhibiting the interplay between competing spontaneously broken symmetries in 
determining the ground state.

\end{abstract}


\maketitle

\runningheads{}{{\em SO(8) Dynamical Symmetry and Quantum 
Hall States in Graphene}---L.-A. Wu, M. Murphy, and M. W. Guidry}

\section{\label{h:intro} Introduction}

Quantum Hall effects are characteristic of two-dimensional (2D) electron gases 
in strong magnetic fields.  The integral quantum Hall effect (IQHE) 
\cite{vonk1980} is explained in terms of weakly-interacting electrons 
filling  quantized Landau levels (LL) produced by application of a 
strong magnetic field perpendicular to the 2D gas \cite{laug1981}.  In contrast, 
the fractional quantum Hall effect (FQHE) \cite{tsui1982} can occur only as a 
result of strong electronic correlations in partially-filled Landau levels 
\cite{laug1983}.

 Because of its massless chiral charge carriers and atomic-monolayer honeycomb 
lattice exhibiting sublattice, valley, and spin degeneracies, 
quantum Hall effects in graphene could be much richer than in the semiconductor 
2D electron gas, where there is no crystal structure and the only degeneracies 
are those of the (non-relativistic) Landau levels and spin. For graphene in 
strong magnetic fields an integral quantum Hall effect \cite{novo2005,zhan2005} 
and a fractional quantum Hall effect \cite{bolo2009,du2009,feld2012,benj2013} 
have been observed,  with anomalous filling factors that reflect the unique 
degeneracies of the graphene electronic structure and the Dirac-like nature of 
its electrons.

The valley isospin and spin degrees of freedom imply that graphene in a strong 
magnetic field is described well by a low-energy Hamiltonian that commutes 
approximately with the generators of an  SU(4) Lie algebra.   This SU(4) 
symmetry allows classification of states in graphene, and can serve as the basis 
for computing explicit breaking of the SU(4) symmetry by small non-symmetric 
terms in the effective Hamiltonian. However, there is growing evidence that 
many states observed in modern experiments cannot be described by 
explicit breaking of SU(4).  For example, the ground state of graphene in a 
magnetic field at low temperature exhibits a rapid increase of the longitudinal 
resistance $R_{xx}$ above a critical magnetic field $B\tsub c$ \cite{chec2008}.  
The value of $B\tsub c$  decreases for increasingly cleaner samples, indicating 
that the resistance is not caused by impurity scattering but instead is 
intrinsic to the state itself \cite{jung2009}.  In quantum Hall systems the 
currents are carried by edge states, so this insulating ground state must 
correspond to an emergent state that does not support edge currents produced by 
{\em spontaneous} (not explicit) breaking of the SU(4) symmetry.

Thus the approximate SU(4) symmetry of graphene can suggest possible low-energy 
collective modes exhibiting spontaneously broken symmetry that are important for 
the properties of graphene in a magnetic field, but the SU(4) symmetry alone 
cannot determine which of these modes is the ground state. Until now, those 
broken-symmetry modes have been addressed quantitatively by numerical 
simulations employing limited numbers of states and particles, or by effective 
low-energy field-theory approximations.  These calculations find various 
possible low-energy collective states resulting from spontaneous 
breaking of the SU(4) symmetry with very similar energies.  Thus they have been 
unable to give a definitive answer to the nature of the insulating ground state.
Let us note that other approaches to the problems addressed here 
have been proposed (for example, Refs.\ \cite{herb07a,herb07b,roy14}).  
However, the present discussion will concentrate  on methods based on 
approximate SU(4) symmetry of the Hamiltonian.

An alternative and potentially more powerful application of symmetries   has 
been employed extensively in both nuclear structure 
and condensed matter physics \cite{clwu86,clwu87,FDSM,guid01,lawu03,su4review}.  
This {\em fermion dynamical symmetry method} truncates  
the Hilbert space to a tractable collective subspace by positing a 
highest symmetry associated with the physical operators for the system, and 
constructing effective Hamiltonians from polynomials in the Casimir invariants 
of the highest symmetry's subgroup chains.  In this approach it is possible not 
only to classify low-energy collective modes, but to solve analytically for the 
properties of these modes and to determine which lie lowest in energy, either 
exactly in particular symmetry limits, or approximately using generalized 
coherent state methods.  

In an earlier Letter the first application of  fermion dynamical symmetry 
methods to graphene  was introduced and applied to determining the ground state 
in strong magnetic fields \cite{wu2016}. This paper develops the full dynamical 
symmetry formalism upon which Ref.\   \cite{wu2016} rests.  It will be shown 
that the highest symmetry is SO(8), with its generators identifiable with 
particle--hole and pairing degrees of freedom that have been discussed 
previously in the physics of graphene.  This symmetry will be shown to be 
isomorphic to an SO(8) symmetry used extensively in nuclear structure physics, 
which permits already-developed mathematics  to be appropriated for the graphene 
problem, and suggests instructive  physical analogies between two very different 
physical systems. A generalized coherent state approximation will be introduced 
that corresponds to a Hartree--Fock-Bogoliubov (HFB) formalism subject to SO(8) 
symmetry constraints.  This permits quantitative evaluation of energy surfaces 
associated with spontaneously-broken symmetry.

The SO(8) highest symmetry will be shown to have an SU(4) subgroup that recovers 
the known physics of SU(4) quantum Hall ferromagnetism as a special case, but 
implies in the more general case  new low-energy physics that transcends SU(4) 
quantum Hall ferromagnetism. Hence,  a solvable and physically-illuminating 
approach to the rich low-energy structure of undoped graphene in strong magnetic 
fields will be proposed that reproduces known physics, but also suggests 
testable new physics in this complex system.

\section{\label{h:grapheneLatticeStructure} Lattice Structure of Graphene}

Comprehensive reviews of graphene physics may be found in Refs.\ 
\cite{cast2009,goer2011}. The presentation here will recall only a select set of 
features that will be relevant for subsequent discussion. Undoped graphene is a 
2-dimensional semiconductor with zero bandgap. It has a bipartite honeycomb 
lattice structure illustrated in \fig{A-B_sublattices}%
\doublefig
{A-B_sublattices}  
{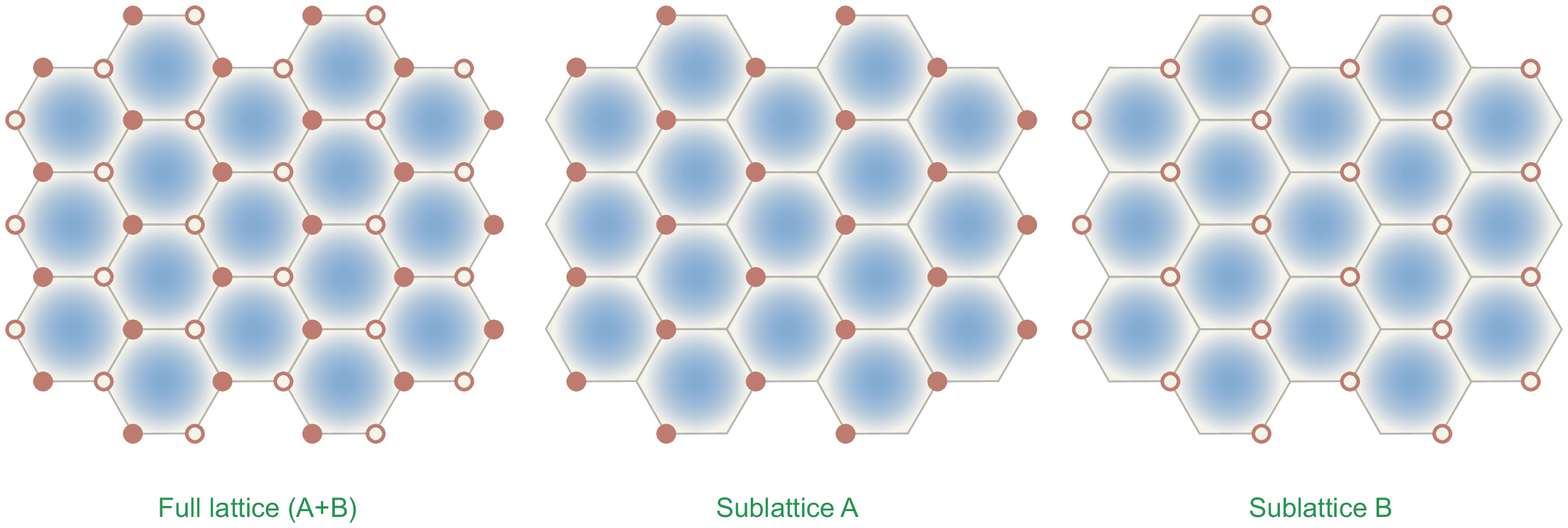}
{\figup} 
{0pt}
{0.55}
{The full bipartite lattice for graphene and the A (solid red points) and B 
(open red points) sublattices.  }
that corresponds to two interlocking triangular sublattices, labeled A and B.  
The two-fold degree of freedom specifying whether an electron is on the A 
sublattice or B sublattice is a spin-like quantum number termed 
the {\em sublattice pseudospin}.  The sublattice pseudospin behaves 
mathematically like the actual spin of the electron (which will be introduced 
later), but it is a separate degree of freedom.

\section{\label{h:structureBrillGraphene} Momentum-Space Structure}

The dispersion of energy with momentum for undoped graphene 
in the absence of a magnetic field  is illustrated in \fig{dispersionGraphene}.%
 \singlefig
          {dispersionGraphene}  
          {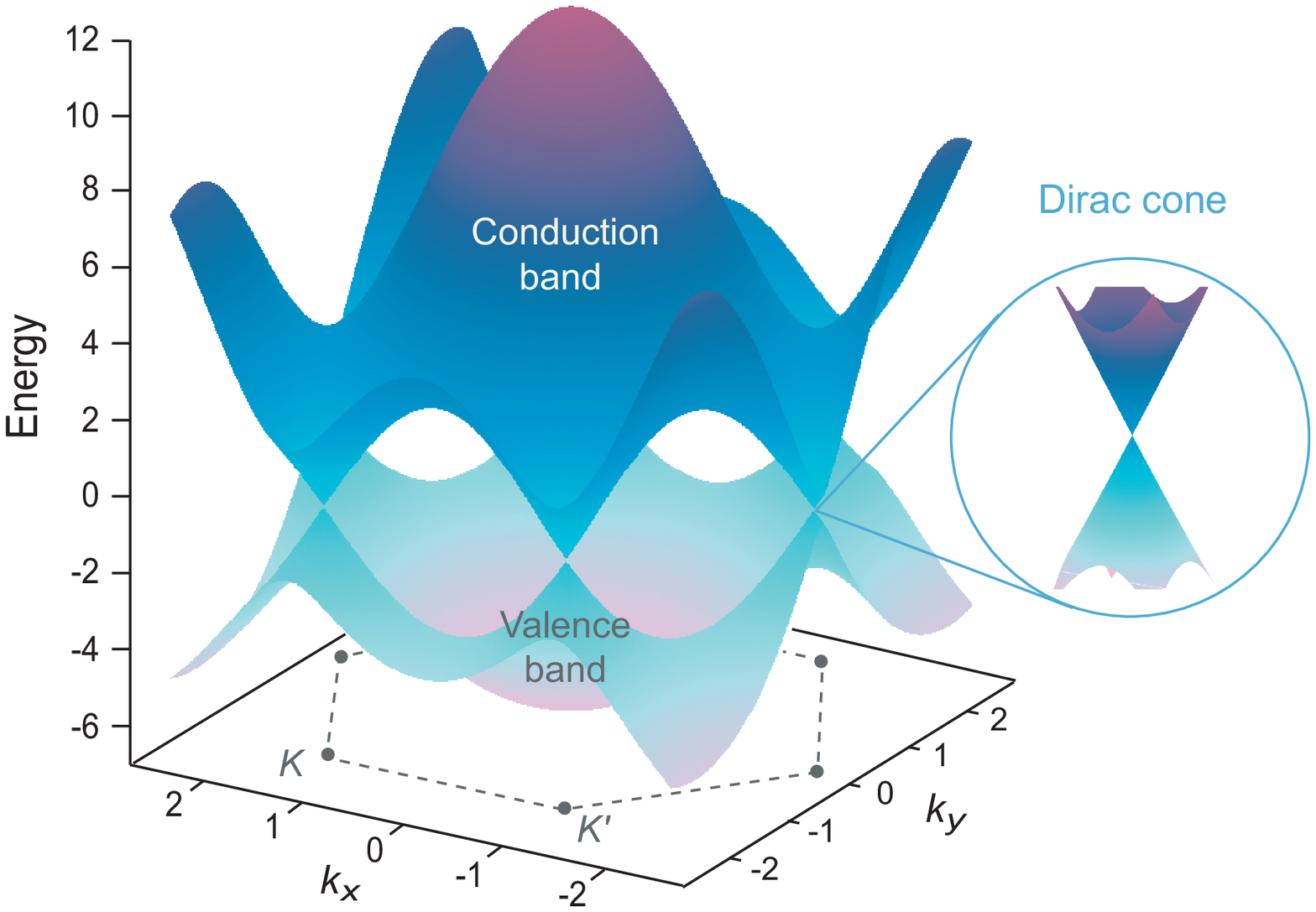}
          {\figup} 
          {0pt}
          {0.44}
          {Electronic dispersion of graphene calculated in a tight-binding model 
with no magnetic field. Two inequivalent points  in the Brillouin zone are 
labeled $K$ and $K'$.  Near these K-points the dispersion becomes linear, 
leading to {\em Dirac cones,} as shown in the expanded view.  For undoped 
graphene the Fermi surface lies at the apex of the cones, where the level 
density vanishes and the effective electronic mass tends to zero. Thus 
low-energy electrons are governed approximately by a Dirac equation for massless 
electrons.
}
The two inequivalent points $K$ and $K'$ are not connected by reciprocal lattice 
vectors.  The corresponding two-fold K degree of freedom may  be viewed 
mathematically as a 2-valued spin-like quantity, commonly termed the {\em valley 
isospin} because of the valley-like structure in the dispersion of 
\fig{dispersionGraphene} around the K-points. For brevity the valley isospin 
will sometimes be termed simply {\em isospin.} Obviously this two-fold 
``isospin'' degree of freedom is a pseudospin that should be distinguished 
physically from the actual spin of the electron, and from standard usage of 
isospin quantum numbers in nuclear and particle physics.

Near the {\em Dirac cones} (inset to \fig{dispersionGraphene}) the dispersion is 
linear, the density of electronic states tends to zero, and the electrons are 
described by a massless Dirac equation in which the Fermi velocity plays the 
role that the speed of light would play in an actual relativistic system. Thus 
the low-energy electrons for undoped graphene in zero magnetic field behave to 
good approximation as {\em massless chiral fermions,} with the chirality 
representing the projection of the sublattice pseudospin on the direction of 
motion, not the projection of the actual spin. The vanishing of the density of 
states at the Dirac point (Fermi surface) implies that the transport properties 
of graphene are different from either a metal or a semiconductor.

\section{\label{h:grapheneMagField} Graphene in a Magnetic Field}

Our interest will be primarily low-energy states in a strong magnetic 
field.  For non-interacting electrons, the quantized levels may be found by 
solving the Dirac equation for massless fermions with a vector potential 
appropriate to the applied magnetic field. The dispersion of energy with 
magnetic field strength for massless Dirac electrons is illustrated in 
\fig{LLgraphene_n0}(a).  Consider the $\nu=0$ state, which corresponds to half 
filling of the fourfold-degenerate $n=0$ LL in graphene, as illustrated in 
\fig{LLgraphene_n0}(b). The graphene honeycomb lattice is bipartite, as 
illustrated in \fig{A-B_sublattices}. The $n=0$ LL is located exactly at the 
Dirac point corresponding to $\epsilon = 0$. For low-energy excitations in each 
valley labeled by $K$ or $K'$, the inter-valley tunneling may be ignored and the 
electrons in the valley reside entirely on either the A or B sublattice. Hence, 
for the $n=0$ LL  labeling with the valley isospin (indicating whether the 
electron is in a $K$ or $K'$ valley) is equivalent to labeling with the 
sublattice pseudospin (indicating whether the electron is on the A or B 
sublattice). This is reminiscent of a N\'eel state with spins on two different 
sublattices, with a N\'eel order defined by the difference in spins on the A and 
B sublattices.

\section{\label{h:quantumHallGraphene} Quantum Hall Effects in Graphene}
 
A quantum Hall effect is signaled by a plateau in the Hall 
conductance $\sigma_{xy}$ having quantized values
\begin{equation}
\sigma_{xy} = \frac{\nu e^2}{h},
\label{sigmaxy}
\end{equation}
where the {\em filling factor} $\nu$ is defined by 
\begin{equation}
\nu = \frac{n_e}{n_B} = \frac{hn_e}{eB},
\label{iqhegraph1.1}
\end{equation}
with $n_e$ the charge-carrier density, $B$ the magnetic field strength, and $n_B 
= B/(h/e)$ the magnetic flux density in units of the fundamental flux quantum 
$h/e$. These plateaus  indicate  formation of an {\em incompressible 
quantum liquid.} This is a compact way to say that the  ground 
state is separated from  excited states by an {\em energy gap,} which inhibits 
compression because of the energy required for excitation
across the 
gap.

\subsection{\label{sh:integralQHEgraphene}Integral Quantum Hall States}

The integral quantum Hall effect (IQHE) in graphene is similar to the integral 
quantum Hall effect for non-relativistic electrons in that it corresponds to the 
formation of incompressible states resulting from the complete filling of Landau 
levels by weakly-interacting electrons.  However, there are two important 
differences between the IQHE states observed in graphene and those observed in 
conventional 2D semiconductor heterostructures:

(1)~In addition to the two-fold spin degeneracy (in the absence of Zeeman 
splitting), there is a two-fold valley degeneracy associated with the distinct 
$K$ and $K^\prime$ points in the first Brillouin zone.  Thus the 
filling factor changes in steps of four between plateaus in the Hall resistance 
for graphene.

(2)~For graphene the filling factor $\nu$ defined in \eq{iqhegraph1.1} vanishes 
at 
the Dirac point for particle--hole symmetric half filling of the graphene 
lattice, since the electron density $n_e$ tends to zero there. Hence, in the 
absence of a Zeeman effect or strong electronic correlations, there is no 
integral quantum Hall effect in graphene for $\nu = 0$.

\singlefig
{LLgraphene_n0}  
{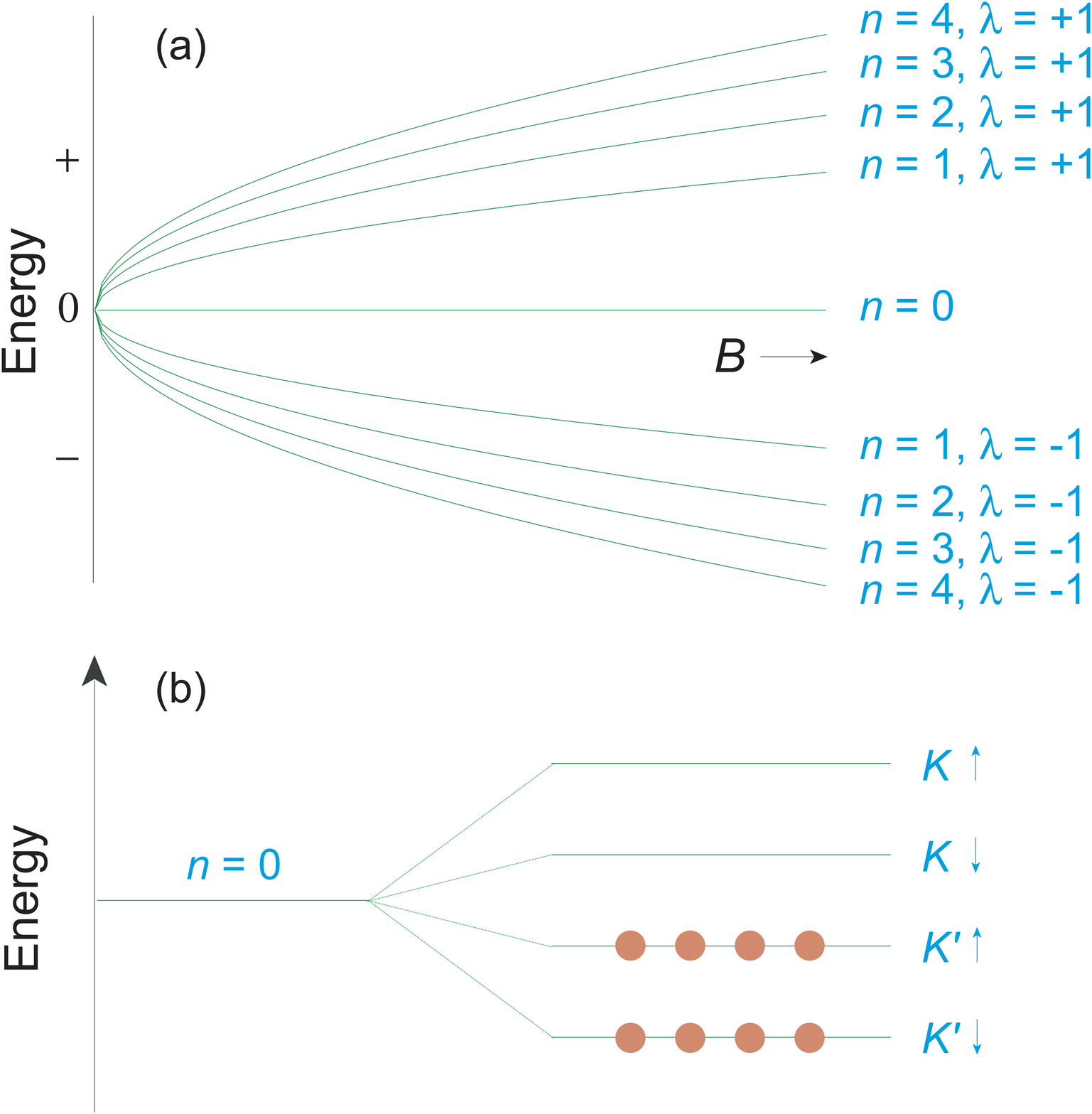}
{\figup} 
{0pt}
{0.13}
{(a)~Relativistic dispersion for massless Dirac electrons as a 
function of magnetic field strength $B$.  States are labeled by a 
principle quantum number $n$ and a quantum number $\lambda$ indicating 
particle-like ($+$) and hole-like ($-$) states.  Each Landau level 
labeled by $n$ has a high orbital degeneracy $\Omega_k$ (see \eq{degen1.1}), 
and an additional 4-fold degeneracy associated with spin and valley isospin.  
(b)~One configuration for occupation 
of the $n=0$ Landau level in monolayer graphene. The splitting and occupation 
are schematic only. The ground state would be a superposition of such 
configurations and in the SU(4) symmetry limit (which obtains for Coulomb-only 
interactions) the 4 levels shown labeled by valley $K$ or $K'$ and spin up or 
down would be degenerate. In realistic cases the splitting often is small, 
suggesting approximate SU(4) symmetry.}

In graphene the analog of the integral quantum Hall effect was first observed 
at  filling factors
\begin{equation}
\nu = \pm 2,\, \pm 6,\, \pm 10,\, \ldots
\label{iqhegraph1.3}
\end{equation}
by sweeping the field or the carrier density through the Landau levels 
\cite{novo2005,zhan2005}. This implies Hall resistance quantization for filling 
factors in the sequence 
\begin{equation}
\nu = \frac{h n_e }{eB} = 4(n+\tfrac12) = 4n + 2,
\label{iqhegraph1.2}
\end{equation}
where $n_e$ is the charge carrier density and $n$ is the Landau level index 
\cite{novo2005,feld2012}. This sequence is quite different from the integral 
quantum Hall effect sequence observed in other 2D electron gases.  However, it 
likely results from the same basic physics as the normal IQHE, modified by 
the 4-fold spin--valley degeneracies of non-interacting, massless Dirac 
electrons.

The period $\Delta\nu = 4$ in \eq{iqhegraph1.2} is is a consequence of the 
approximate four-fold degeneracy of the graphene Landau level, and the added 
$\ttfrac12$ (which is not present in non-relativistic 2D systems) is a Berry 
phase effect that results from the special status of the $n=0$ state for 
massless Dirac fermions: a quantum phase arises at the band degeneracy point 
associated with precession of the pseudospinor describing the 2-fold sublattice 
degree of freedom, which modifies the quantization condition for electronic 
orbits \cite{zhan2005,ando1998,miki1999}.  The four-fold near-degeneracy of the 
Landau level follows because the Zeeman energy is small compared with the 
interaction energy, and the pseudospin degree of freedom representing the two 
inequivalent Dirac cones at the corners of the Brillouin zone (K and K$^\prime$) 
does not couple to external fields if the two sublattices are equivalent.

\subsection{\label{sh:fractionalQHEgraphene}Fractional Quantum Hall States}

Because of the Dirac-cone dispersion with the Fermi level located at the apex of 
the cones (see \fig{dispersionGraphene}), low-energy excitations in graphene 
occur in regions of reduced electron density, which disfavors electron 
correlations.  But by placing a strong perpendicular magnetic field on the 
system the resulting Landau quantization (corresponding semiclassically to 
requiring that an integral number of deBroglie wavelengths wrap around a 
cyclotron orbit, and implying that an integral number of magnetic flux quanta 
pass through the area bounded by the orbit) leads to a bunching of levels into 
regions of locally high degeneracy separated by gaps.  These regions of locally 
high level density may exhibit conditions  more favorable for the development of 
strong correlations between electrons.

Landau levels (LL) become strongly correlated when inter-LL excitations are of 
sufficiently high energy that they may be neglected and  the low-energy 
excitations involve only intra-LL transitions.  Then the kinetic energy of the 
LL is a constant that can be omitted. This limit of strong electronic 
correlations has two important physical implications: (1)~The approximate 4-fold 
spin--valley degeneracy of the graphene Landau levels leads to quantum Hall 
ferromagnetic states that will be discussed further below. (2)~The strong 
correlations can produce incompressible states at partial filling of the LL that 
are reminiscent of the fractional quantum Hall effect (FQHE) in semiconductor 
devices.

After the discovery of the integral quantum Hall effect in graphene, experiments 
performed at higher magnetic field strengths observed fragile quantum Hall 
states at filling factors $0,\,\pm 1, \pm 4$.  The $\pm4$ states are thought to 
be the result of single-particle Zeeman splitting of the Landau levels but the 
states at filling factors $0, \pm1$ are thought to be caused by 
electron--electron interactions breaking degeneracies of the $n=0$ Landau level 
\cite{bolo2009}.

Quantum Hall incompressible states having actual fractional filling were later 
observed at filling factors such as $\nu = \tfrac13, \tfrac23, 
\tfrac43, \ldots$   \cite{bolo2009,du2009,feld2012}. The fractional states  
follow the standard composite fermion model \cite{jain1989} sequence for filling 
factors $\nu = 0-1$, but only even-numerator fractions are seen for $\nu = 1-2$. 
These sequences, and the energy gaps for the corresponding incompressible 
states, are thought to reflect the interplay of strongly-correlated chiral 
electrons and the characteristic internal symmetries of graphene that will be 
discussed further below \cite{feld2012}.

\subsection{\label{sh:classificationQHEgraphene}Classification of Quantum Hall 
States in Graphene}

As has been seen, in the quantum Hall effect for 2D electron gases produced in 
semiconductor devices, two fundamental classes of incompressible states are 
found:  

\begin{enumerate}
\item 
Those where the requisite gaps are produced by complete filling of 
Landau levels (Fermi energy lying between Landau levels) that are explained 
by weakly-interacting electrons subject to impurity-scattering localization.  
\item
Those where the gaps are produced by strong electron correlations 
within a partially-filled Landau level.
\end{enumerate}

\noindent
For the normal semiconductor quantum Hall effect, the first case is commonly 
termed the integral quantum Hall effect (IQHE), because it leads uniquely to 
quantum Hall plateaus at integral filling factors, and the second case is termed 
the fractional quantum Hall effect (FQHE), because it is characterized uniquely 
by quantum Hall plateaus at fractional values of the filling factor.

In graphene, the anomalous counting implied by the four-fold spin--valley 
degeneracy for massless chiral fermions modifies this correspondence.  One may 
again assume incompressible states divided into those that have gaps produced by 
weakly-interacting electrons filling Landau levels completely and those that 
have gaps produced by correlations within a partially filled level.  However 
(1)~states of the first class occur at integral fillings but the integers of 
\eq{iqhegraph1.2} are not those of the normal IQHE because of the anomalous 
counting implied by graphene's internal structure.  (2)~States of the second 
class may have either integral or fractional filling factors because of the 
anomalous counting.  

Thus IQHE and FQHE may be used as shorthand labels in graphene, but it should be 
understood that what is meant by ``IQHE'' in graphene is the formation of 
incompressible states by weakly-interacting electrons completely filling Landau 
levels, and by ``FQHE''  the formation of incompressible states by 
strongly-correlated electrons in a partially-filled level, irrespective of 
whether the observed filling factors are integers or rational fractions.   Our 
primary interest in this paper lies in those incompressible states formed by 
electron--electron and electron--lattice correlations in partially-filled Landau 
levels, and thus in graphene ``FQHE'' states.

\section{\label{h:symmetriesFQHEgraphene}Quantum Hall Symmetries in 
Graphene}

In the normal two-dimensional electron gas produced in semiconductor devices the 
Landau levels can contain $eB/h$ states, where $e$ is the electronic charge and 
$B$ is the magnetic field. As has been seen, in graphene there is an additional 
4-fold degeneracy associated with the spin and valley degrees of freedom. It is 
common to unite these four degrees of freedom  in terms of an SU(4) symmetry 
that is termed {\em quantum 
Hall ferromagnetism (QHFM).}

\subsection{\label{sh:su4FQHE}SU(4) Quantum Hall States}

SU(4) symmetry for graphene in a strong magnetic field is expected when all four 
spin and valley levels are degenerate.  Two conditions must be satisfied 
for this condition to be fulfilled.
\begin{enumerate}
\item 
Landau-level mixing caused by inter-LL electronic transitions must be 
negligible.
\item
Perturbations within a single LL that break the 4-fold spin--isospin symmetry 
must be small.
\end{enumerate}
The resulting theory predicts quantum Hall states that have no analog in 
semiconductor heterostructures.  Let us study these states by introducing an 
effective low-energy Hamiltonian exhibiting approximate SU(4) symmetry.

\subsubsection{\label{ss:spin-valleyOperators}Spin and Valley Isospin Operators}

The two-dimensional electronic spin degree of freedom and the two-dimensional 
electronic valley (K) degree of freedom are most elegantly expressed in terms of 
independent spin and valley isospin states. For the spin, introduce the Pauli 
matrix vector $\vec\sigma = (\sigma_x,\sigma_y,\sigma_z)$, with the standard 
representation in terms of the $2\times 2$ matrices
\begin{equation}
\sigma_x = \twomatrix0110
\quad
\sigma_y = \twomatrix{0}{-i}{i}{0}
\quad
\sigma_z = \twomatrix{1}{0}{0}{-1} ,
\label{algebra1.5}
\end{equation}
which obey the SU(2) Lie algebra
\begin{equation}
\comm{\frac{\sigma_i}{2}}{\frac{\sigma_j}{2}}
= i\epsilon_{ijk} \frac{\sigma_k}{2},
\label{su2comm}
\end{equation}
where $\epsilon_{ijk}$ is the completely antisymmetric 3rd-rank tensor. The 
matrices \eqnoeq{algebra1.5} are assumed to operate on a spinor basis of spin-up 
and spin-down electrons
\begin{equation}
\ket\uparrow = \onematrix 10
\qquad
\ket\downarrow = \onematrix 01 .
\label{algebra1.6}
\end{equation}

A set of equations in the valley isospin space completely analogous to Eqs.\ 
\eqnoeq{algebra1.5}--\eqnoeq{algebra1.6} for the spin space results if one 
defines the SU(2) Pauli-matrix representation for the valley isospin operators
$\vec\tau = (\tau_x,\tau_y,\tau_z)$,
\begin{equation}
\tau_x = \twomatrix0110
\quad
\tau_y = \twomatrix{0}{-i}{i}{0}
\quad
\tau_z = \twomatrix{1}{0}{0}{-1} ,
\label{algebra1.7c}
\end{equation}
which operate on the valley isospinor basis
\begin{equation}
\ket K = \onematrix 10
\qquad
\ket{K'} = \onematrix 01 .
\label{algebra1.7e}
\end{equation}
The operators $\vec \sigma$ and $\vec \tau$ may now be used to define an 
effective low-energy Hamiltonian having Landau-level and internal spin and 
valley isospin degrees of freedom.

\subsubsection{\label{ss:effectiveHam}Effective Low-Energy Hamiltonian}

The two largest energy scales for graphene in a strong magnetic field are the 
Landau-level  separation and the Coulomb energy.  At the charge neutral point 
(Fermi energy for undoped graphene) the LL separation is approximately three 
times larger than the Coulomb energy, which is in turn considerably larger than 
any additional terms in the interaction.  Therefore, a strategy is adopted here 
of ignoring excitations between Landau levels and projecting onto the $n=0$ LL. 
At a quantitative level the Landau level mixing cannot be ignored (see the 
discussion in Ref.\ \cite{khar2012}), but such an approximation gives the 
correct qualitative physics and the effect of excluded Landau levels can be 
included to some degree by parameter renormalization, which will be sufficient 
for our purposes.  

Within this single Landau level the Hamiltonian is assumed to be dominated by a 
long-range Coulomb interaction that is SU(4) symmetric, with shorter-range 
interactions in spin and valley degrees of freedom (originating in both screened 
electron--electron interactions and electron--phonon interactions) causing SU(4) 
symmetry breaking. To implement this  a graphene Hamiltonian projected onto the 
$n=0$ Landau level is adopted that was proposed in Ref.\ \cite{khar2012} and 
employed further in Ref.\ \cite{wufe2014},%
\begin{equation}
 H = 
 H\tsub C + H\tsub v + H\tsub Z ,
 \label{so5_1.1a}
\end{equation}
where the valley-independent Coulomb interaction $H\tsub C$ may be expressed as
\begin{equation}
 H\tsub C =
 \tfrac12 \, \sum_{i \ne j} \, \frac{e^2}{\epsilon|\bm r_i - \bm r_j|} ,
 \label{so5_1.1b}
\end{equation}
$H\tsub v$ is the short-range, valley-dependent interaction, 
\begin{equation}
 H\tsub v =
 \tfrac12 \sum_{i\ne j}\left[
 g_z\tau_z^i \tau_z^j + g_{\perp} (\tau_x^i \tau_x^j + \tau_y^i \tau_y^j)
 \right]
 \delta(\bm r_i - \bm r_j) ,
 \label{so5_1.1c}
\end{equation}
where the Pauli matrices $\tau_\alpha$ operate on the valley isospin 
and $g_z$ and $g_\perp$ are coupling constants, and the Zeeman energy $H\tsub 
Z$ is given by
\begin{equation}
 H\tsub Z = -\mu\tsub B B \sum_i \sigma_z^i,
 \label{so5_1.1d}
\end{equation}
where $\mu\tsub B$ is the Bohr magneton, $B$ is the magnetic field strength, 
the Pauli matrices $\sigma_\alpha$ operate on the electronic spin degrees 
of freedom, and the $z$ direction for the spin space is chosen to be aligned 
with $B$.

\subsubsection{\label{ss:symmetrySU4} Symmetries of the Effective Hamiltonian}

Letting $\alpha = (x, y, z)$ and $\beta=(x,y)$, the set of 15 operators
\begin{subequations}
\begin{align}
 \spin_\alpha &=
 \sum_{m_k}
\sum_{ \tau \sigma \sigma'} \mel{\sigma'}{\sigma_\alpha}{\sigma}
c^\dagger_{\tau\sigma'm_k} c_{\tau\sigma m_k}
\label{algebra1.4a}
\\
T_\alpha &=
\sum_{m_k}\sum_{\sigma\tau\tau'} \mel{\tau'}{\tau_\alpha}{\tau}
c^\dagger_{\tau'\sigma m_k} c_{\tau\sigma m_k}
\label{algebra1.4b}
\\
N_\alpha &=
\tfrac12 \sum_{m_k}\sum_{\sigma \sigma' \tau }
\mel{\tau}{\tau_z}{\tau}
\mel{\sigma'}{\sigma_\alpha}{\sigma} 
c^\dagger_{\tau\sigma' m_k}
c_{\tau\sigma m_k}
\label{algebra1.4c}
\\
\Piop\alpha\beta &= 
\tfrac12 \sum_{m_k}\sum_{ \sigma \sigma' \tau \tau'}
\mel{\tau'}{\tau_\beta}{\tau}
\mel{\sigma'}{\sigma_\alpha}{\sigma}
c^\dagger_{\tau' \sigma' m_k}
c_{\tau\sigma m_k} 
\label{algebra1.4d}
\end{align}
\label{algebra1.4}%
\end{subequations}
is closed under commutation, defining an SU(4) Lie algebra that commutes with 
the Coulomb interaction $H\tsub C$ \cite{khar2012,wufe2014}. Thus, if
$H\tsub v$ and $H\tsub Z$  are small compared with  $H\tsub C$  in 
\eq{so5_1.1a}, 
the Hamiltonian will have an approximate SU(4) invariance.  The operator 
$\boldspin$ represents the total spin and the operator $\vec T$ represents 
the total valley pseudospin. In the $n=0$ Landau level for graphene there is an 
equivalence between valley and sublattice degrees of freedom, so $\vec N$ can 
be viewed as a N\'eel vector in the $n=0$ Landau level measuring the 
difference in spins on the A and B sublattices. The operators 
$\Piop\alpha\beta$ 
coupling  spin and valley isospin will be discussed further below.

\subsubsection{\label{ss:symmetryBreakingSU4} Explicit Symmetry Breaking}

The occurrence of SU(4) symmetry and its explicit symmetry-breaking pattern 
depend on the values of the effective coupling parameters $g_z$ and $g_\perp$. 
For the lattice spacings found in graphene, each can be estimated to be 
considerably smaller than the SU(4)-symmetric Coulomb term, so one may expect 
SU(4) to be broken only weakly by explicit terms in realistic systems. Four 
basic explicit symmetry-breaking patterns have been identified 
\cite{khar2012,wufe2014}:

1.~For arbitrary non-zero values of $g_z$ and $g_\perp$, 
the symmetry is broken 
to 
\begin{equation}
{\rm SU(4)} \supset {\rm SU(2)}\tsub s \times 
{\rm U(1)}\tsub v \supset {\rm U(1)}\tsub s \times {\rm U(1)}\tsub v,
\label{so5_1.13}
\end{equation}
where $SU(2)\tsub s$ is associated with global conservation of spin and 
$U(1)\tsub s$ with conservation of its $z$ component, and
$U(1)\tsub v$ is associated with conservation of the $T_z$ component of the 
valley isospin. (Conservation of $T_z$ implies physically that the difference 
in electronic densities between the $K$ and $K'$ sites is invariant, which 
might be expected to be true for low-energy states having minimal scattering 
between valleys.)  In the absence of Zeeman splitting spin is conserved, but 
only the $z$ component of the valley isospin is conserved.  The full 
Hamiltonian \eqnoeq{so5_1.1a} including the Zeeman term 
conserves only the $z$ components of the spin and valley isospin.

2.~If $g_\perp=0$ but $g_z \ne 0$, the symmetry is broken to
\begin{align}
{\rm SU(4)} &\supset {\rm SU(2)}_{\rm\scriptstyle s}^{K} \times 
{\rm SU(2)}_{\rm\scriptstyle s}^{K'} \times U(1)\tsub v
\nonumber
\\
&\supset {\rm U(1)}_{\rm\scriptstyle s}^{K} \times
{\rm U(1)}_{\rm\scriptstyle s}^{K'} \times {\rm U(1)}\tsub v.
 \label{so5_1.14}
\end{align}
In the absence of Zeeman splitting this corresponds to conserving 
independent spin rotations for the wavefunction in each valley labeled by $K$ 
and $K'$, but breaking of the valley isospin  to U(1)$\tsub v$. The 
full 
Hamiltonian \eqnoeq{so5_1.1a} including the Zeeman term conserves only the $z$ 
components of the spin in each valley separately, and the $z$ component of the 
valley isospin.

3.~If $g\tsub z = g_\perp \ne 0$, the symmetry is broken to
\begin{equation}
{\rm SU(4)} \supset
 {\rm SU(2)}\tsub s \times {\rm SU(2)}\tsub v
 \supset {\rm U(1)}\tsub s \times {\rm SU(2)}\tsub v,
 \label{so5_1.15}
\end{equation}
corresponding to full spin and valley isospin rotational symmetry in the absence 
of Zeeman splitting. The complete Hamiltonian \eqnoeq{so5_1.1a} including the 
Zeeman term conserves the  SU(2) isospin symmetry but only the $z$ component 
of the spin.

4.~If $g_\perp = -g_z \ne 0$, the Hamiltonian commutes with  
$\Piop\alpha\beta$, $\boldspin$, and $T_z$, and these 10 operators generate 
the 
Lie group SO(5), so \cite{wufe2014}
\begin{equation}
{\rm SU(4)} \supset {\rm SO(5)} \supset {\rm U(1)}\tsub s \times {\rm SU(2)}_z,
 \label{so5_1.16}
\end{equation}
where the $\sutwo_z$ symmetry is generated by $(T_z, \Piop zx, \Piop zy)$. Thus, 
in the absence of Zeeman splitting the system exhibits an SO(5) symmetry 
involving both spin and valley isospin. The full Hamiltonian \eqnoeq{so5_1.1a} 
including the Zeeman term conserves the $z$ component of spin and the $\sutwo_z$ 
symmetry. The SO(5) subgroup plays the role of a transitional symmetry 
connecting the N\'eel-like states associated with $N_\alpha$ and the states 
associated with valley degrees of freedom $T_x$ and $T_y$.

The subgroup structure for these four patterns of explicit SU(4) symmetry 
breaking is illustrated in \fig{dynamicalChains_QHF}.%
\doublefig
{dynamicalChains_QHF}  
{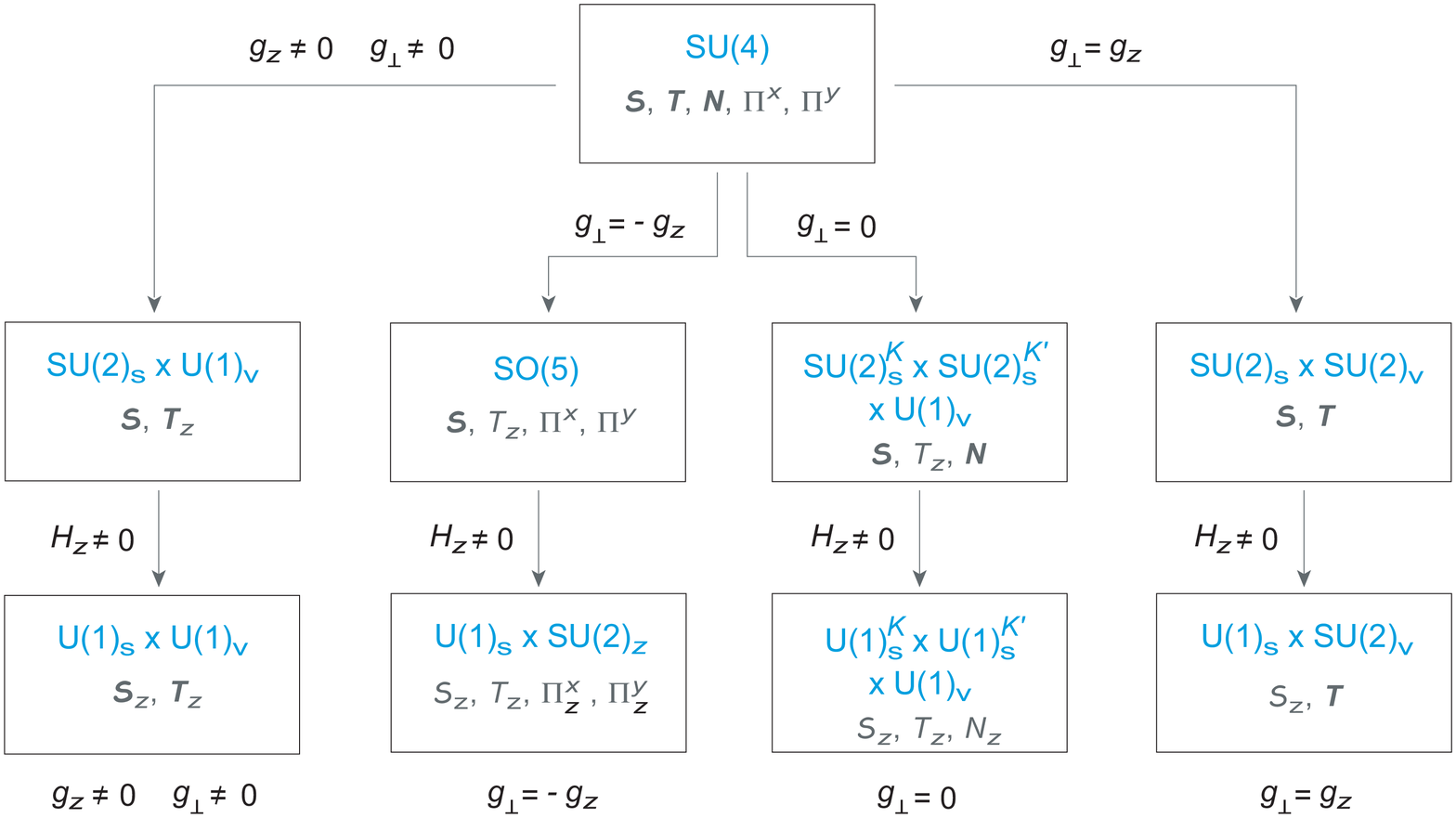}
{\figup}
{0pt}
{0.40}
{Explicit symmetry-breaking structure for SU(4) quantum Hall ferromagnetism 
described by the Hamiltonian \eqnoeq{so5_1.1a}.  
}
These symmetries and explicitly-broken symmetries have proven extremely useful 
in understanding the states of graphene in a strong magnetic field 
\cite{khar2012,wufe2014}.

\section{\label{h:fdsGraphene} Fermion Dynamical Symmetries}

Let us now consider using symmetries in an even more powerful way than that 
discussed in the preceding section.  Specifically, let us attempt to describe 
the quantum Hall ferromagnetic behavior of graphene using {\em fermion dynamical 
symmetries} of an effective Hamiltonian operating in a highly-truncated 
collective subspace. This has the potential to prescribe dynamics as well as 
taxonomy and conservation laws, within a many-body model having analytical 
solutions that illuminate the physics of quantum Hall states in graphene.

\subsection{\label{sh:symmetryGenerators} Symmetry Generators}

For clarity of discussion a valence space will be assumed corresponding to a 
single Landau level.  To avoid cluttered notation the index $n$ 
labeling the Landau level is suppressed and the fermion creation 
operator 
$c^\dagger_{\tau \sigma m_k}$ and the corresponding hermitian conjugate 
$c_{\tau \sigma m_k}$ are introduced. The index $\tau$ takes one of two values 
($\pm$) labeling 
the valley isospin projections corresponding to valleys $K$ or $K'$, the 
electron spin polarization $\sigma$ takes one of two values 
($\uparrow\downarrow$) labeling spin-up or spin-down, and $m_k$ is a quantum 
number distinguishing degenerate states within a given Landau level (typically 
an angular momentum in symmetric gauge or a linear momentum in Landau gauge). 
The fermionic operators $c^\dagger$ and $c$ are assumed to obey
\begin{equation}
 \{ c_\alpha^{\vphantom{\dagger}}, c^\dagger_\beta  \}
= \delta_{\alpha\beta} 
\qquad
\{c_\alpha^\dagger c_\beta^\dagger \}
=
\{c_\alpha^{\vphantom{\dagger}} c_\beta^{\vphantom{\dagger}} \} = 0
,
\label{anticommutator}
\end{equation}
where $\{a, b \} \equiv ab + ba$. By virtue of this anticommutator, the 
$c^\dagger_\alpha$ create a fermion in the state labeled by $\alpha$, the 
$c\phantomdagger_\alpha$ annihilate a fermion in the same state, and 
$n_\alpha=c^\dagger_\alpha c\phantomdagger_\alpha$ counts 
the number of fermions in the state labeled by $\alpha$.

The four states representing possible combinations of $\tau$ and 
$\sigma$ are displayed in \tableref{quantumNumberMapping},%
{\renewcommand\arraystretch{1.0}
\begin{table}[tpb]
  \vspace*{\tabup}
  \centering
  \caption{Quantum numbers}
  \label{tb:quantumNumberMapping}
    \begin{centering}
      \setlength{\tabcolsep}{6 pt}
      \vspace{\tabtitlesep}
      \begin{tabular}{ccccc}
      
        \tableLineOne
        
            Valley &
            $\tau$ &
            $\sigma$ &
            $m_i$ &
            $a$

        \tableLineTwo
        
            $K$ &
            $+$ &
            $\uparrow$ &
            $+\tfrac32$ &
            $1$

        \\ [\tablerowMore]  
        
            $K$ &
            $+$ &
            $\downarrow$ &
            $+\tfrac12$ &
            $2$

        \\  [\tablerowMore]     
        
            $K'$ &
            $-$ &
            $\uparrow$ &
            $-\tfrac12$ &
            $3$

        \\  [\tablerowMore]  
        
            $K'$ &
            $-$ &
            $\downarrow$ &
            $-\tfrac32$ &
            $4$

        \tableLineThree
        
      \end{tabular}
    \end{centering}
\end{table}
}%
and correspond physically to the four possible combinations 
of the electron being in either the $K$ or $K'$ valley with either spin up or 
spin down (see \fig{LLgraphene_n0}b).  For brevity  the label 
$a = 1,2,3,4$ displayed in the last column of the table will often be used to 
distinguish these 
states.  The four basis states labeled by $a$ 
are illustrated graphically in \fig{grapheneBasis}.
\singlefig
{grapheneBasis}       
{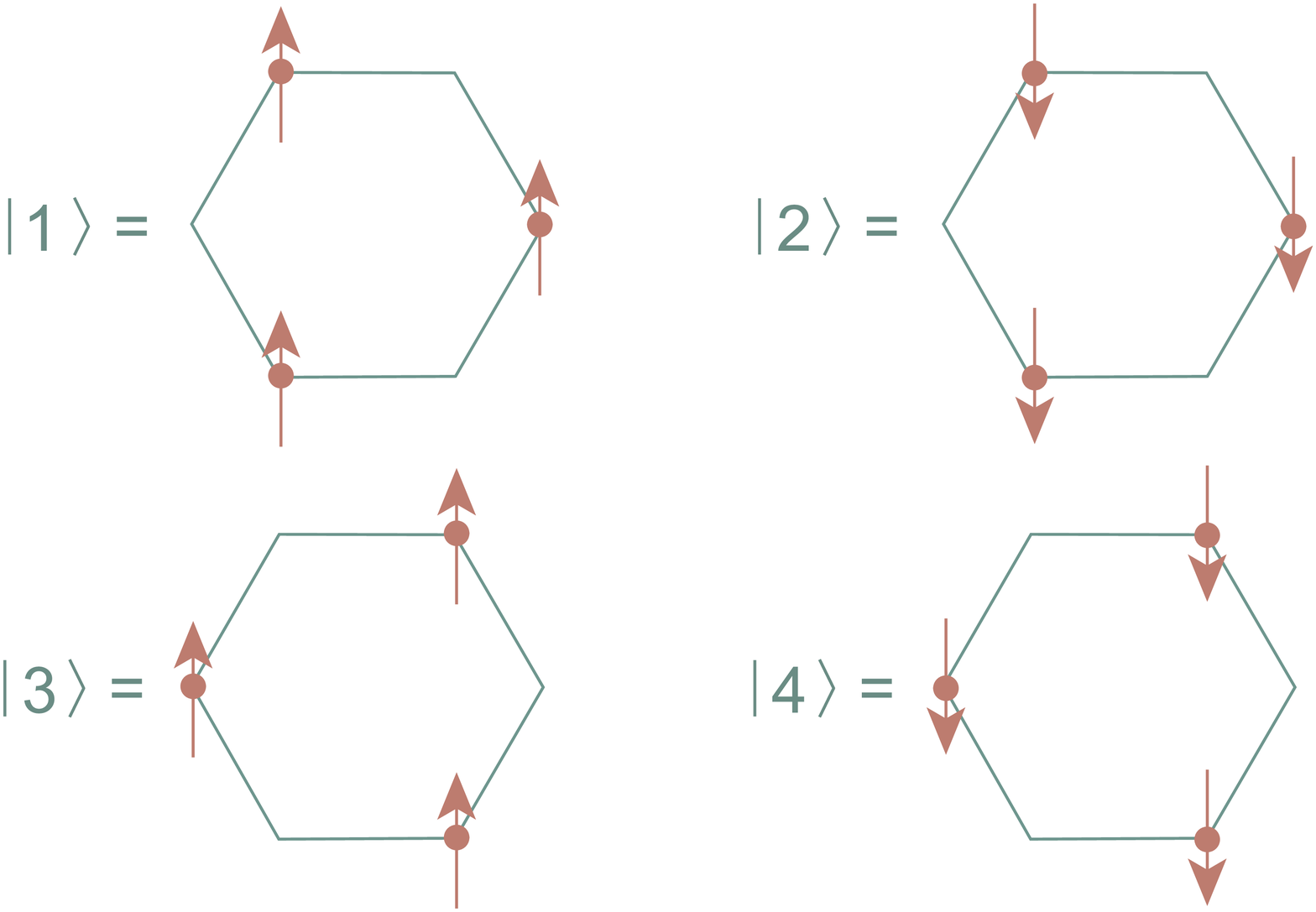}    
{\figup}         
{0pt}         
{0.16}         
{The four isospin--spin basis vectors $\ket a$ of 
\tableref{quantumNumberMapping}.}
\tableref{quantumNumberMapping} also displays a unique mapping of these four 
states to a label $m_i$ that takes the four possible projection quantum numbers 
$\left\{ \pm \ttfrac12, \pm \ttfrac32 \right\}$ of a fictitious angular momentum 
$i=\ttfrac32$; the motivation for this mapping will become apparent later.

Let us now introduce an operator $A^\dagger_{ab}$ that creates a pair of 
electrons, one in the $a=(\tau_1,\sigma_1)$ level and one in the 
$b=(\tau_2,\sigma_2)$ level, with the total $m_k$ of each pair coupled to zero 
term by term,
\begin{equation}
A^\dagger_{ab}  = \sum_{m_k} c^\dagger_{a m_k} c^\dagger_{b 
-m_k} ,
\label{algebra1.1}
\end{equation}
and its hermitian conjugate $A_{ab}$, which annihilates a corresponding electron 
pair.  Each index $a$ or $b$ ranges over four values, implying  16 
components in \eq{algebra1.1}.  However, the pair wavefunction must satisfy the 
Pauli principle which, upon expanding the indices $a$ and $b$ using
\tableref{quantumNumberMapping}, eliminates the four diagonal $(a=b)$ 
possibilities.  
Furthermore, because of the antisymmetry requirement 
the pair 
creation operators are constrained by
$
A^\dagger_{ab} = - A^\dagger_{ba},
$
implying that only half of the remaining 12 operators $A^\dagger$ are 
independent.  Thus \eq{algebra1.1} defines six independent operators 
$A^\dagger$, with six independent hermitian conjugates $A$. Let us introduce in 
addition to these pairing operators the 16 particle--hole operators $B_{ab}$ 
through 
\begin{equation}
B_{ab} = \sum_{m_k} c^\dagger_{am_k} c\phantomdagger_{bm_k} - \tfrac14 
\delta_{ab}\phantomdagger \Omega ,
\label{algebra1.2}
\end{equation}
where $\delta_{ab}$ is the Kronecker delta and $\Omega$ is the total degeneracy 
  the single Landau level (see \eq{degen1.2}). The 
commutators for 
the 28 operators $A$, $A^\dagger$, and $B$  are found to be \cite{chen86}
\begin{subequations}
\begin{align}
\comm{A_{ab}}{A^\dagger_{cd}} &=
-B_{db} \delta_{ac} - B_{ca}\delta_{bd}
+ B_{cb} \delta_{ad} + B_{da} \delta_{bc}
\label{algebra1.3a}
\\
\comm{B_{ab}}{B_{cd}} &=
\delta_{bc} B_{ad} - \delta_{ad}B_{cb}
\label{algebra1.3b}
\\
\comm{B_{ab}}{A^\dagger_{cd}} &=
\delta_{bc} A^\dagger_{ad} + \delta_{bd} A^\dagger_{ca}
\label{algebra1.3c}
\\
\comm{B_{ab}}{A_{cd}} &=
-\delta_{ac} A_{bd} - \delta_{ad} A _{cb} ,
\label{algebra1.3d}
\end{align}
\label{algebra1.3}%
\end{subequations}
which is isomorphic to an SO(8) Lie algebra. Thus, the 28 members of the 
operator set $(A, A^\dagger, B)$ exhibit SO(8) symmetry under commutation. The 
ultimate justification for introducing this SO(8) Lie algebra will lie in the 
results to which it will lead.  However, a general discussion of why the known 
physics of graphene suggests the efficacy of such an algebra in understanding 
the possible collective modes of the system may be found in Appendix 
\ref{h:appendixA}.

\subsection{\label{sh:relationSU4Symmetry} Relationship with Standard Graphene 
SU(4) Symmetry}
 
The SU(4) generators \eqnoeq{algebra1.4} may  
be 
expressed in terms of the operators \eqnoeq{algebra1.2} by employing Eqs.\ 
\eqnoeq{algebra1.5}--\eqnoeq{algebra1.7e} in \eq{algebra1.4} and utilizing the 
equivalences in \tableref{quantumNumberMapping} to expand the 
indices for $B_{ab}$ from \eq{algebra1.2}.  For example, consider the spin 
operator $\spin_y$. From \eq{algebra1.4a} one may write
\begin{align*}
\spin_y &=
 \sum_{m_k}
\sum_{ \tau \sigma \sigma'} \mel{\sigma'}{\sigma_y}{\sigma}
c^\dagger_{\tau\sigma'm_k} c_{\tau\sigma m_k}\phantomdagger
\\
&=
\sum_{m_k}
\left(
-i c^\dagger_{+\uparrow m_k} c_{+\downarrow m_k}\phantomdagger
+ i c^\dagger_{+\downarrow m_k} c_{+\uparrow m_k}\phantomdagger
\right.
\\[-1ex]
&\left. \qquad\qquad\qquad -i c^\dagger_{-\uparrow m_k} c_{-\downarrow 
m_k}\phantomdagger
+ i c^\dagger_{-\downarrow m_k} c_{-\uparrow m_k}\phantomdagger
\right)
\\
&= -iB_{12} + i B_{21} -i B_{34} +i B_{43}.
\end{align*}
By such methods one finds that the spin operators 
\eqnoeq{algebra1.4a} may be expressed in terms of the $B_{ab}$ generators of 
SO(8) as
\begin{subequations}
\begin{align}
\spin_x &=
B_{12} + B_{21} + B_{34} + B_{43}
\label{algebra1.8a}
\\
\spin_y &=
-i\left(B_{12} - B_{21} + B_{34} - B_{43} \right)
\label{algebra1.8b}
\\
\spin_z &=
B_{11} - B_{22} + B_{33} - B_{44} ,
\label{algebra1.8c}
\end{align}
\label{algebra1.8}
\end{subequations}
the valley isospin operators \eqnoeq{algebra1.4b} as
\begin{subequations}
\begin{align}
T_x &=
B_{13} + B_{31} + B_{24} + B_{42}
\label{algebra1.9a}
\\
T_y &=
-i\left(B_{13} - B_{31} + B_{24} - B_{42} \right)
\label{algebra1.9b}
\\
T_z &=
B_{11} + B_{22} - B_{33} - B_{44} ,
\label{algebra1.9c}
\end{align}
\label{algebra1.9}
\end{subequations}
The N\'eel vector of \eq{algebra1.4c} as
\begin{subequations}
\begin{align}
N_x &=
\tfrac12 \left( B_{12} + B_{21} - B_{34} - B_{43} \right)
\label{algebra1.10a}
\\
N_y &=
-\tfrac i2 \left(B_{12} - B_{21} - B_{34} + B_{43} \right)
\label{algebra1.10b}
\\
N_z &=
\tfrac12 \left( B_{11} - B_{22} - B_{33} + B_{44} \right),
\label{algebra1.10c}
\end{align}
\label{algebra1.10}
\end{subequations}
and the operators $\Piop\alpha\beta$ of \eq{algebra1.4d} as
\begin{subequations}
\begin{align}
\Piop xx &=
\tfrac12 \left( B_{14} + B_{41} + B_{23} + B_{32} \right)
\label{algebra1.11a}
\\
\Piop yx &=
-\tfrac i2 \left(B_{23} - B_{32} + B_{41} - B_{14} \right)
\label{algebra1.11b}
\\
\Piop zx &=
\tfrac12 \left( B_{13} + B_{31} - B_{24} - B_{42} \right)
\label{algebra1.11c}
\\
\Piop xy &=
-\tfrac i2 \left( B_{32} - B_{23} + B_{41} - B_{14} \right)
\label{algebra1.11d}
\\
\Piop yy &=
-\tfrac 12 \left(B_{41} + B_{14} - B_{23} - B_{32} \right)
\label{algebra1.11e}
\\
\Piop zy &=
-\tfrac i2 \left( B_{31} - B_{13} - B_{42} + B_{24} \right) .
\label{algebra1.11f}
\end{align}
\label{algebra1.11}%
\end{subequations}
The inverse transformations expressing the $B_{ab}$ in terms of the 
$\{\spin_\alpha,\,  T_\alpha,\, N_\alpha, \,\Piop \alpha x, \,\Piop 
\alpha y\}$ are given in Appendix \ref{h:appendixB}.

Hence the SU(4) algebra generated by the operators in \eq{algebra1.4} is a 
subalgebra of the SO(8) algebra, with its generators corresponding to particular 
linear combinations of the subset of SO(8) generators defined by the 
particle--hole operators $B_{ab}$ in \eq{algebra1.2}. In fact, the present 
formalism also could be constructed by starting with the SU(4) algebra generated 
by the $B_{ab}$ operators and expanding that to the SO(8) algebra of 
\eq{algebra1.3} by adding pairing operators, using the motivation discussed in 
Appendix \ref{h:appendixA}.

\section{\label{h:pairRepresentations} Pair Representations}

One wishes to investigate possible collective modes for graphene electrons 
undergoing strong correlations within a single Landau level.  Pairs of fermions 
often afford a convenient basis for discussing collective states, so let us 
consider some possible configurations involving pairs of electrons  in graphene.

\subsection{\label{sh:degeneraciesFilling}Degeneracies and Level Filling}

Let us first consider the degeneracies of undoped graphene placed in a strong 
magnetic field, confining the discussion to the case of a single 
Landau level for simplicity .  The single-particle states within the 
Landau 
level will be assumed  labeled by the quantum numbers  $(n,\llm)$, where 
$n$ is the 
principle quantum number labeling the Landau level and $\llm$ is a quantum 
number distinguishing the degenerate states within the Landau level. 
In the absence of spin and valley degrees of freedom, the states 
$(n,\llm)$ of the Landau level are assumed to hold a maximum of $2\kdegen$ 
electrons (for consistency with earlier applications of the SO(8) 
algebra in nuclear physics,  $\kdegen$ will be defined to be the electron pair 
degeneracy, so that $2\kdegen$ is the electron degeneracy).  From the solution 
of the Dirac equation in a magnetic field
\begin{equation}
2\kdegen  = \frac{BS}{(h/e)} ,
\label{degen1.1}
\end{equation}
where $B$ is the strength of the magnetic field, $S$ is the area of the 
two-dimensional sample and
$
h/e  = 4.136 \times 10^{-15} \units {Wb}
$
defines the magnetic flux quantum.
But graphene has in addition $2\times 2 = 4$ 
internal degrees of freedom associated with the $\ket{\rm spin}\otimes\ket{\rm 
isospin}$ space.  Thus there are four copies of each Landau level in 
graphene and  the total electron degeneracy $2\Omega$ is given by
\begin{equation}
2\Omega = 4(2\kdegen) = \frac{4BS}{(h/e)}.
\label{degen1.2}
\end{equation}
Some pair degeneracies calculated from \eq{degen1.2} as a 
function of domain size and magnetic field strength 
are displayed in \tableref{degeneracyTable},%
{\renewcommand\arraystretch{1.0}
\begin{table}[tpb]
  \vspace*{\tabup}
  \centering
  \caption{Pair degeneracies $\Omega$ for  magnetic field strengths $B$}
  \label{tb:degeneracyTable}
    \begin{centering}
      \setlength{\tabcolsep}{8 pt}
      \vspace{\tabtitlesep}
      \begin{tabular}{cccc}
      
        \tableLineOne
        
            Domain size &
            $B=2\units T$ &
            $B=10\units T$ &
            $B=50\units T$

        \tableLineTwo
        
            $5\ \mu{\rm m}\times 5\ \mu{\rm m}$ &
            $24,150$ &
            $120,750$ &
            $603,750$

        \\ [\tablerowMore]  
        
            $10\ \mu{\rm m}\times 10\ \mu{\rm m}$ &
            $96,660$ &
            $483,000$ &
            $2.145 \times 10^6$

        \tableLineThree
        
      \end{tabular}
    \end{centering}
\end{table}
}%
where it is assumed that the collective wavefunction is delocalized over the 
entire domain size.

The {\em fractional occupation} $\shellfill$ of the single Landau level  [not to 
be confused with the filling factor $\nu$ given in Eqs.\ 
\eqnoeq{iqhegraph1.1} and \eqnoeq{fillingFactor}] may  be defined as
\begin{equation}
\shellfill \equiv \frac{n}{2\Omega} = \frac{N}{\Omega},
\label{fillingFraction}
\end{equation}
where $n$ is the number of electrons and $N=n/2$ is the number of electron 
pairs. For half-filling of the $n=0$ Landau level located at the Fermi surface 
(corresponding to the ground state of undoped graphene) the electron number 
$n\tsub{gs}$ is then
\begin{equation}
n\tsub{gs}=\Omega = \frac{2BS}{(h/e)}.
\label{degen1.3}
\end{equation}
These degeneracies and occupation numbers are just the standard results for 
relativistic Landau levels in a 2D electron gas subject to a strong 
perpendicular magnetic field, but modified by the graphene spin and valley 
degeneracies.

Graphene exhibits both integral and fractional quantum Hall effects but the 
filling factors are anomalous relative to those for standard quantum Hall 
effects in semiconductor heterostructures. This is because of

\begin{enumerate}
 \item 
 The 4-fold degeneracy associated with the spin and valley degrees of 
freedom, which introduces factors of four in the counting.
\item
The nature of the Dirac solution, illustrated in \fig{LLgraphene_n0}, for which 
the negative-energy solutions may be interpreted as electron holes, the 
positive-energy solutions as electrons, and the $n=0$ level is unique, being 
half-filled in the neutral ground state (equivalently, it may be thought of as 
being shared equally by particles and holes).
\end{enumerate}

\noindent
Because of the particle-hole symmetry,  the charge carriers change sign near 
the Dirac points
and the Hall conductivity vanishes at charge neutrality (the electron number 
density tends to zero at a Dirac point). For this reason, the filling factor for 
graphene must be defined {\em relative to the charge-neutral state.} At charge 
neutrality the $n=0$ Landau level is half filled and when the $n=0$ LL is 
completely full the filling factor is  $4\times (1/2) = 
2$, from \eq{iqhegraph1.2}. The quantum Hall filling factor $\nu$ may be related 
to the Landau level fractional occupation  $\shellfill$ employed in the present 
formalism by
\begin{equation}
\nu = 4(\shellfill - \tfrac12)  = 4\left( \frac{n}{2\Omega} - \frac12 \right) .
\label{fillingFactor}
\end{equation}
Therefore, half filling of the $n=0$ Landau level corresponds to a fractional 
occupation $\shellfill = \tfrac12$ but to a filling factor $\nu=0$,
$\nu=-2$ 
corresponds to $f=0$ (completely empty), $\nu=-1$ corresponds to 
$f=\tfrac14$ filling, $\nu=+1$ corresponds to $f=\tfrac34$ filling, and $\nu=+2$ 
corresponds to $f=1$ (completely full).
%
%

\subsection{\label{sh:multipairStates}Many-Pair States}

Consider the states created by repeated application of the pair creation 
operator $A^\dagger_{ab}$ defined by \eq{algebra1.1} to the pair vacuum. It is 
useful to classify states according to a seniority-like quantum number $u$ 
defined to be the number of  particles in the system not coupled to one of the 
pairs defined in \eq{algebra1.1}.  The $u=0$ subspace will be of particular 
interest since it will contain states of maximal collectivity with respect to 
the pairs \eqnoeq{algebra1.1}.  An $N$-pair state in the $u=0$ subspace is given 
by
\begin{equation}
\AdagPower12 \AdagPower13 \AdagPower14 \AdagPower23 \AdagPower24 \AdagPower34 
\, \ket0  ,
\label{u0pairs1.0}
\end{equation}
where the total pair number $N$ is
\begin{equation}
N = \tfrac12 n = N_{12} + N_{13} + N_{14} + N_{23} + N_{24} + N_{34},
\label{u0pairs1.1}
\end{equation}
with $n$ giving the total number of electrons and $N_{ab}$ giving the number 
of electron pairs created by $\Adag ab$ operating on the vacuum state.  For our 
discussion here it will always be assumed that one is dealing with $u=0$ 
states, 
corresponding physically to no broken pairs.

\subsection{\label{sh:so8su4irreps} States in SO(8) $\supset$ SU(4) 
Irreducible Representations}

From \eq{algebra1.3}, the 16 operators $B_{ab}$ are closed under 
commutation and form a 
$
\ufour \supset 
\uone \times \sufour
$ 
subalgebra of the SO(8) algebra \eqnoeq{algebra1.3}. 
Let us now investigate the irreducible representations (irreps) that are 
associated with the 
$\soeight \supset \sufour$ subgroup chain in the $u=0$ representations.

\subsubsection{\label{ss:hwsState} The Highest-Weight State}

For $u=0$ at half filling, the number of pairs is $ N=\tfrac12 \Omega = 2k+1 $ 
and the highest-weight U(4) representation is  given by $(\tfrac\Omega2, 
\tfrac\Omega2, 0, 0)$.  Let us define a highest-weight (HW) state in the $u=0$ 
space, and choose it to correspond to the pair state with maximal value of $m_i$ 
from \tableref{quantumNumberMapping}, which results from placing one electron in 
the $a=1$ state and one electron in the $a=2$ state. Thus, for $N=2k+1$ pairs 
the highest weight state is given by 
\begin{align}
\ket{\rm HW} &=
\frac{1}{(2k+1)!} \left( \Adag12 \right)^{2k+1} \ket0
\nonumber
\\
&= \frac{1}{(2k+1)!} \left( \sum_{m_k}c^\dagger_{1 m_k}
c^\dagger_{2,-m_k}
\right)^{2k+1} \ket0
,
\label{hwstate}
\end{align}
where the sum runs over the $2k+1$ states in the Landau level labeled by the 
$m_k = (-k, -k+1, \ldots, k-1, k)$ quantum number.
The other states of the irreducible representation may then be created by the 
Cartan--Dynkin algorithm, which consists of using raising and lowering operators 
in the weight space to construct successively all the other states beginning 
with the highest weight state \cite{wyb74}.

The state in \eq{hwstate} appears to have a quite complex form, involving a sum 
with number of terms equal to the pair degeneracy of the Landau level raised to 
a power equal to the pair degeneracy (with the pair degeneracy typically a large 
number).  However, the actual structure of this state is considerably simpler 
than \eq{hwstate} would suggest because of the Pauli principle. As an 
illustrative example of this assertion, let's construct explicitly the 
highest-weight state for the case $k=1$, corresponding to $2k+1 = 3$ pairs in a 
single Landau level.  Writing the sum over $m_k = (-1,0,+1)$ in \eq{hwstate} out 
term by term gives
\begin{align*}
\ket{\rm HW} &=  \frac{1}{3!} \left( 
c_{1,-1}^\dagger
c_{21}^\dagger
+ 
c_{10}^\dagger
c_{20}^\dagger
+ 
c_{11}^\dagger
c_{2,-1}^\dagger
\right)^3 \ket0
\nonumber
\\
&=
c_{10}^\dagger \, c_{20}^\dagger \, c_{11}^\dagger \,
c_{21}^\dagger \, c_{1,-1}^\dagger \, c_{2,-1}^\dagger \,
\ket 0
=
\prod_{m_k=-k}^{m_k =+k}
c_{1m_k}^\dagger c_{2m_k}^\dagger \ket 0,
\end{align*}
where in raising the sum of operators 
inside the parentheses to the $2k+1=3$ power, all products containing two 
or more creation operators with the same index vanish because of the Pauli 
principle.  Similar considerations apply for arbitrary values of $k$ and in 
general the highest-weight state is given by
\begin{align}
\ket{\rm HW} &= \frac{1}{N!} 
(\Adag12)^{N} \ket 0
=
\frac{1}{N!}
\left(\sum_{m_k} c^\dagger_{1m_k} c^\dagger_{2 -m_k}\right)^{N}
\ket0 
\nonumber
\\
&=
 \prod_{m_k = -k}^{m_k=+k} c^\dagger_{1m_k} c^\dagger_{2m_k} \ket 0
,
\label{u0pairs1.4}
\end{align}
where the simplification in going from the second to the third line is a 
consequence of the antisymmetry  of the fermion creation operators (the Pauli 
principle) implied by \eq{anticommutator}. Thus the highest-weight state is a 
product state of pairs, one pair for each of the $N=2k+1$ levels labeled by 
$m_k$ in the Landau level.

\subsubsection{\label{sh:otherStates}Other SO(8) $\supset$ SU(4) States}

By the Cartan--Dynkin algorithm, other states in the $u=0$ subspace can be 
constructed by applying successively to the highest-weight state appropriate 
lowering and raising operators.  These will  be functions  
of the generators $B_{ab}$, so for an arbitrary state $\ket\psi$ in the 
weight space one has schematically
$
\ket\psi = F(B_{ab}) \ket{\rm HW}  ,
$
where the function $F(B_{ab})$ is specified by the Cartan--Dynkin procedure.
As an example, consider the action of the valley isospin lowering operator 
$T_-$ on the highest-weight state.  From Eqs.\ \eqnoeq{algebra1.9} and 
\eqnoeq{algebra1.2}, 
$$
T_- \equiv \tfrac12 (T_x - iT_y)
=
\sum_{m_k} (c_{{\scriptscriptstyle K'\uparrow} m_k}^\dagger 
c_{{\scriptscriptstyle K\uparrow} m_k}^{\vphantom{\dagger}} + 
c_{{\scriptscriptstyle K'\downarrow} m_k}^\dagger c_{{\scriptscriptstyle 
K'\downarrow} m_k}^{\vphantom{\dagger}} ) . 
$$
Thus the state $\ket{\psi}$ created by applying $T_-$ to $\hws$ is
\begin{align}
\ket{\psi} &=
\prod_{m_k}
\left[ 
\sum_{n_k} (c_{{\scriptscriptstyle K'\uparrow} n_k}^\dagger 
c_{{\scriptscriptstyle K\uparrow} n_k}^{\vphantom{\dagger}} + 
c_{{\scriptscriptstyle K'\downarrow} n_k}^\dagger c_{{\scriptscriptstyle 
K'\downarrow} n_k}^{\vphantom{\dagger}} )
\right]
c_{{\scriptscriptstyle K\uparrow} m_k}^\dagger
c_{{\scriptscriptstyle K\downarrow} m_k}^\dagger
\, \ket 0
\nonumber
\\
&=
\prod_{m_k} 
\left(
c_{3 m_k}^\dagger
c_{2 m_k}^\dagger
+
c_{4 m_k}^\dagger
c_{1 m_k}^\dagger
\right)
\ket0 
,
\label{u0pairs1.7}
\end{align}
where the simplifications are 
because the only terms that survive correspond to those where an annihilation 
operator in a factor inside the square brackets is exactly balanced by a 
creation operator from the factor outside the square brackets. Likewise,  the 
other states of the $u=0$ representation can be constructed by 
using 
successive applications of raising and lowering operators fashioned from the 
generators defined in in Eqs.\ \eqnoeq{algebra1.8}--\eqnoeq{algebra1.11}.

\subsection{\label{sh:pair-product} Equivalence of Pair and Product  
Wavefunctions}

From the preceding discussion, for $N=\tfrac12 \Omega$ the states may be 
written as
\begin{align}
\ket\psi &= 
F(B_{ab}) \hws
\nonumber
\\
&= 
\prod_{m_k} \left[
\sum_{\tau\sigma\tau'\sigma'} \Phi^*_{\tau\sigma\tau'\sigma'}
c_{\tau\sigma m_k}^\dagger c_{\tau' \sigma' m_k}^\dagger
\right] \ket0 ,
\label{upairs1.8}
\end{align}
where $\tau,\tau'$ denote valley isospin projection quantum numbers and 
$\sigma,\sigma'$ denote spin projection quantum numbers. This is the same form 
as the most general collective pair state used  by Kharitonov  \cite{khar2012} 
in his classification of possible broken symmetry states for the $n=0$ Landau 
level in graphene (see \eq{basis1.0} in the Appendix).  Thus, for undoped 
graphene the general pairing wavefunction \eqnoeq{u0pairs1.0} characteristic of 
the SO(8) $\supset$ SU(4) dynamical symmetry
is in fact {\em equivalent to the product form \eqnoeq{upairs1.8}} employed in 
standard discussion of quantum Hall ferromagnetism, for which the summations are 
over the internal $(\tau,\sigma)$ rather than Landau $(m_k)$ degrees of freedom.
The equivalence of Eqs.\ \eqnoeq{upairs1.8} and \eqnoeq{u0pairs1.0}, 
despite their superficially very different forms, is a fundamental consequence 
of the Pauli principle acting in the collective fermionic pair subspace, which 
greatly restricts allowed pair configurations. 

The equivalence established above implies 
that the present SO(8) formalism can be used to derive the framework used in 
Ref.\ \cite{khar2012} to describe possible collective states in graphene, which 
establishes an essential connection between the standard discussion of SU(4) 
quantum Hall magnetism in graphene and the present more general SO(8) 
dynamical symmetry formalism.  As will be demonstrated further below, this 
permits the  SO(8) formalism to 
(1)~encompass the established physics of SU(4) quantum Hall ferromagnetism,
(2)~extend quantum Hall ferromagnetism to suggest possible additional 
collective 
graphene physics beyond the SU(4) limit, and 
(3)~provide {\em analytical solutions} for the states suggested by spontaneous 
breaking of SU(4) symmetry that cannot be obtained using SU(4) symmetry and 
must 
be addressed numerically, or with effective field theory approximations, within 
standard SU(4) quantum Hall ferromagnetism.

Furthermore, it will be seen below that the equivalence of product and 
paired 
forms for the wavefunction implies a deep formal 
connection between the collective states resulting from strong electron 
correlations in graphene Landau levels and the collective states produced by 
strong nucleon correlations in nuclear structure physics, and a suggestive 
formal connection to the properties of the strongly-correlated electrons 
responsible for high-temperature superconductivity.

\section{\label{h:beyondQHF} Beyond Quantum Hall Ferromagnetism}

The preceding discussion has established that the fermion dynamical symmetry 
method applied to undoped graphene in a strong magnetic field has one dynamical 
symmetry chain $\soeight \supset \sufour$ that recovers exactly 
SU(4)-symmetric quantum Hall ferromagnetism. Since the $B_{ab}$ operators 
introduced in 
\eq{algebra1.2} form an SU(4) subgroup of SO(8) that is in one-to-one 
correspondence with the operators used to formulate the effective low-energy 
Hamiltonian \eqnoeq{so5_1.1a}, \S\ref{h:fdsGraphene} implies that all of the 
physics associated with this effective Hamiltonian that has been discussed in 
the prior literature (see \cite{khar2012,wufe2014} and references cited therein) 
is implicit in the present formalism.  

Furthermore, the discussion of \S\ref{h:pairRepresentations} shows that the pair 
basis \eqnoeq{u0pairs1.0} of the truncated collective subspace for the SO(8) 
fermion dynamical symmetry is in fact identical to the most general wavefunction 
\eqnoeq{upairs1.8} that has been proposed \cite{khar2012} for collective states 
breaking the SU(4) symmetry, despite its superficially very different form.  
Thus the Hilbert-space truncation implied by the collectively-paired SO(8) 
subspace \eqnoeq{u0pairs1.0} recovers the  understanding in the existing 
literature of the classes of states to be expected from spontaneous breaking of 
the SU(4) symmetry by valley-dependent correlations.  

However, the existing discussions of these collective states in terms of broken 
SU(4) symmetry have been largely qualitative, and have turned to numerical 
simulations to discuss the actual structure and energy of the states.  It will 
now be 
demonstrated that the present formalism is capable not only of classifying, 
but also of addressing the quantitative nature of those collective states {\em 
in analytical fashion.}  Furthermore, it will be shown that the SO(8) 
highest symmetry implies subgroup chains in addition to $\soeight \supset 
\sufour$ that are associated with spontaneous breaking of the 
symmetry by correlations and have not been 
discussed in the 
previous literature and that may play a role in graphene. 

Let us begin that discussion by first transforming to a more convenient 
representation of the 
SO(8) generators.  This new representation will be physically equivalent to the 
original representation, but will offer some advantages in 
interpretation, and will expose an unexpected relationship between
graphene physics and that of a very different field, nuclear structure physics.

\section{\label{h:coupledrepresentations} Coupled Representations}

 For the pair creation operators defined in \eq{algebra1.1}, each electron 
creation  operator $c^\dagger$ carries both spin and valley isospin; hence the 
products $c^\dagger c^\dagger$ correspond to a Clebsch--Gordan series 
representing sums of terms having different values of total spin and total 
isospin.    Likewise, in the particle--hole operators of \eq{algebra1.2} each 
creation operator $c^\dagger$ and annihilation operator $c$ carries spin and 
isospin, so the product $c^\dagger c$ in \eq{algebra1.2} represents a 
superposition of states carrying different total spin and total valley isospin.  
These representations with indefinite spin and isospin will be termed {\em 
uncoupled representations.} 

On physical grounds, the spin is expected to be conserved (If the Zeeman term in 
the Hamiltonian is neglected) and the valley isospin is expected to be 
approximately conserved for low-energy excitations.  Thus,  it is desirable to 
use the uncoupled representation of the pairing and particle--hole operators to 
construct new {\em coupled representations}  that have good total spin and good 
total valley isospin quantum numbers for bilinear operators.

\subsection{\label{sh:coupledPairRep} Coupled Representation for Pairing 
Operators}

Using standard angular momentum coupling theory \cite{angularMomentum}, an 
electron pair creation operator coupled to good spin and valley isospin may be 
defined by
\begin{equation}
\Adagcoupled{M_S}{M_T}{S}{T} \equiv
\sum _{m_1m_2n_1n_2}
\clebsch{\scriptscriptstyle\frac12}{m_1}{\scriptscriptstyle\frac12}{m_2}{S}{M_S}
\clebsch{\scriptscriptstyle\frac12}{n_1}{\scriptscriptstyle\frac12}{n_2}{T}{M_T}
c^\dagger_{m_1 n_1 m_k}
c^\dagger_{m_2 n_2 -m_k},
\label{coupled1.0}
\end{equation}
where $S$ is the total spin of the pair with $M_S$ its projection, $T$ is the 
total valley isospin of the pair with $M_T$ its projection, and 
$\clebsch{j_1}{m_1}{j_2}{m_2}{J}{M}$ are Clebsch--Gordan coefficients for the 
angular momentum sum $\vec j_1 + \vec 
j_2 = \vec J$ that couple the pair to good total spin or total valley isospin.
Antisymmetry implies that the the pair wavefunction can have only $S=1$, 
$T=0$\,; or $S=0$, $T=1$ (spin-triplet, isospin-singlet; or spin-singlet, 
isospin-triplet pairs). 
Explicitly the possibilities are
\begin{equation}
\begin{array}{c}
 \Adagcoupled0010 =  
A_{14}^\dagger - A_{23}^\dagger
\quad
\Adagcoupled1010 = \sqrt 2 A_{12}^\dagger
\qquad
\Adagcoupled{-1}010 = \sqrt2 A_{34}^\dagger
\\[3pt]
\Adagcoupled0001 = A_{14}^\dagger + A_{23}^\dagger
\quad
\Adagcoupled0101 = \sqrt 2 A_{13}^\dagger
\qquad
\Adagcoupled0{-1}01 = \sqrt2 A_{24}^\dagger ,
\end{array}
\label{coupled1.3}
\end{equation}
with the  hermitian conjugates of \eq{coupled1.3} giving the six corresponding 
pair annihilation 
operators in coupled representation.
It is useful to define an alternative set of six coupled pairing operators 
$S^\dagger$ and $D_\mu^\dagger (\mu = 0, \pm 1, \pm 2)$
according to
\begin{equation}
 \begin{array}{c}
 S^\dagger = \tfrac{1}{\sqrt2} \,\Adagcoupled0010 
= \tfrac{1}{\sqrt2} \left( A_{14}^\dagger - A_{23}^\dagger \right)
\\[7pt]
D_0^\dagger = \tfrac{1}{\sqrt2} \,\Adagcoupled0001
= \tfrac{1}{\sqrt2} \left( A_{14}^\dagger + A_{23}^\dagger \right)
\\[7pt]
D_1^\dagger = \tfrac{1}{\sqrt2} \, \Adagcoupled0101
= A_{13}^\dagger
\quad
D_{-1}^\dagger = \tfrac{1}{\sqrt2} \,\Adagcoupled0{-1}01
= A_{24}^\dagger
\\[7pt]
D_2^\dagger = \tfrac{1}{\sqrt2} \, \Adagcoupled1010 
= A^\dagger_{12}
\quad
D_{-2}^\dagger = \tfrac{1}{\sqrt2} \,\Adagcoupled{-1}010
= A_{34}^\dagger
 \end{array}
 \label{coupled1.4}
\end{equation}
and the six corresponding hermitian conjugates $S$ and $D_\mu$.
The physical meaning of these pairs may be deduced by constructing the 
corresponding electronic configurations.  Consider $D_2^\dagger$.  From Eqs.\ 
\eqnoeq{coupled1.4} and \eqnoeq{algebra1.1}
\begin{align*}
D_2^\dagger &= \frac{1}{\sqrt2}\, \Adagcoupled1010 = A^\dagger_{12}
=
\sum_{m_k} c^\dagger_{1 m_k} c^\dagger_{2 -m_k}
= \sum_{m_k} c^\dagger_{K\uparrow m_k} c^\dagger_{K \downarrow -m_k},
\end{align*}
where in the last step the correspondence between the index $a=1,2,3,4$ and the 
valley ($K$ or $K'$) and spin ($\uparrow \downarrow$) labels in  
\tableref{quantumNumberMapping} has been invoked.  This implies that 
$D^\dagger_2 \ket 0$ creates a state with one spin-up and one spin-down electron 
on each equivalent site $K$ in the Brillouin zone, as illustrated schematically 
in \fig{actionDdagger2}.
\singlefig
{actionDdagger2}       
{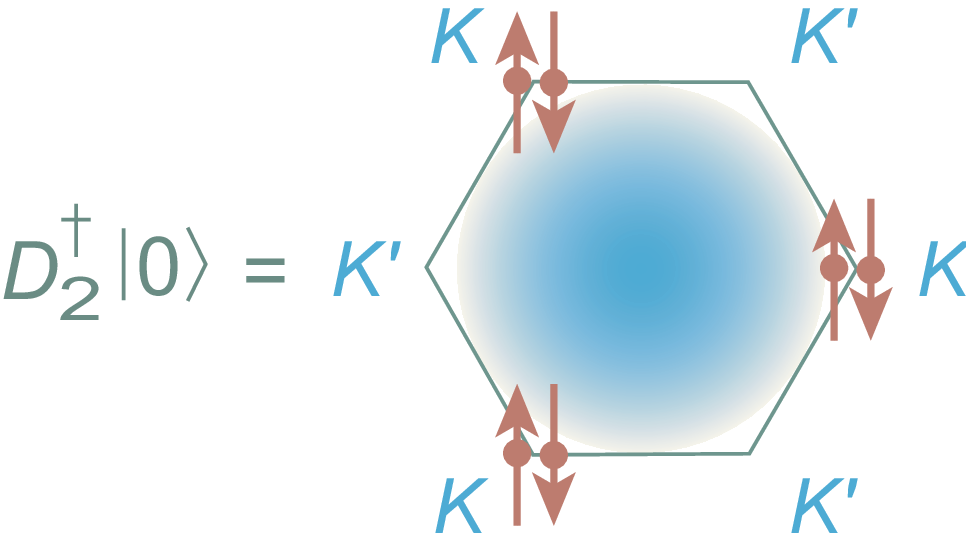}    
{\figup}         
{0pt}         
{0.35}         
{The action of the pair creation operator $D^\dagger_2$ on the vacuum state is 
to create a charge density wave  with a spin-singlet pair on each site $K$ 
and no electrons on the $K'$ sites.}
This is a component of a lattice-scale charge density wave, since the charge 
differs by two 
electronic units between adjacent sites. Likewise, one finds that 
$D^\dagger_{-2}\ket0$ creates a charge density wave as in \fig{actionDdagger2}, 
but with the spin-singlet pairs on the $K^\prime$ sites.
The pair configurations  produced by all generators of \eq{coupled1.4} operating 
on the pair vacuum $\ket0$  are summarized in \fig{S_D_pairs_brillouin}.%
\singlefig
{S_D_pairs_brillouin}       
{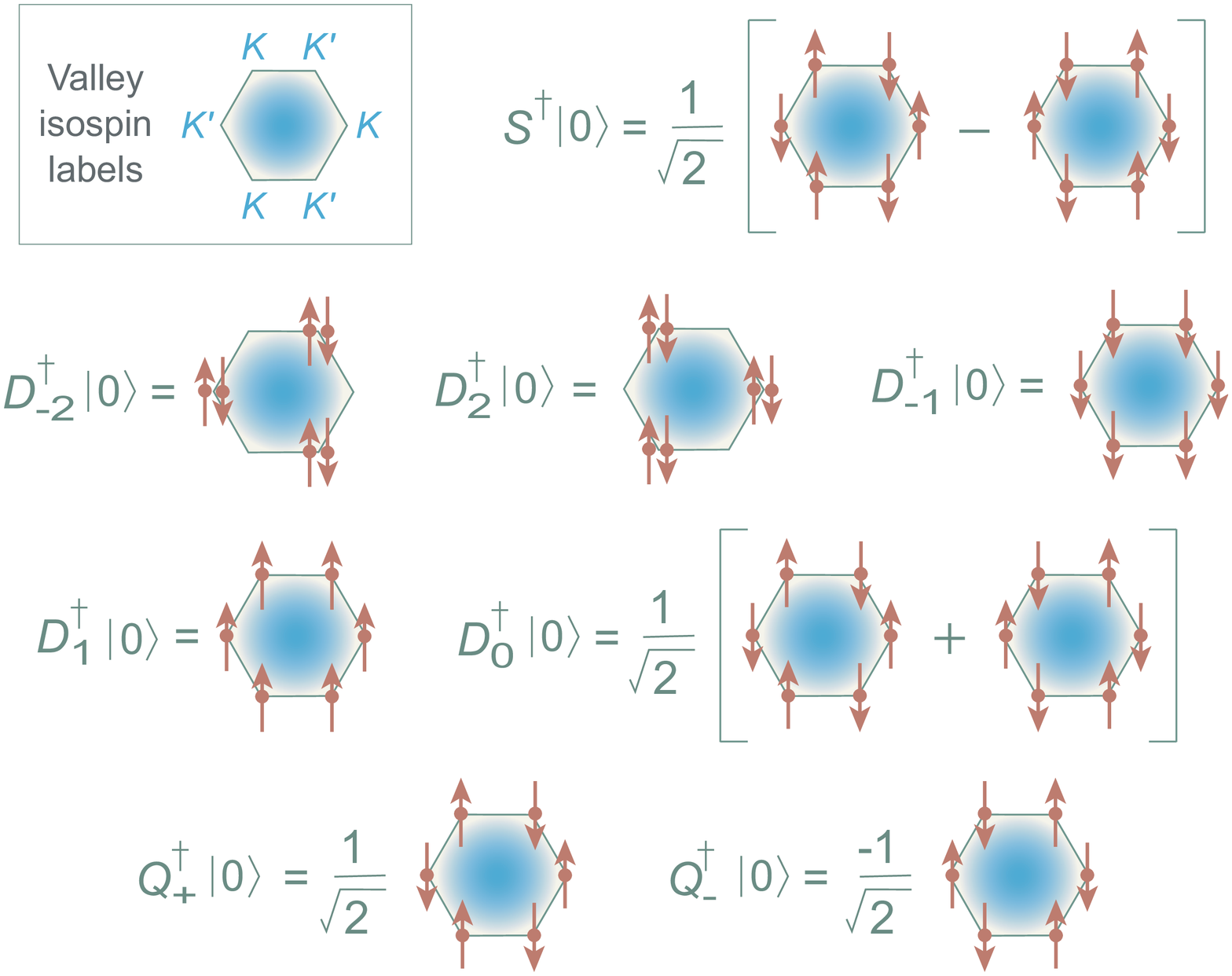}    
{\figup}         
{0pt}         
{0.30}         
{Configurations  created by the $S^\dagger$ and $D^\dagger_\mu$ pair creation 
operators of \eq{coupled1.4}, and for the linear combinations $Q_\pm$ of 
\eq{Qdef}, operating on the pair vacuum $\ket0$.   The upper left box 
illustrates the valley isospin labeling. A total of six valence 
electrons as assumed to be distributed on sites in the Brillouin zone. In each 
pair 
configuration, location of the dots ($K$ or $K'$ site) indicates the valley 
isospin, spin-up electrons are indicated by up arrows, and spin-down electrons 
are indicated by down arrows.}
Also shown are the configurations generated by the linear combinations
\begin{align}
 \ket{Q_\pm} = Q_\pm^\dagger\ket0 \equiv 
 \tfrac12 \left(S^\dagger \pm D_0^\dagger \right) \ket0
=\tfrac12 \left(\ket S \pm \ket{D_0}\right) ,
 \label{Qdef}
\end{align}
which will be useful in later discussion.

Kharitonov has given a general classification of low-lying collective modes for 
the $n=0$ Landau level of graphene in terms of collective pairs \cite{khar2012}. 
 The collective pairs created by the SO(8) pair generators in 
\fig{S_D_pairs_brillouin} are similar physically to the pairs identified by 
Kharitonov, as will now be described.

(1)~The configuration generated by $S^\dagger\ket0$ is to the 
difference of two terms, each with alternating 
spin-up and spin-down on adjacent sites, implying that all spins on the A 
sublattice (identified with valley $K$) point in one direction and all 
spins on the B sublattice (identified with  valley $K^\prime$) point in 
the opposite direction. Each term corresponds to a spin density wave 
(AF order), with a N\'eel vector defined by the difference in 
total spins on the two sublattices serving as an order parameter, but because 
of the difference of the two terms the net AF order for this configuration is 
zero (see \S\ref{sh:orderParameters} and \tableref{orderParms}).

(2)~The configurations generated by $D^\dagger_{\pm 2} \ket0$ are
spin-singlet charge density waves, with alternating charges of two and zero 
units on adjacent sites. An appropriate order parameter is the 
difference in charge between the A and B sublattices.

(3)~The configurations generated by $D^\dagger_{\pm 1} \ket0$ have one 
spin on each site, all pointing in the same direction;  this
is a ferromagnetic state, with the net spin as an order parameter.

(4)~The configuration generated by $D^\dagger_{0} \ket0$ is the same as that 
generated by $S^\dagger\ket 0$, except for a positive sign for the second term.  
This also implies alternating spins on adjacent 
sites and AF order for each term, but the total AF order vanishes because of 
the contribution of the two terms. 

(5)~The configurations corresponding to $Q_\pm^\dagger\ket0$ are states 
with AF order characterized by the difference in spins on the two sublattices 
labeled by $K$ and $K'$.

Thus the coupled-representation pairs  carrying good spin and 
valley isospin quantum numbers in \eq{coupled1.4} represent physical degrees of 
freedom already discussed in the literature as candidate collective modes 
representing spontaneous breaking of the SU(4) graphene 
symmetry by interactions in a single partially-filled Landau level.

\subsection{\label{sh:coupledPHRep} Coupled Representation for Particle--Hole
Operators}

It is desirable to express the particle--hole generators of \eq{algebra1.2} in 
coupled representation. Let us begin by introducing a set of operators
\begin{equation}
P_\mu^r = \sum_{m_j m_l} 
(-1)^{{\ttfrac32} + m_\ell}
\clebsch{\ttfrac32}{m_j}{\ttfrac32}{m_\ell}{
\,r } {\,\mu }
B_{m_j -m_\ell} ,
\label{multipole1.1}
\end{equation}
with the definition
\begin{equation}
B_{m_j -m_\ell} \equiv \sum_{m_k} c^\dagger_{m_j m_k} 
c^{\vphantom{\dagger}}_{-m_\ell 
m_k}
-\tfrac14 \delta_{m_j -m_\ell} \Omega
\label{multipole1.2}
\end{equation}
where $m_j$ and $m_\ell$ take the values of the fictitious angular 
momentum projection $m_i$ in  \tableref{quantumNumberMapping}, providing a 
labeling equivalent to that of $a$ and $b$ in $B_{ab}$, with $m_j$ or $m_\ell$ 
values $\left\{\ttfrac32, \ttfrac12, -\ttfrac12, -\ttfrac32 \right\}$ mapping 
to $a$ or $b$ values 
$\left\{ 1,2,3,4\right\}$, respectively.  For example, from  
\tableref{quantumNumberMapping}, 
$
B_{ab} = B_{12}$ and $ B_{m_j m_\ell} = 
B_{\scriptscriptstyle 3/2,1/2}
$
label the same quantity, which is defined in \eq{algebra1.2}.
From the selection rules for angular momentum coupling in 
\eq{multipole1.1}, the index $r$ can take the values $r = 0, 1, 2, 3$, with 
$2r+1$ projections $\mu$ for each possibility, which gives a total of 16 
operators $P^r_\mu$. 
By inserting the explicit values of the Clebsch--Gordan coefficients the 16 
independent $P_\mu^r$ may be evaluated in terms of the $B_{ab}$.
 \begin{align}
P^0_0 &=
\tfrac12 (B_{11} + B_{22} + B_{33} + B_{44}) 
\nonumber
\\
&=
\tfrac12(n_1 + n_2 + n_3 + n_4 -\Omega) =
\tfrac12 (n-\Omega)
\nonumber
\\[2pt]
P^1_0 &=
 \sqrt{\tfrac{9}{20}}(B_{11} - B_{44}) + \sqrt{\tfrac{1}{20}}(B_{22} -B_{33})
 \nonumber
 \\
 &=
 \sqrt{\tfrac{9}{20}}(n_1 - n_4) + \sqrt{\tfrac{1}{20}}(n_2 -n_3)
\nonumber
 \\[2pt]
P^1_1 &= -\sqrt{\tfrac{3}{10}} \,B_{12} - \sqrt{\tfrac{4}{10}} \,B_{23}
-\sqrt{\tfrac{3}{10}}\, B_{34}
\nonumber
\\[2pt]
 P^1_{-1} &=
 \sqrt{\tfrac{3}{10}} \,B_{21} + \sqrt{\tfrac{4}{10}} \,B_{32}
+\sqrt{\tfrac{3}{10}}\, B_{43}
\nonumber
 \\[2pt]
 P^2_0 &=
 \tfrac{1}{2} ( B_{11} - B_{22} + B_{44} - B_{33})
 \nonumber
 \\
 &= \tfrac12(n_1 - n_2 + n_4 - n_3)
\nonumber
 \\[2pt]
   P^2_1 &=
 \tfrac{1}{\sqrt{2}} (B_{34} - B_{12})
 \qquad
 P^2_{-1} = \tfrac{1}{\sqrt2} (B_{21} - B_{43})
 \label{PofB}
 \\[2pt]
 P^2_2 &=
 -\tfrac{1}{\sqrt{2}} (B_{13} + B_{24})
 \qquad
 P^2_{-2} =
 -\tfrac{1}{\sqrt2} (B_{31} + B_{42})
\nonumber
 \\[2pt]
  P^3_0 &=
 \sqrt{\tfrac{1}{20}} (B_{11} - B_{44}) 
 + \sqrt{\tfrac{9}{20}} (B_{33}-B_{22})
 \nonumber
 \\
 &=
 \sqrt{\tfrac{1}{20}} (n_1 - n_4) 
 - \sqrt{\tfrac{9}{20}} (n_2 - n_3)
\nonumber
 \\[2pt]
  P^3_1 &=
-\sqrt{\tfrac15} B_{12} + \sqrt{\tfrac35} B_{23} - \sqrt{\tfrac15} B_{34}
\nonumber
 \\[2pt]
 P^3_{-1} &=
 \sqrt{\tfrac15} B_{21} - \sqrt{\tfrac35} B_{32}
 + \sqrt{\tfrac15} B_{43}
\nonumber
 \\[2pt]
  P^3_2 &=
 \sqrt{\tfrac12} (B_{24} -  B_{13})
 \qquad
  P^3_{-2}=
 \sqrt{\tfrac12} (B_{42} - B_{31})
\nonumber
 \\[2pt]
 P^3_3 &=
 -B_{14}
 \qquad
 P^3_{-3} =
 B_{41},
\nonumber
\end{align}
where the
\begin{equation}
n_i = B_{ii} = \sum_{m_k} c^\dagger_{im_k} c_{im_k} - \tfrac14 \Omega
\label{numberOperators}
\end{equation}
are number operators for each of the four states and
the total particle number $n$ is the sum over the four states labeled by
$a$ in  \tableref{quantumNumberMapping},
$
 n = n_1 + n_2 + n_3 + n_4 = {\rm total\ particle\ number} .
$
It will be convenient notationally to sometimes replace the operator $P_0^0$ 
with the operator $S_0$, according to the relationship
\begin{equation}
S_0 \equiv \tfrac12 (n-\Omega) = P^0_0,
 \label{p0n}
\end{equation}
where  $2\Omega$ is 
the degeneracy of 
the space for the particles that participate in the SO(8) symmetry.
Physically $S_0 = \tfrac12 (n-\Omega)$ is one half 
the particle number measured from half filling (which corresponds to 
$n=\Omega$).

\subsection{\label{sh:lieCoupled} Lie Algebra for Coupled Operators}

Because the six operators defined by \eq{coupled1.4}, their six hermitian 
conjugates, and the 16 operators defined by \eq{multipole1.1} are independent 
linear combinations of the SO(8) generators defined in Eqs.\ \eqnoeq{algebra1.1} 
and \eqnoeq{algebra1.2}, the 28 operators $\{S, \, S^\dagger, \, D_\mu, \,
D^\dagger_\mu, \, P^\ell_\mu \}$ also close an SO(8) algebra under commutation. 
The 
SO(8) commutation relations for the coupled representation $\{S, \,S^\dagger, \,
D_\mu, \, D^\dagger_\mu, \, P^\ell_\mu \}$ are given explicitly by 
\cite{gino80,FDSM}
\begin{subequations}
\begin{align}
 \comm{S}{S^\dagger} &=
 -2S_0 
 \label{commcoupled1.1}
 \\
 \comm{D_{\mu'}}{D^\dagger_{\mu}} &=
 -2\delta_{\mu\mu'} S_0 + \sum_{t {\rm\ odd}} (-1)^{\mu'}
 \nonumber
 \\
 &\qquad \times \
 \clebsch{2,}{\scriptscriptstyle-\mu'}{2}{\mu}{\,t,}{\scriptscriptstyle\mu-\mu'}
 \sixj{2}{2}{t}{\tfrac32}{\tfrac32}{\tfrac32}
 P_{\scriptscriptstyle\mu,-\mu'}^t
\label{commcoupled1.2}
\\
\comm{D_\mu^\dagger}{S} &= P_\mu^2
\label{commcoupled1.3}
\\
\comm{P^r_\mu}{S^\dagger} &=
2 \delta_{r2}D^\dagger_\mu + 2 \delta_{r0} \delta_{\mu0} S^\dagger
\label{commcoupled1.4}
\\
\comm{P_{\mu'}^r}{D_\mu^\dagger} &=
2(-1)^{\mu'} \delta_{r2} \delta_{-\mu\mu'}
-4\sqrt{5(2r+1)}
\nonumber
\\
&\qquad\times\ 
\clebsch{r}{\scriptscriptstyle\mu'}{2}{\mu}{2,}{\scriptscriptstyle\mu+\mu'}
\sixj{2}{2}{r}{\tfrac32}{\tfrac32}{\tfrac32}
D_{\scriptscriptstyle\mu+\mu'}^\dagger
\label{commcoupled1.5}
\\
\comm{P^r_{\mu'}}{P^s_\mu} &=
2 (-1)^{r+s} \sqrt{(2r+1)(2s+1)} 
\,\,\sum_t 
\clebsch{r}{{\scriptscriptstyle\mu'}}{s}{{\scriptscriptstyle\mu}}{\,t,}{{
\scriptscriptstyle \mu+\mu'} }
\nonumber
\\
&\times 
\left[1-(-1)^{r+s+t}\right]
\sixj{r}{s}{t}{\tfrac32}{\tfrac32}{\tfrac32}
P^{\,t}_{\scriptscriptstyle\mu+\mu'} 
\label{commcoupled1.6}
\end{align}
 \label{commcoupled}%
 \end{subequations}
where $S_0$ is defined in \eq{p0n} and
$\{\,\}$
denotes the Wigner 6-$j$ symbol \cite{angularMomentum} for the recoupling of 
three 
angular momenta to good total angular 
momentum.

\section{\label{h:collectiveSubspace}Collective Subspace}

The action of the SO(8) pair creation operators on the pair vacuum $N$ times 
creates a $2N$-particle state \cite{gino80},
\begin{equation}
\ket{N_s N_d} = (S^\dagger)^{N_s} (D^\dagger)^{N_d}
\ket 0 ,
\label{collective1.1}
\end{equation}
where the total number of pairs is $N = N_S + N_D$.  The portion of the full 
Hilbert space that is spanned by the states \eqnoeq{collective1.1} will be 
termed the {\em collective subspace.}  It will play an important role in  
subsequent discussion where it will be shown that the SO(8) symmetry may be used 
to construct effective Hamiltonians that are diagonal in this space, and that 
the generators of SO(8) do not couple the subspace to the remainder of the 
space.

\section{\label{h:isoFDSM} Analogy with SO(8) Symmetry 
 in Nuclei}

The reason for our alternative labeling of the states in  
\tableref{quantumNumberMapping} in terms of  the index $m_i$, and our 
particular choices of phases and normalizations in equations, can now be made 
clear. With these labelings and choices the six coupled particle--particle 
operators $S^\dagger$ and $D_\mu^\dagger$  defined in \eq{coupled1.4}, their six 
hermitian conjugates $S$ and $D_\mu$, and the 16 coupled particle--hole 
operators $P^\ell_\mu$ defined in \eq{multipole1.1}, are mathematically in 
one-to-one correspondence with the 28 generators for the Ginocchio SO(8) model 
\cite{gino80} and the SO(8) Fermion Dynamical 
Symmetry Model \cite{clwu86,clwu87,weimin1989}. These have found broad 
application in nuclear structure physics 
\cite{FDSM} and may be viewed as a microscopic justification for the 
Interacting Boson Model (IBM) \cite{IBM}, which is one of the most commonly used 
phenomenological models in nuclear structure physics. 

This correspondence has three important implications: (1)~Mathematically,  the 
group-theoretical methodology obtained for SO(8) already in 
nuclear physics applications may be appropriated for use in the graphene 
problem. (2)~Physically the 
nature of the generators for the nuclear physics and graphene SO(8) symmetries 
are fundamentally different, but analogs of physical interpretations applied 
already for nuclear physics SO(8) symmetries may shed light on the graphene 
problem. (3)~Philosophically, the SO(8) correspondence between graphene and 
nuclear structure physics implies a satisfying convergence of mathematical 
reasoning and physical abstraction in two completely different scientific 
subfields. This convergence will be elaborated further in \S\ref{h:analogy_su4}.

\section{\label{h:subgroupChainsNuclear}Nuclear Analog Subgroup Chains}

The SO(8) group has various subgroups (subsets of generators closed under 
commutation) and these in turn may have other subgroups.  These sequences of 
subgroups define subgroup chains.  These chains will be discussed first in terms 
of the nuclear physics basis $\{S, \, S^\dagger,\, D_\mu,\, D^\dagger_\mu,\, 
P^\ell_\mu \}$, and then in terms of a new basis that is mathematically 
equivalent but is physically better suited to describing the physics of 
graphene.  

In the nuclear physics basis $P^0$ is the particle number and generates a group 
$\uonecharge$, while $P^1$ is proportional to the total angular momentum and 
generates a group $\sothree_L$.  In the nuclear physics context the total 
angular momentum and the particle number are expected to be conserved exactly 
for all physical states. Thus one seeks subgroup chains of SO(8) that end in the 
subgroup  $\sothree_L \times \uonecharge$ corresponding to charge and angular 
momentum conservation. Three SO(8) subgroup chains satisfy these conditions.

\subsection{\label{sh:so5chain}The Nuclear Analog SO(5) $\times$ SU(2)$\tsub p$ 
Subgroup Chain}

From \eq{commcoupled1.1} the quasispin generators $(S, S^\dagger, S_0)$ close an 
$\sutwo\tsub p$ algebra that is a subalgebra of SO(8), and from 
\eq{commcoupled1.4} the operators $P^r_\mu$ with $r=1,3$ close an SO(5) algebra 
and commute with these SU(2) quasispin generators.  Thus, one subgroup of SO(8) 
is
\begin{equation*}
\soeight \supset \sofive \times \sutwo\tsub p ,
\end{equation*}
where SO(5) is generated by the 10 operators $(P^1_\mu, P^3_\mu)$, where $\mu$ 
takes the $2r+1$ values 
$
\mu = (-r, -r+1, \ldots, r-1, r)
$ 
and the {\em quasispin group} $\sutwo\tsub p$  is generated by $(S, S^\dagger, 
S_0)$, where $S_0 = P^0$. Furthermore, the three generators $P^1_\mu$ are 
components of the total angular momentum $L$ and generate an $\sothree_L$ 
subgroup of SO(5), and $S_0$ generates a $\uonecharge$ subgroup of $\sutwo\tsub 
p$ corresponding to conservation of charge.  Thus one subgroup chain is
\begin{align}
\soeight &
\supset \underset{\scriptscriptstyle \{P^1, \,
P^3\} }{\sofive}\times
\underset{\scriptscriptstyle \{ S, \, S^\dagger,\, S_0\} 
}{\sutwo\tsub p}
\supset
\underset{\scriptscriptstyle 
\{P^1\} }{\sothree_L}\times\underset{\scriptstyle \{ 
S,\, S^\dagger, \, S_0 \} }{\sutwo\tsub p}
\nonumber
\\[1pt]
&\supset
\underset{\scriptscriptstyle\{ 
P^1\} }{\sothree_L}\times\underset{\scriptscriptstyle \{ S_0 
\} }{\uonecharge} ,
\label{chains1.2n}
\end{align}
where the generators of each subgroup are indicated in brackets below the 
subgroup and the product group on the last line corresponds to conservation of 
total angular momentum and particle number.  

\subsection{\label{sh:su4chain}The Nuclear Analog SO(6) $\sim$ SU(4) Subgroup 
Chain}

The groups SU(4) and SO(6) share the same Lie algebra.  In nuclear physics it is 
more common to refer to this group as SO(6), but to maintain a parallel with the 
ensuing treatment of graphene it will be labeled SU(4) in the present 
discussion. From \eq{commcoupled1.6}, the 16 operators $P_\mu^r(r=0, 1, 2, 3)$ 
are closed under commutation, corresponding to 
$$
\soeight \supset {\rm U(4)} \supset \uonecharge 
\times {\rm SU(4)},
$$
where the generator of $\uonecharge$ is $P^0 = S_0$ and the 15 
operators $P^r_\mu(r=1,2,3)$ are the generators of SU(4).  
Furthermore, the subset of $P^r$ with odd $r$ are generators of the SO(5) 
symmetry discussed above and so generate an SO(5) subgroup of this SU(4) group. 
 Hence a second subgroup chain is
\begin{align}
\soeight &\supset
\underset{\scriptscriptstyle\{ P^0, \, P^1, \, p^2, \, 
P^3\}}{\rm U(4)}
\supset
\underset{\scriptscriptstyle \{ S_0\}}{\uonecharge} 
\times 
\underset{\scriptscriptstyle\{ P^1, \, p^2, \, P^3\}}{\sufour}
\nonumber
\\[1pt]
&\supset
\underset{\scriptscriptstyle \{ P^1, \, P^3\}}{\sofive}
\times
\underset{\scriptscriptstyle \{ S_0\}}{\uonecharge}
\supset
\underset{\scriptscriptstyle\{ 
P^1\} }{\sothree_L}\times\underset{\scriptscriptstyle \{ S_0 
\} }{\uonecharge} ,
\label{chains1.3n}
\end{align}
where \eq{p0n} has been used to replace  $P^0_0$ with $S_0$.

\subsection{\label{sh:s07}The Nuclear Analog SO(7) Subgroup Chain}

From Eqs.\ \eqnoeq{commcoupled1.2}, \eqnoeq{commcoupled1.5}, and 
\eqnoeq{commcoupled1.6}, the 21 operators $\{ S_0,\, D^\dagger_\mu,\, 
D^{\vphantom{\dagger}}_\mu, \, P^1_\mu, \, P^3_\mu\}$ close an 
$
\soeight \supset {\rm SO(7)} 
$
subalgebra of SO(8) and 
the subset  $\{ P^1, \, P^3, \, S_0 \}$ closes an $\sofive \times 
\uonecharge$ subalgebra of SO(7). Thus a third subgroup chain is given by
\begin{align}
\soeight &\supset
\underset{\scriptscriptstyle \{S_0,\, D^\dagger,\, D,\, P^1,\, P^3 
\}}{\soseven}
\supset
\underset{\scriptscriptstyle \{ P^1, \, P^3\}}{\sofive}
\times
\underset{\scriptscriptstyle \{ S_0\}}{\uonecharge}
\nonumber
\\
&\supset
\underset{\scriptscriptstyle\{ 
P^1\} }{\sothree_L}\times\underset{\scriptscriptstyle \{ S_0 
\} }{\uonecharge} ,
\label{chains1.4n}
\end{align}
The relationships of the nuclear analog SO(8) subgroup chains described above 
are summarized in \fig{dynamicalChains}.%
 \singlefig
{dynamicalChains}  
{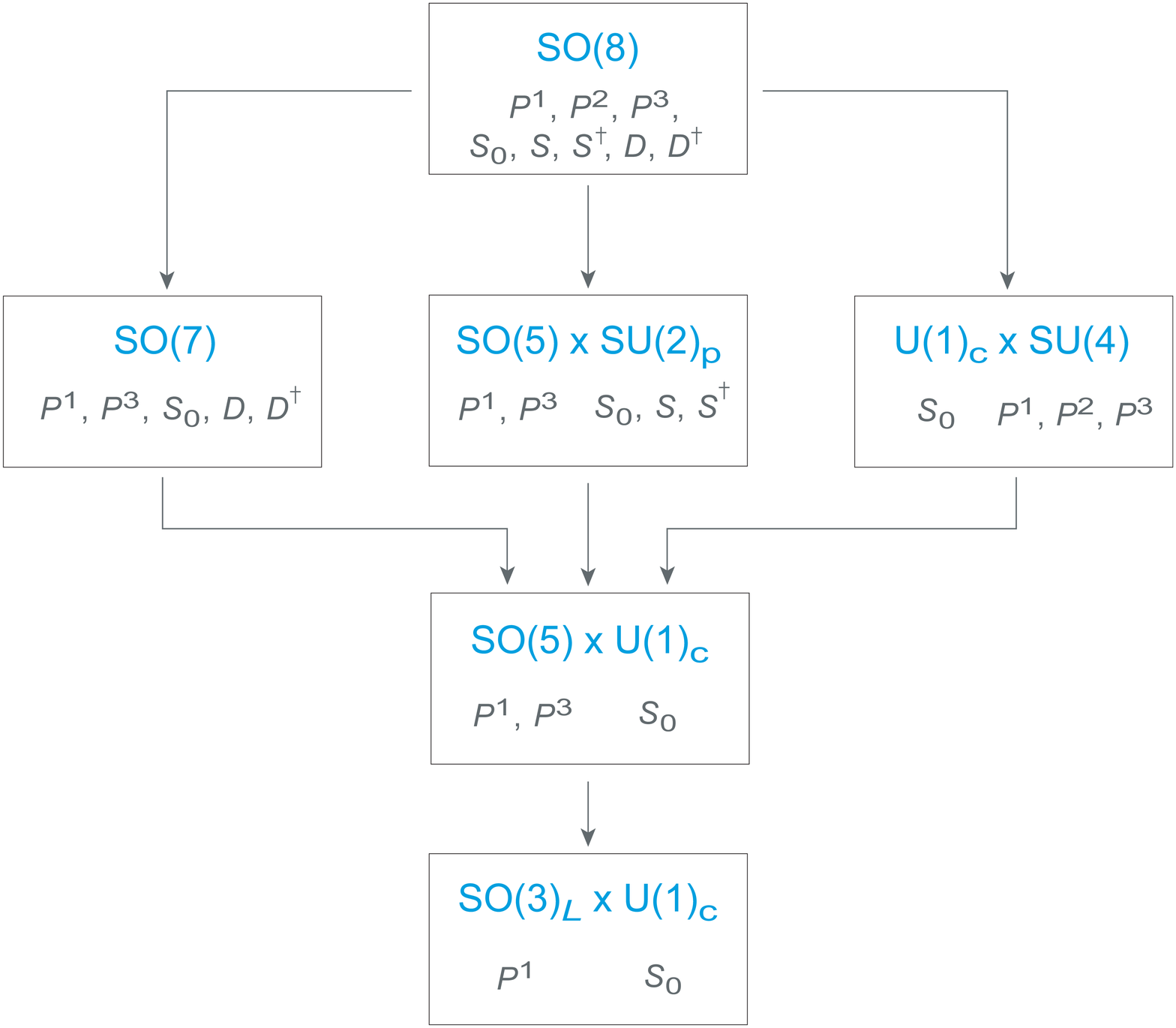}
{\figup}
{0pt}
{0.25}
{Nuclear analog SO(8) subgroup chains with generators in the coupled 
representation $\{ P^0, \, P^1, \, P^2, \, P^3,\, S, \,S^\dagger,\, D_\mu,\, 
D^\dagger_\mu \}$ given by Eqs.\ \eqnoeq{multipole1.1} and \eqnoeq{coupled1.4} 
and obeying the Lie algebra of \eq{commcoupled}. Generators are indicated below 
each group factor and $S_0$ and $P^0\equiv P_0^0$ may be interchanged using 
\eq{p0n}. The subgroup  structure expressed in this basis is in one-to-one 
correspondence with the SO(8) Fermion Dynamical Symmetry Model \cite{clwu87} of 
nuclear structure physics. However, for the description of graphene it is more 
useful to transform this basis to the new one employed in 
\fig{dynamicalChains_graphene} using Eqs.\ \eqnoeq{ngx}, which gives a more 
direct physical interpretation of quantities important in graphene physics.}

\section{\label{h:subgroupChainsGraphene}Graphene SO(8) Subgroup Chains}

The subgroup chains in \S\ref{h:subgroupChainsNuclear} were expressed in terms 
of the nuclear physics basis $\{S, \,S^\dagger,\, D_\mu,\, D^\dagger_\mu, \, 
P^\ell_\mu \}$.  This basis demonstrates the deep connection between graphene 
quantum Hall physics and nuclear structure physics, and is suitable 
mathematically to describe graphene quantum Hall effects.  However,  it is not 
well suited physically to interpreting the graphene quantum Hall effects for 
three reasons.

(1)~The relationship between the generators of the $\soeight \supset \sufour$  
subgroup in the nuclear physics basis and those of the SU(4) quantum 
Hall ferromagnetism basis defined in \eq{algebra1.4} is not clear, which 
hinders interpretation of the present results in terms of preceding results 
found in the graphene literature.

(2)~Charge and electronic spin are expected to be conserved in graphene (if the 
Zeeman term is neglected), but none of the generators $\{S,\, S^\dagger, \,
D_\mu, \,D^\dagger_\mu,\, P^\ell_\mu \}$ can be interpreted as spin in the 
application to graphene (instead, spin is a linear combination of these 
generators).  For physical applications it is desirable to employ a basis in 
which the relevant conservation laws are manifest.

(3)~In addition to the exact conservation laws expected for charge and spin in 
graphene, it is expected on physical grounds that for low-energy excitations 
the scattering between valleys is strongly suppressed and the difference in 
electron densities between neighboring valleys should be very nearly conserved. 
 This difference is expressed by the $z$ component of the valley isospin $T_z$, 
and the corresponding approximate invariance is reflected in a $\uone\tsub v$ 
symmetry generated by $T_z$.  But $T_z$ is not proportional to any of the $P^r$ 
generators (it is a linear combination of these generators), so this 
approximate invariance is not manifest in the nuclear SU(4) basis.

Thus a new basis will be employed for the SO(8) generators in application to 
graphene for which the particle number (charge) operator $n$ or $S_0$ 
and the 12 pairing operators $\{D_\mu,\, D^\dagger_\mu,\, S,\, S^\dagger\}$ 
are retained, but 
the 15 SU(4) generators $\{ P^1, \,P^2, \,P^3\}$ in the nuclear 
representation are replaced with the 15 SU(4) generators
$\{\spin_\alpha,\,  T_\alpha,\,  N_\alpha,\, \Piop\alpha x,\, \Piop \alpha y\}$ 
with $\alpha=x,y,z $ defined in the graphene representation given in
\eq{algebra1.4},
\begin{equation*}
\begin{array}{l}
 \{P^1, \,P^2, \,P^3, \,S_0,\, S, \,S^\dagger,\, D\phantomdagger_{\mu}, \,
D^\dagger_\mu \} \tsub{Nuclear SO(8)}
\quad\longrightarrow
\\[5pt]
\{ \spin_\alpha,\,  T_\alpha,\,  N_\alpha,\, \Piop \alpha x, \,\Piop \alpha y, 
\,
S_0, \,
S, \,S^\dagger, \,D\phantomdagger_{\mu},\, D^\dagger_\mu \} \tsub{Graphene 
SO(8)} .
\end{array}
\end{equation*}
The transformation from the  $\{ P^1, \,P^2,\, P^3\}$ 
generators to the 
$\{\spin_\alpha, \, T_\alpha, \, N_\alpha, \,\Piop \alpha x, \,
\Piop \alpha y \}$ 
generators is given in Appendix \ref{h:appendixB}.

\subsection{\label{sh:orderParameters}Order Parameters}

In the new basis it will be convenient to take as order parameters
\begin{subequations}
\begin{align}
\ev {\spin_z} &= 
\ev {\hat n_1} - \ev{\hat n_2} + \ev{\hat n_3} - \ev{\hat n_4}
\label{orderp1.1a}
\\
\ev{T_z} &=
\ev {\hat n_1} + \ev{\hat n_2} - \ev{\hat n_3} - \ev{\hat n_4}
 \label{orderp1.1b}
\\
\ev{N_z} &=
\ev {\hat n_1} - \ev{\hat n_2} + \ev{\hat n_4} - \ev{\hat n_3}
 \label{orderp1.1c}
\end{align}
 \label{orderp1.1}%
\end{subequations}
where $\hat n_i$ is the number operator counting particles in basis state 
$\ket i$ and the expectation value is taken with respect to the collective 
wavefunction. Physically
(1)~The net spin is measured by $\ev{\spin_z}$, which characterizes 
{\em ferromagnetic} order.
(2)~The difference in charge between the A and B sublattices is measured by
$\ev{T_z}$, which characterizes {\em charge density wave} order.
(3)~The difference in spins between the A and B sublattices is measured by 
$\ev{N_z}$, which characterizes {\em antiferromagnetic} (N\'eel or spin density 
wave) order.
The order parameters evaluated for the states of \fig{S_D_pairs_brillouin} are 
displayed in
\tableref{orderParms}.
{\renewcommand\arraystretch{1.0}
\begin{table}[tpb]
  \vspace*{\tabup}
  \centering
  \caption{Order parameters for states in \fig{S_D_pairs_brillouin}}
  \label{tb:orderParms}
    \begin{centering}
      \setlength{\tabcolsep}{8 pt}
      \vspace{\tabtitlesep}
      \begin{tabular}{lccc}
      
        \tableLineOne
        
            State &
            $\ev{\spin_z}$ &
            $\ev{T_z}$ &
            $\ev{N_z}$

        \tableLineTwo
        
            $\ket S = S^\dagger \ket0$ &
            $0$ &
            $0$ &
            $0$

        \\ [\tablerowMore]  
        
            $\ket{D_{-2}} = D^\dagger_{-2}\ket0$ &
            $0$ &
            $-1$\hphantom{$-$} &
            $0$

        \\  [\tablerowMore]     
        
            $\ket{D_{2}} = D^\dagger_{2}\ket0$ &
            $0$ &
            $1$ &
            $0$

        \\  [\tablerowMore]  
        
            $\ket{D_{-1}} = D^\dagger_{-1}\ket0$ &
            $-1$\hphantom{$-$} &
            $0$ &
            $0$ 
            
        \\  [\tablerowMore]  
        
            $\ket{D_{1}} = D^\dagger_{1}\ket0$ &
            $1$ &
            $0$ &
            $0$ 
            
        \\  [\tablerowMore]  
        
            $\ket{D_{0}} = D^\dagger_{0}\ket0$ &
            $0$ &
            $0$ &
            $0$ 
            
        \\  [\tablerowMore]  
        
            $\ket{Q_+} = \tfrac12 (\ket S + \ket{D_0})$ &
            $0$ &
            $0$ &
            $1$ 
            
        \\  [\tablerowMore]  
        
            $\ket{Q_-} = \tfrac12(\ket S - \ket{D_0})$ &
            $0$ &
            $0$ &
            $-1$\hphantom{$-$}

        \tableLineThree
        
      \end{tabular}
    \end{centering}
\end{table}
}%

\subsection{\label{sh:conservedQuantities}Conserved Quantities}

In the new basis it will be assumed that both charge and spin are exactly 
conserved for the physical states of the model in the absence of the Zeeman term 
$H\tsub Z$, that the charge and the $z$-component of spin are exactly conserved 
if the Zeeman term is included in the Hamiltonian, and that  $T_z$ is conserved, 
where appropriate. Neglecting the Zeeman term, the spin--charge symmetry 
corresponds to a group structure $\suspin \times \uonecharge$, where $\suspin$ 
is generated by the spin operators and $\uonecharge$ is generated by the 
particle number operator.  Thus one seeks subgroup chains of SO(8) that end in 
the subgroup  $\suspin \times \uonecharge$ corresponding to charge and spin 
conservation, and in some of these chains a U(1)$\tsub v$ 
subgroup implying conservation of $T_z$ will also be required.

The group and subgroup structure in the new basis is illustrated in 
\fig{dynamicalChains_graphene},%
 \doublefig
{dynamicalChains_graphene}  
{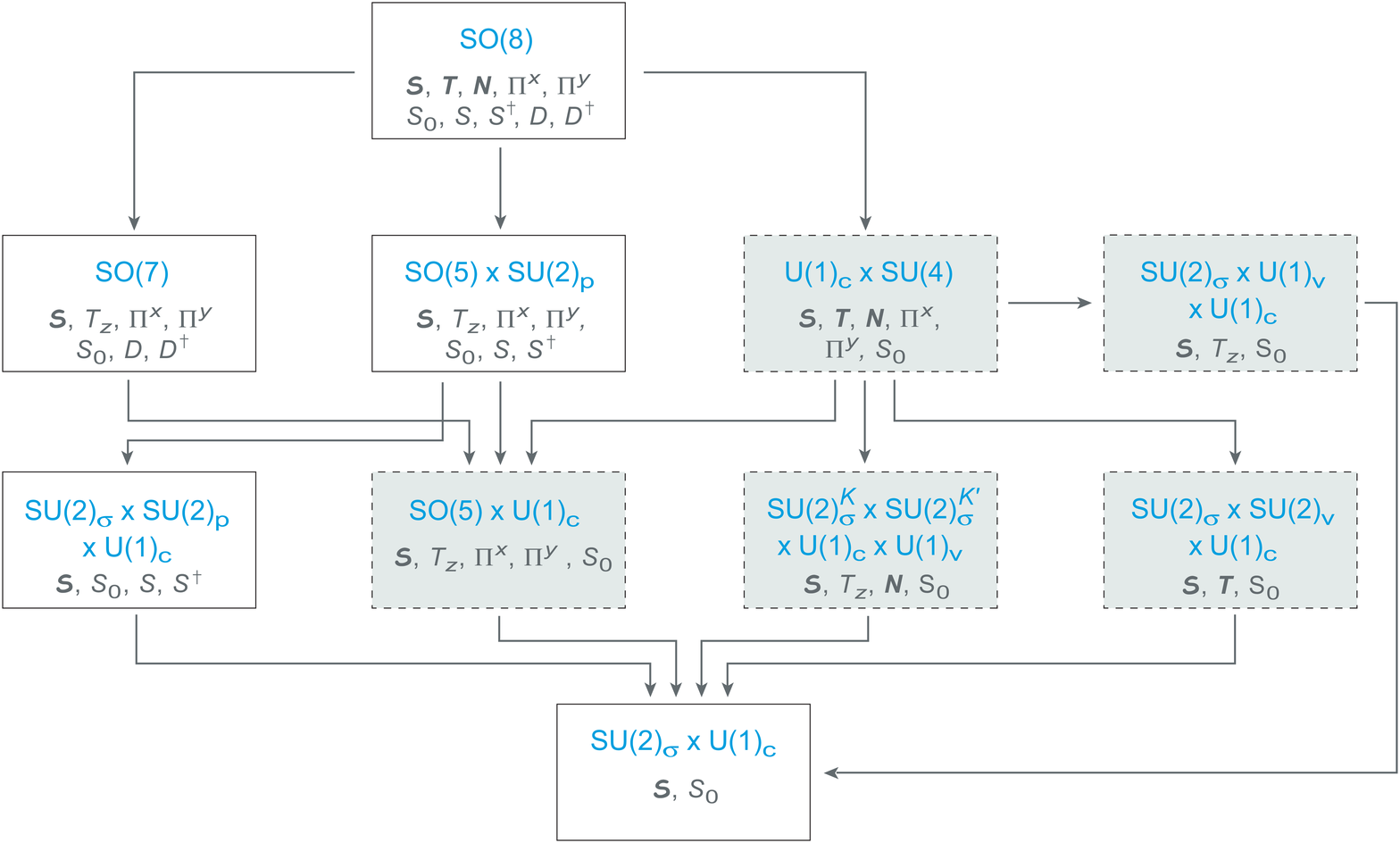}
{\figup}
{0pt}
{0.26}
{SO(8) subgroup chains with generators in a representation more suitable than 
that of \fig{dynamicalChains} for interpreting graphene physics.  Group 
generators are indicated in each box. The boxes shown with shading and dashed 
boundaries and  are relevant for interpreting the SU(4) quantum Hall 
ferromagnetism illustrated in  \fig{dynamicalChains_QHF}.  In this 
classification  the Zeeman term has been neglected so all subgroup chains end in 
the group $\sutwo_\sigma \times \uone\tsub c$ corresponding to the conservation 
of total spin and total charge. If the Zeeman term is included it will influence 
directly only the spin sector and break all $\suspin$ factors down to 
$\uone_\sigma$ generated by the $z$ component of the spin, $\spin_z$.}
where seven nontrivial subgroup chains may be identified 
that begin with SO(8) and end with the symmetry $\uone\tsub c \times \uone\tsub 
s$ corresponding to conservation of charge and $z$-component of the spin in the 
magnetic field.
\begin{subequations}
\begin{align}
 \soeight  &\supset
 \sufour \supset \sofive \supset \sutwo
 \label{GroupChains1}
 \\
 \soeight &\supset
 \sufour \supset \sutwo_\sigma^K \times \sutwo_\sigma^{K'} \supset 
\sutwo
 \label{GroupChains2}
 \\
 \soeight  &\supset
 \sufour \supset \sutwo_\sigma \times \sutwo\tsub v \supset 
\sutwo
\label{GroupChains3}
 \\
 \soeight  &\supset
 \sufour  \supset 
\sutwo
\label{GroupChains4}
 \\
 \soeight  &\supset
 \sofive \times \sutwo\tsub p \supset \sofive 
\supset \sutwo
\label{GroupChains5}
 \\
 \soeight  &\supset
 \sofive \times \sutwo\tsub p 
 \nonumber
 \\
&\supset \sutwo_\sigma \times \sutwo\tsub p
\supset \sutwo
\label{GroupChains6}
 \\
 \soeight  &\supset
 \soseven
 \supset \sofive
\supset \sutwo
\label{GroupChains7}
\end{align}
 \label{GroupChains}%
\end{subequations}
where for brevity all  U(1) factors are dropped in the notation and SU(2) means 
SU(2)$_\sigma$ corresponding to conservation of spin.
Each of these corresponds to a different dynamical symmetry 
that is realized for particular choices of parameters for the SO(8) 
Hamiltonian, and that 
yields {\em exact manybody solutions} using the dynamical symmetry methodology.
Let us now discuss in more detail 
three subgroup chains of SO(8).

\subsection{\label{sh:so5chainG}The Graphene SO(5) $\times$ SU(2) 
Subgroup Chains}

The quasispin generators $(S, \,S^\dagger, \,S_0)$ close an $\sutwo\tsub p$ 
algebra 
that is a subalgebra of SO(8), and the operators $\{\spin_\alpha,\, 
\Piop \alpha x, \,
\Piop \alpha y,\, T_z\}$ close an SO(5) algebra and commute with these 
SU(2) 
quasispin generators.  Putting this together, one subgroup of SO(8) is
$
\soeight \supset \sofive \times \sutwo\tsub p ,
$
Furthermore, the three generators $\spin_\alpha$ are components of the total 
spin and generate an $\suspin$ subgroup of SO(5), and $S_0$ generates a 
$\uonecharge$ subgroup of $\sutwo\tsub p$ corresponding to conservation of 
charge.  Thus one subgroup chain is
\begin{align}
\soeight &\supset \sofive 
\times
\sutwo\tsub {p}
\supset
\suspin\times\sutwo\tsub p
\nonumber
\\[2pt]
&\supset
\suspin
\times\uonecharge ,
\label{chains1.2}
\end{align}
where the product group on the last line corresponds to conservation of spin and 
charge.  This subgroup chain with its corresponding generators is illustrated in 
\fig{dynamicalChains_graphene}.
Alternatively, SO(5) may be broken according to the pattern
\begin{align}
\soeight &\supset\sofive\times
\sutwo\tsub p
\supset
\sofive \times
\uonecharge
\nonumber
\\[2pt]
&\supset
\suspin\times 
\uonecharge ,
\label{chains1.2b}
\end{align}
which also conserves spin and charge, and is illustrated in 
\fig{dynamicalChains_graphene}.

\subsection{\label{sh:su4chainG}The Graphene SU(4) Subgroup Chains}

A $\ufour \supset \uone\tsub c \times \sufour$ subgroup of SO(8) 
may be obtained by removing the 12 pairing operators from the SO(8) generator 
set.  The 
$\uone\tsub c$ subgroup is generated by the particle number (charge) and the 
SU(4) subgroup is generated by the 15 remaining operators, which are defined in 
\eq{algebra1.4} in the current basis. There are several options 
for chains corresponding to further subgroups.

(1)~The subset  $\{ S_\alpha, \,\Piop \alpha x, \,\Piop \alpha y,\, T_z\}$
defines generators of the SO(5) symmetry
discussed above and so forms an SO(5) subgroup of this SU(4) group.  Hence 
one SU(4) subgroup chain is
\begin{align}
\soeight &\supset
{\rm U(4)}
\supset
\uonecharge 
\times 
\sufour
\nonumber
\\
&\supset
\sofive
\times
\uonecharge
\supset
\suspin \times\uonecharge ,
\label{chains1.3}
\end{align}
which is displayed in \fig{dynamicalChains_graphene}.

(2)~Physically, the total spin is conserved.  If there is little inter-valley 
scattering one may also assume the spin within each $K$ valley and each $K'$ 
valley to be separately conserved, corresponding to a $\sutwo_\sigma^K \times 
\sutwo_\sigma^{K'}$ symmetry. Thus a second SU(4) subgroup chain corresponds to
\begin{align}
\soeight &\supset
{\rm U(4)}
\supset
\uonecharge 
\times 
\sufour
\nonumber
\\
&\supset \sutwo_\sigma^K \times 
\sutwo_\sigma^{K'} \times \uonecharge \times \uone\tsub v
\nonumber
\\[3pt]
&\supset \suspin \times \uonecharge ,
\label{chainy1.1}
\end{align}
where $\uone\tsub v$ is generated by $T_z$.
This chain also is displayed in \fig{dynamicalChains_graphene}.

(3)~Finally, one can imagine that SU(4) is broken into subgroups corresponding 
to simultaneous conservation of both spin and valley isospin, giving a third 
SU(4) subgroup chain
\begin{align}
\soeight &\supset
{\rm U(4)}
\supset
\uonecharge 
\times 
\sufour
\label{chainy1.2}
\\
&\supset
\suspin \times \sutwo \tsub v \times \uonecharge
\supset \suspin \times \uonecharge ,
\nonumber
\end{align}
as illustrated in \fig{dynamicalChains_graphene}.

Comparing \fig{dynamicalChains_graphene} with \fig{dynamicalChains_QHF}, it is 
apparent that the three 
$\soeight\supset\sufour$ subgroup chains defined in this section 
correspond to the three symmetry-breaking patterns described in 
\S\ref{h:symmetriesFQHEgraphene} and discussed in Refs.\ 
\cite{khar2012,wufe2014} for the SU(4) quantum Hall ferromagnetism model. 
The portion of the SO(8) subgroup structure leading to SU(4) quantum Hall 
ferromagnetism is indicated by the  shaded boxes with dashed outer 
boundaries in \fig{dynamicalChains_graphene}.
Thus 
it is seen explicitly that the special case corresponding to the 
$\soeight\supset\sufour$  subgroup chains of the present  model imply the 
 results of Refs.\ \cite{khar2012,wufe2014}.

\subsection{\label{sh:s07G}The Graphene SO(7) Subgroup Chain}

The 21 operators $\{ \spin_\alpha, \, \Piop \alpha x, \,\Piop \alpha y, \, 
T_z,\, S_0, \, D^\dagger_\mu, D^{\vphantom{\dagger}}_\mu\}$ close an $ \soeight 
\supset {\rm SO(7)} $ subalgebra of SO(8), and the subset of generators $\{ 
\spin_\alpha, \, \Piop \alpha x, \, \Piop \alpha y,\, T_z, \, S_0 \}$ close an 
$\sofive \times \uonecharge$ subalgebra of SO(7). Thus a third subgroup chain is 
given by
\begin{align}
\soeight &\supset
{\rm U(4)}
\supset
\uonecharge 
\times 
\sufour
\label{chains1.4}
\\
&\supset
\soseven
\supset
\sofive
\times
\uonecharge
\supset
\suspin \times \uonecharge ,
\nonumber
\end{align}
as illustrated in \fig{dynamicalChains_graphene}.  This 
subgroup chain is of particular interest because it will define a {\em 
critical dynamical symmetry} that represents an entire phase 
exhibiting critical behavior and interpolating between two other phases.

\section{\label{h:dynamicalSymm} Dynamical Symmetry Limits}

Let us use the subgroup structure of the preceding section to obtain exact 
solutions of the correlated many-body problem in these dynamical 
symmetry 
limits. The basic idea is to use the Casimir invariants of the  subgroup chains 
like those described in 
\S\ref{h:subgroupChainsGraphene} and illustrated in 
\fig{dynamicalChains_graphene} to label states.  Then model Hamiltonians 
constructed only from the Casimir invariants of a single subgroup chain permit 
analytical solution of the effective Schr\"odinger equation in that symmetry 
limit  \cite{FDSM,su4review}. Specifically, if a Hamiltonian $H = f(C_1, C_2, 
\ldots, C_n)$ can be 
expressed  as a function of the Casimir invariants of some subgroup chain $G_1 
\supset G_2 \supset \cdots \supset G_n$,
where the $C_i$ represent Casimir operators 
for the groups $G_i$, then the system is said to possess a dynamical symmetry 
associated with the subgroup chain. 

The discussion will be simplified by restricting to the lowest-order Casimir 
invariant for each group, which corresponds physically to omitting $n$-body 
interactions with $n>2$. Elementary properties of Lie groups then permit the 
eigenvalues $E$ and eigenfunctions $\Psi$ of this Hamiltonian to be expressed in 
closed form as
$$
E =
f(C_1(\nu_1), C_2(\nu_2), \ldots, C_n(\nu_n))
\quad
\Psi = 
\ket{\nu_1, \nu_2, \ldots, \nu_n} ,
$$
where the $\nu_i$ stand for the quantum numbers required to specify the 
irreducible representations (irreps) of the groups $G_i$.
The physical properties of the corresponding states can then be elucidated by 
using the methods of Lie groups and Lie algebras to evaluate matrix elements for 
observables.  In this way, one generally finds that the dynamical symmetries 
associated with subgroup chains of some highest symmetry define collective 
(emergent) states that correspond to particular patterns of spontaneous 
symmetry breaking.

\subsection{\label{sh:casimirInvariants} Casimir Invariants}

In terms of the generators \eqnoeq{coupled1.4}, \eqnoeq{multipole1.1}, and 
\eqnoeq{algebra1.4}, the quadratic Casimir operator $C_g$ for the \sofive\ 
subgroup is
\begin{align}
\casimir\sofive &= \sum _{r=1,3} \vec P^r \tightdot \vec 
P^r
=
\vec\Pi_x \tightdot \vec\Pi_x
+
\vec\Pi_y \tightdot \vec\Pi_y
+
\tfrac14\, \boldspin \tightdot \boldspin + \tfrac14 \tau_z^2
\nonumber
\\
&=
\tfrac14 \sum  \vec\sigma^i \tightdot \vec \sigma^j
(\tau_x^i \tau_x^j + \tau_y^i \tau_y^j)
\nonumber
\\
&\quad
+ \tfrac14 \sum \vec\sigma^i \tightdot \vec \sigma^j
+ \tfrac14  \sum \tau_z^i \tau_z^i ,
\label{casimir_so5}
\end{align}
where $\vec\Pi_\beta \equiv \Piop \alpha\beta$.
The corresponding Casimir operator for the $\sosix \sim \sufour$ subgroup is
\begin{align}
\casimir\sufour &= \hspace{-7pt}
\sum _{r=1,2,3} \hspace{-3pt}\vec P^r \tightdot \vec P^r
=
\vec\Pi_x \tightdot \vec\Pi_x
+
\vec\Pi_y \tightdot \vec\Pi_y
+
\vec N \tightdot \vec N + \tfrac14 (\boldspin \tightdot \boldspin + \vec \tau 
\tightdot \vec \tau)
\nonumber
\\
&=
\tfrac14 \sum (\vec\tau^i \tightdot \vec\tau^j) (\vec \sigma^i \tightdot 
\vec\sigma^j)
+ \tfrac14 \sum (\vec\sigma^i \tightdot \vec\sigma^j
+ \vec\tau^i \tightdot \vec\tau^j),
\label{casimir_su4}
\end{align}
the quadratic Casimir operator for the \soseven\ subgroup is
\begin{equation}
\casimir\soseven = \tfrac12 \, \vec D^\dagger \tightdot \vec D + S_0(S_0 -5) + 
\casimir\sofive ,
\label{casimir_so7}
\end{equation}
the Casimirs for the SU(2) subgroups are
\begin{equation}
\casimir{\sutwo_\sigma} = \tfrac14 \boldspin \tightdot \boldspin
\qquad
\casimir{\sutwo\tsub p} = S^\dagger S + S_0(S_0 -1)
\label{casimir_su2}
\end{equation}
and for the U(1) charge subgroup the invariant operator is trivially the single 
generator,
\begin{equation}
\casimir{\uone\tsub c} = S_0 = \tfrac12 (n-\Omega),
\label{casimir_u1}
\end{equation}
or some power of it. Finally, the quadratic Casimir operator for the full SO(8) 
group may be expressed as
\begin{align}
\casimir{\soeight} &= 
\tfrac12 (S^\dagger S + D^\dagger \tightdot D)
+\casimir\sufour 
+ S_0(S_0-6) .
\label{casimir_so8}
\end{align}
The Casimir operators that appear in each of these subgroups chains and the 
relevant quantum numbers labeling the states for each dynamical symmetry are 
summarized in \tableref{so8Subgroups}.%
%
%
%
{\renewcommand\arraystretch{1.0}
\begin{table*}[tpb]
\vspace*{\tabup}
  \centering
  \caption{Properties of SO(8) and its subgroups$^\dagger$}
  \label{tb:so8Subgroups}
    \begin{centering}
      \setlength{\tabcolsep}{3 pt}
      \vspace{\tabtitlesep}
      \begin{tabular}{llllll}
      
        \tableLineOne
        
            Group &
            Dim &
            Generators &
            Quantum numbers$^*$ &
            Casimir operators $C\tsub g$ &
            Casimir eigenvalues$^{**}$

        \tableLineTwo
        
            $\soeight$ &
            $28$ &
            $P^1,P^2,P^3,S_0,S,S^\dagger, D, D^\dagger$ &
            $\rho_1, \rho_2, \rho_3, \rho_4$ &
            $\tfrac12 (S^\dagger S + D^\dagger \tightdot D)
                +C_\sufour + S_0(S_0-6)$ &
            $\tfrac14(\Omega-u)(\Omega-u+12) + \phi(\rho_i)$

        \\ [\tablerowMore]  
        
            $\soseven$ &
            $21$ &
            $\{P^1, P^3, S_0, D, D^\dagger\}$ or&
            $\theta_1, \theta_2, \theta_3$ &
            $\tfrac12 \, \vec D^\dagger \tightdot \vec D + S_0(S_0 -5) + 
            C_\sofive$ &
            $\tfrac12 (\Omega-w)(\Omega-w+10) + \zeta(\theta_i)$
            
        \\ [\tableLineShift]  
        
            \ &   
            \ &
            $\{ \vec\Pi_x, \vec\Pi_y, \boldspin, \tau_z, S_0, D, D^\dagger \}$&
            \ &
            \ &
            \

        \\  [\tablerowMore]       
        
            $\sofive$ &
            $10$ &
            $\{ P^1, P^3 \}$ or &
            $\tau, \omega$ &
            $ \vec P^1 \tightdot \vec P^1 +   \vec P^3 \tightdot \vec P^3
            =$ &
            $\tau(\tau+3)+\tfrac12\omega(\omega+4) + \tau\omega$
            
        \\  [\tableLineShift] 
        
            \  &  
            \  &
            $\{\vec\Pi_x, \vec\Pi_y, \boldspin, \tau_z\}$ &
            \  &
            $ \vec\Pi_x \tightdot \vec\Pi_x + \vec\Pi_y \tightdot \vec\Pi_y
            + \tfrac14 \, \boldspin \tightdot \boldspin +\tfrac14 \tau_z$&
            \
            
        \\  [\tablerowMore] 
        
            U(4) &
            $16$ &
            $P^0, P^1, P^2, P^3$ &
            $n,\sigma_1,\sigma_2,\sigma_3$ &
            $ \vec P^0\tightdot \vec P^0 + \vec P^1 \tightdot \vec P^1 
            + \vec P^2 \tightdot \vec P^2
            + \vec P^3 \tightdot \vec P^3 $ &
            $n^2 + \sigma(\sigma+4)$

        \\  [\tablerowMore] 
        
            $\sufour$ &
            $15$ &
            $\{P^1, P^2, P^3\}$ or&
            $\sigma_1, \sigma_2, \sigma_3$ &
            $ \vec P^1 \tightdot \vec P^1 + \vec P^2 \tightdot \vec P^2
            + \vec P^3 \tightdot \vec P^3 =$ &
            $\sigma(\sigma+4)$
            
         \\  [\tableLineShift] 
        
            \ &   
            \ &
            $\{\vec\Pi_x, \vec\Pi_y, \vec N, \boldspin, \vec\tau\}$ &
            \ &
            $\vec\Pi_x \tightdot \vec\Pi_x + \vec\Pi_y \tightdot \vec\Pi_y
            +\vec N \tightdot \vec N + \tfrac14 (\boldspin \tightdot 
            \boldspin +\vec\tau \tightdot \vec\tau)$ &
            \
            
         \\ [\tablerowMore] 
        
            $\sutwo\tsub p$ &
            $3$ &
            $S_0, S, S^\dagger$ &
            $\nu$ &
            $S^\dagger S + S_0(S_0 -1)$ &
            $\tfrac14 (\Omega-\nu)(\Omega -\nu +2)$

        \\ [\tablerowMore] 
        
            $\suspin$ &
            $3$ &
            $ \boldspin$ &
            $s$ &
            $\tfrac14 \, \boldspin \tightdot \boldspin$ &
            $s(s+1)$

        \\ [\tablerowMore] 
        
            $\suiso$ &
            $3$ &
            $\vec T$ &
            $\tau$ &
            $\tfrac14 \, \vec T \tightdot \vec T$ &
            $T(T+1)$

        \\  [\tablerowMore] 
        
            $\uonecharge$ &
            $1$ &
            $S_0$ &
            $n$ &
            $S_0$ &
            $\tfrac12 (n-\Omega)$

        \tableLineThree
        \noalign{\vskip 3pt}
            \multicolumn{6}{l}{\footnotesize 
            $^\dagger\sufour \sim \sosix$ (they share the same 
Lie algebra).  $S_0 = P^0= \tfrac12 (n-\Omega)$, where $n$ is particle 
number. Components of spin
$\boldspin$ are functions of $P^1$ and $P^3$.}
\\[-2pt]
\noalign{\vskip 0pt}
            \multicolumn{6}{l}{\footnotesize 
            $^*$The  quantum numbers $\{\rho_i\}$, $\{\sigma_i\}$, 
$\{\theta_i\}$, and $(\tau,\omega)$ are Dynkin labels \cite{wyb74} for the 
irreps of SO(8), SU(4), SO(7), and SO(5), respectively; see Ref.\ 
\cite{chen86}.}
\\[-2pt]
\noalign{\vskip -1.5pt}
            \multicolumn{6}{l}{\footnotesize 
            \hphantom{$^*$}The SO(7) quantum number $w$ appearing in the 
Casimir eigenvalue is the number of particles that do not form $D$ pairs (see 
the $D_\mu^\dagger$ creation operators }
\\[-2pt]
\noalign{\vskip -1.5pt}
            \multicolumn{6}{l}{\footnotesize 
            \hphantom{$^*$}defined in \eq{coupled1.4}.  The seniority quantum 
number 
$\nu$ is the number of particles that do not form $S$ pairs.}
\\[-3pt]
\noalign{\vskip 0pt}
\multicolumn{6}{l}{\footnotesize 
            $^{**}$The number of particles not coupled to $S$ or $D$ pairs is 
$u$.  The functions $\phi(\rho_i)$ and $\zeta(\theta_i)$ are given by 
\cite{FDSM}}
\\[2pt]
\noalign{\vskip -1.5pt}
            \multicolumn{6}{l}{\footnotesize 
            \hphantom{$^{**}$}$\qquad\phi(\rho_1, \rho_2, 
\rho_3) = \tfrac12(\rho_1^2 + \rho_2^2) + \tfrac14(\rho_1 + \rho_3)
(\rho_1 + \rho_3 + 4\rho_2 + 12) + \rho_2(\rho_2+4)$ }
\\[2pt]
\noalign{\vskip -1.5pt}
            \multicolumn{6}{l}{\footnotesize 
            \hphantom{$^{**}$}$\qquad \zeta(\theta_2, \theta_3) = 
            \theta_2(\theta_2+3) +\tfrac12 \theta_3(\theta_3+4) 
+\theta_2\theta_3$ }
\\[2pt]
\noalign{\vskip -1.5pt}
            \multicolumn{6}{l}{\footnotesize 
            \hphantom{$^{**}$}They are non-zero only if $u\ne 0$. }
      \end{tabular}
    \end{centering}
\end{table*}
}
In the next section these results for the Casimir operators will be used to 
construct the most general Hamiltonian permitted in the truncated collective 
space for specific dynamical symmetries.

\subsection{\label{sh:mostGeneralH}Most General Dynamical Symmetry Hamiltonian}

As has been seen, in a particular dynamical symmetry limit the most general 
Hamiltonian can be constructed from a sum of Casimir invariants for the groups 
contained in the corresponding subgroup chain.  For SO(8) dynamical symmetries 
the most general Hamiltonian in the absence of the Zeeman term is represented 
by the linear combination
\begin{align*}
H &= 
H_0 + a \casimir{\soeight}
+ b \casimir{\sufour}
+
 c \casimir{\sofive} 
+d \casimir{\sutwo\tsub p}
+ e \casimir{\sutwo_\sigma}
 \label{mostGeneralHamCasimir}
\end{align*}
where $H_0$ is assumed constant in the symmetry limit, the Casimir operators 
$C_g$ have been summarized in the preceding section, and $\casimir{\soseven}$ 
does not appear explicitly because it has been eliminated by the constraint 
\cite{clwu87}:
\begin{equation}
\casimir{\soseven} = \casimir{\soeight} -\casimir{\sufour} + \casimir{\sofive}
-S^\dagger S + S_0.
 \label{so7dependence}
\end{equation}
Hamiltonians representing specific dynamical symmetry limits then correspond to 
particular choices of the coefficients $a, b, \ldots$ in this general 
expression. It may be shown that the most general Hamiltonian can also be 
expressed in the compact form (see Eqs.\ (4.1) of Ref.\ \cite{clwu87})
\begin{equation}
H = H_0' + G_0 S^\dagger S + G_2 D^\dagger \tightdot D + \sum_{r=1,2,3} B_r 
P^r 
\tightdot P^r ,
\label{mostGeneralHam}
\end{equation}
where $H_0'$ is assumed constant in a symmetry limit and where the different 
dynamical symmetry limits correspond to specific choices for the values of the 
parameters $G_0$, $G_2$, and $B_r$. The last term is expressed in terms of the 
$P^r$ from the nuclear basis.  It can be converted to the graphene basis by 
inverting Eqs.\ \eqnoeq{ngx} of Appendix \ref{h:appendixB} to solve 
for the $P^r$.

Let us now discuss each of the SO(8) dynamical symmetries and their physical 
interpretations. For brevity, let us refer to

\begin{enumerate}
\item
The dynamical symmetry structure 
associated with Eqs.\ \eqnoeq{chains1.2} and \eqnoeq{chains1.2b} as the {\em 
$SO(5)\times SU(2)$ dynamical symmetry,} 
\item
The dynamical symmetry structure associated with Eqs.\ 
\eqnoeq{chains1.3}--\eqnoeq{chainy1.2} as the {\em SU(4) dynamical symmetry, } 
and 
\item
The dynamical symmetry structure associated with \eq{chains1.4} as the {\em 
SO(7) dynamical symmetry.}
\end{enumerate} 
Initially the role of the Zeeman term (which would break the full spin symmetry 
down to conservation of its $z$ component) will be ignored and it will be 
assumed that the chains end in the subgroup $\sutwo_\sigma \times \uone\tsub c$ 
corresponding to the physical requirement that spin and charge be conserved 
exactly.

\subsection{\label{sh:so5dynamical} The SO(5) $\times$ SU(2) Dynamical Symmetry}

The dynamical symmetry chains given in Eqs.\ 
\eqnoeq{chains1.2}--\eqnoeq{chains1.2b} and illustrated in 
\fig{dynamicalChains_graphene} correspond to two alternative ways of choosing 
subgroups of $\sofive\times\sutwo\tsub p$:
\begin{equation*}
\begin{array}{l}
\hphantom{\sofive\times\sutwo\tsub p\ 
}\raisebox{-0.8ex}{$\supsetne$} \ \suspin \times 
\sutwo\tsub p\ \raisebox{-0.8ex}{$\supsetse$}
\\ 
\sofive\times\sutwo\tsub p
\hphantom{\supsetne\ \suspin\times\suiso\supsetse} \ \  \suspin \times 
\uonecharge
\\
\hphantom{\sofive\times\suiso\ }\raisebox{1.0ex}{$\supsetse$}  
\makebox[\widthof{$\ \suspin\times\suiso\ $}][c]{$\sofive \ \times \ 
\uonecharge$} \raisebox{1.0ex}{$\supsetne$} .
\end{array}
\label{so5dyn1.1}
\end{equation*}
In the upper branch of the middle step the SO(5) 
symmetry is broken to its $\sutwo$ spin subgroup, with $\sutwo\tsub p$ unbroken. 
Physically this corresponds to conservation of the spin associated with the 
$S_\alpha$ angular momentum algebra and the pseudospin associated with the 
$\{S,\, S^\dagger, \, S_0\}$ pair algebra, but not the full SO(5) symmetry.  In 
the lower branch of the middle step, the SO(5) symmetry remains intact and 
$S$-pair pseudospin $\sutwo$ is broken to $\uone$ charge.  In the final subgroup 
of both chains, only the spin and charge remain as conserved quantities.

The Hamiltonian in the $\sofive\times\sutwo$ dynamical symmetry limit 
corresponds to \eq{mostGeneralHam} with the restriction that $G_2 = B_2 = 0$, 
\begin{align}
H\tsub{SO(5)} &= 
G_0 S^\dagger S +  \sum_{r=1,3} 
B_r \, P^r \tightdot P^r 
\nonumber
\\
&=
\vec\Pi_x \tightdot \vec\Pi_x
+
\vec\Pi_y \tightdot \vec\Pi_y
+
\tfrac14\, \boldspin \tightdot \boldspin + \tfrac14 \tau_z^2
\\
&=
\tfrac14 \sum  \vec\sigma^i \tightdot \vec \sigma^j
(\tau_x^i \tau_x^j + \tau_y^i \tau_y^j)
\nonumber\\
&+ \tfrac14 \sum \vec\sigma^i \tightdot \vec \sigma^j
+ \tfrac14  \sum \tau_z^i \tau_z^i ,
\nonumber
\label{ham_so5}
\end{align}
where  terms that are constant within a given representation have been omitted.

The most general \soeight\ state in the $u=0$ collective subspace is given by 
\eq{collective1.1} and corresponds to a superposition of $S$ and 
$D_\mu$ pairs. Schematically,
\begin{equation}
\ket{\soeight} = (S^\dagger)^{N-N_d} (D^\dagger)^{N_d}
\ket 0 ,
\label{so8PairState}
\end{equation}
with $S^\dagger$ and $D^\dagger$ defined in \eq{coupled1.4} and $N_d \le N$, 
where $N$ is the total pair number. On the other hand, the most general states 
corresponding to the various subgroup chains illustrated in 
\fig{dynamicalChains_graphene} correspond to pair superpositions having specific 
constraints on the relative contribution of $S$ and $D$ pairs. The collective 
subspace for the $\sofive \times \sutwo\tsub p$ subgroup of SO(8) is of the 
form (see Eq.\ (9.13) of Ref.\ \cite{gino80})
\begin{equation} \ket{\sofive \times \sutwo\tsub p} = (S^\dagger)^N 
\ket0, 
\label{so5PairState} 
\end{equation} 
implying that it is a superposition of $S$ pairs. Conceptually, the 
wavefunction of the $\sofive \times \sutwo\tsub p$ subgroup chain for $u=0$ is 
obtained from the most general state in the collective subspace by converting 
all of its $D$ pairs to $S$ pairs.

As was seen in \S\ref{sh:coupledPairRep} and \fig{S_D_pairs_brillouin}, the $S$ 
and $D$ pairs correspond to coherent superpositions of particular electronic 
distributions in spin and valley pseudospin.  Thus, specific $S$ and $D$ pair 
content for SO(8) dynamical symmetry subgroup chains implies specific collective 
modes associated with coherent distribution of the electrons in spin and valley 
space.  It has been noted above that in the $\sofive \times \sutwo$ dynamical 
symmetry limit the ground states correspond to a superposition of $S$ pairs. The 
nature of this collective state may be inferred from \fig{S_D_pairs_brillouin} 
and is illustrated in \fig{S_D_pairs_brillouin_so5}.%
\singlefig
{S_D_pairs_brillouin_so5}  
{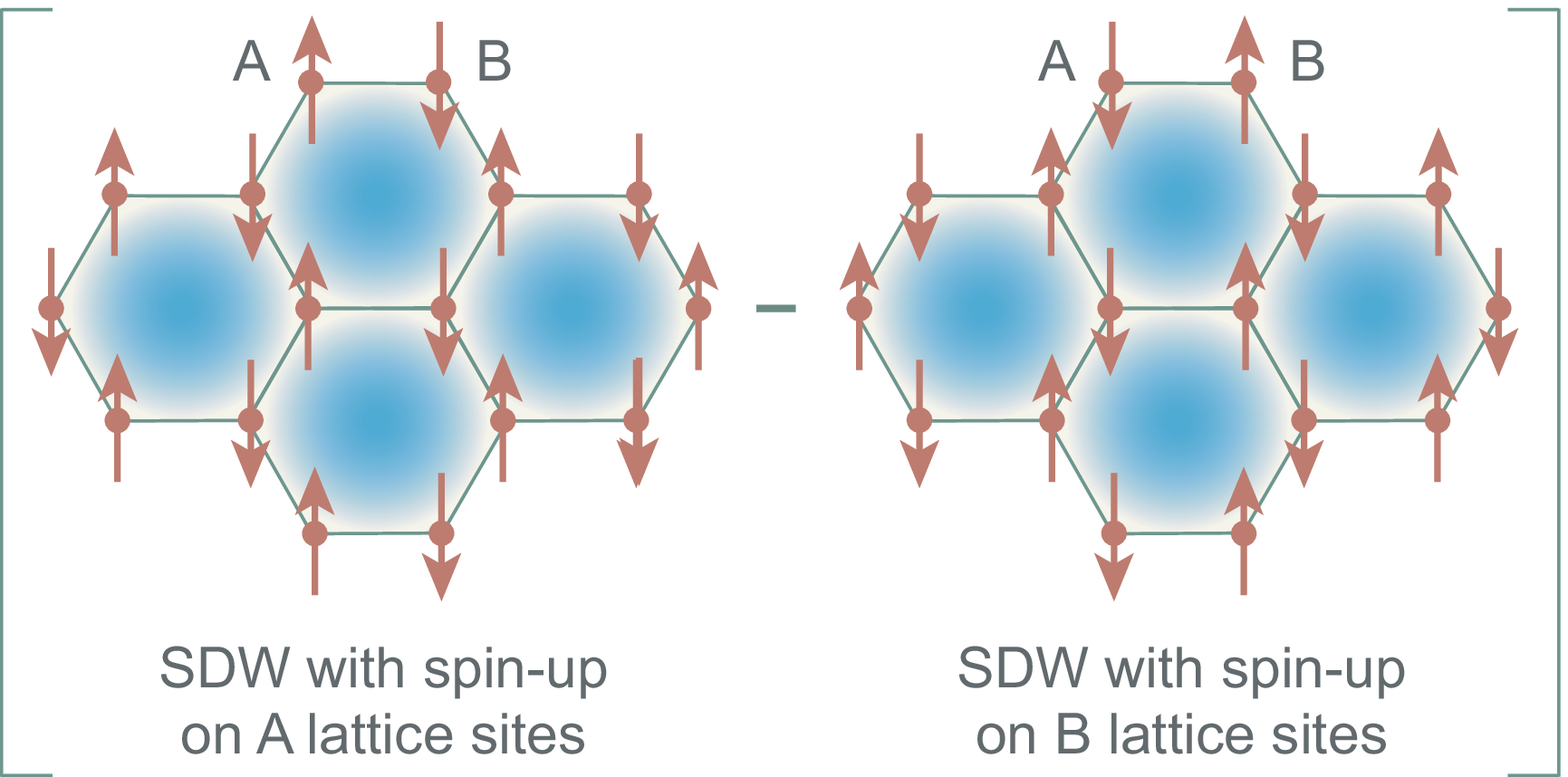}
{\figup}
{0pt}
{0.43}
{The collective state corresponding to the $\soeight \supset \sofive \times 
\sutwo$ ground state. The state is an antisymmetric combination of a spin 
density wave with spin-up on the A lattice sites and spin-down on the B sites, 
and a spin density wave with spin-up on the B sites and spin-down on the A 
sites. }

\subsection{\label{sh:su4dynamical} The SU(4) Dynamical Symmetry}

The SU(4) dynamical symmetry corresponds to the three SO(8) subgroup chains
\begin{align*}
\uonecharge\times\sufour &\supset
\sofive\times\uonecharge
\supset \suspin\times\uonecharge
\nonumber
\\
\uonecharge 
\times 
\sufour
&\supset \sutwo_\sigma^K \times 
\sutwo_\sigma^{K'} \times \uonecharge \times \uone\tsub v
\nonumber
\\
& \supset \suspin\times\uonecharge
\\
\uonecharge 
\times 
\sufour
&\supset
\suspin \times \sutwo \tsub v \times \uonecharge
\nonumber
\\
&\supset \suspin \times \uonecharge ,
\label{three_su4_chains}
\end{align*}
that were introduced in Eqs.\ \eqnoeq{chains1.3}--\eqnoeq{chainy1.2} and 
\fig{dynamicalChains_graphene}. As already noted, these three 
dynamical symmetry chains are in one to one correspondence with the explicit 
symmetry breaking patterns that have been identified for SU(4) quantum Hall 
ferromagnetism.

The most general SU(4) wavefunction for $N$ pairs in the $u=0$ collective SO(8) 
 subspace is given by
\cite{gino80}
 \begin{equation}
 \ket\sufour = \sum_{p=1}^{N/2} \beta_p \left( S^\dagger\right)^{N-2p}
 \left[\left( S^\dagger\right)^2 - D^\dagger\tightdot 
D^\dagger \right]^p \ket0,
 \label{su4PairState}
 \end{equation}
which corresponds physically to a restriction of the general SO(8) wavefunction 
\eqnoeq{so8PairState} to a specific superposition of $S$ and $D$ pairs. The 
wavefunction for the $\sofive \times \uonecharge$ subgroup of SU(4) is given by 
\eq{so5PairState}.  The wavefunction of the parent SU(4) group is a 
superposition of $S$ and $D$ pairs but the $\sofive\times\sutwo$ subgroup has a 
wavefunction containing only $S$ pairs.

\subsection{\label{sh:so7} The SO(7) Dynamical Symmetry}

The SO(7) dynamical symmetry corresponds to the SO(8) subgroup chain 
\begin{equation*} \soseven \supset \sofive\times\uonecharge \supset 
\suspin\times\uonecharge \end{equation*} that was introduced in \eq{chains1.4} 
and displayed in \fig{dynamicalChains_graphene}. The Hamiltonian in the SO(7) 
dynamical symmetry limit corresponds to \eq{mostGeneralHam} with the restriction 
that $G_0 = B_1 = B_2 = B_3 = 0$, 
\begin{equation}
H\tsub{SO(7)} = G_0 S^\dagger S +  \sum_{r=1,3} 
B_r \, P^r 
\tightdot P^r ,
\label{ham_so7}
\end{equation}
where terms have been dropped that are constant within a given 
representation.

From the nuclear physics analog SO(8) symmetry \cite{FDSM}, one may surmise that 
SO(7) will play the role of a {\em critical dynamical symmetry} interpolating 
smoothly between the collective states corresponding to the $\sufour$ dynamical 
symmetry and the collective states corresponding to $\sofive$ dynamical 
symmetry. Such critical dynamical symmetries have been discussed previously in 
both nuclear physics \cite{wmzha87,zhang88,FDSM} and for the strongly-correlated 
electrons leading to cuprate and iron-based high-temperature superconductivity 
\cite{guid01,lawu03,su4review}.  They may be viewed as the generalization of a 
quantum critical point to an entire {\em quantum critical phase,} and may 
represent a fundamental organizing principle for quantum critical behavior.  The 
physical implications of this SO(7) critical dynamical symmetry for graphene 
quantum Hall physics will be discussed further below.

\section{\label{h:so8CoherentStates}Generalized Coherent States}

The dynamical symmetry limits discussed above represent special solutions 
resulting from particular choices of the coupling parameters appearing in the 
Hamiltonian.  For arbitrary choices of the coupling parameters the solutions 
will correspond generally to superpositions of the different symmetry-limit 
solutions and will not have exact analytical forms. In this more general case it 
is quite feasible to obtain solutions numerically, since the collective subspace 
is highly truncated relative to the full Hilbert space.  However, there is a 
powerful alternative approach:  the {\em generalized coherent state 
approximation,} which permits {\em analytical solutions} for arbitrary choices 
of the coupling parameters in the Hamiltonian.

For the SO(8) Lie algebra introduced in this paper for graphene the 
Gilmore--Perelomov algorithm \cite{arec1972,gilm1972,pere1972,gilm1974,zhan90} 
may be implemented to obtain solutions in terms of a set of generalized coherent 
states. These solutions represent the most general Hartree--Fock--Bogoliubov 
theory that can be formulated in the space, subject to a dynamical symmetry 
constraint \cite{weimin1989}.  The solutions of this Symmetry-Constrained 
Hartree--Fock--Bogoliubov (SCHFB) theory correspond to determining the stable 
points of energy surfaces, which represent the coherent-state expectation values 
of the effective Hamiltonian on the coset space. Thus the coherent state 
solutions also represent a microscopically-derived implementation of 
Ginzburg--Landau theory.  These coherent state solutions are uniquely well 
suited to study the interplay of competing spontaneous symmetry breaking in 
determining the ground state of the system and its properties.

\subsection{\label{sh:so8Coherent} Constructing SO(8) Coherent States}

The coherent states associated with the full set of subgroup chains in 
\fig{dynamicalChains_graphene} will be discussed in future work.  In this paper 
the power of the method will be illustrated succinctly by restricting to 
the coherent states associated with the subgroup chains of SO(8) that contain 
the SO(5) subgroup, as illustrated in \fig{dynamicalChains_coherentSO5}.%
\singlefig
{dynamicalChains_coherentSO5}  
{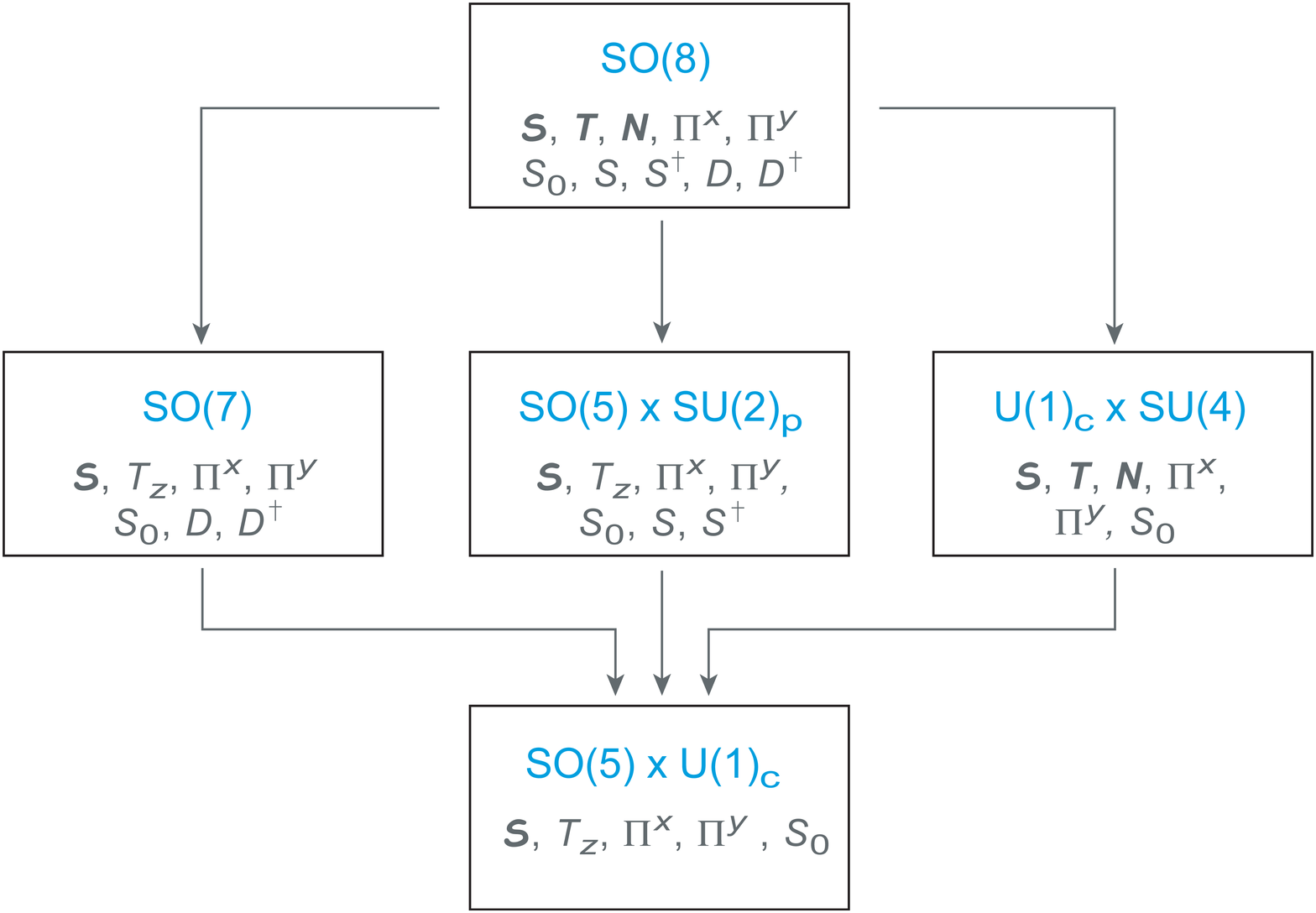}
{\figup}
{0pt}
{0.20}
{Subgroup chains included in the present coherent state analysis.}
Thus the corresponding coherent state solutions will represent a superposition 
of the symmetry-limit solutions for the 
$$
\begin{array}{c}
\soeight\supset\sofive\times\sutwo\tsub p\supset\sofive
\\[5pt]
\soeight\supset\sufour\supset\sofive
\qquad
\soeight\supset\soseven\supset\sofive
\end{array}
$$
dynamical symmetries.  These coherent state solutions will be seen to have the 
following properties: 

\begin{enumerate}

\item
The $\soseven$ dynamical symmetry will play the role of a critical dynamical 
symmetry interpolating between $\sufour$ and $\sofive$ symmetry-limit 
solutions.

\item
Because of fundamental symmetries obeyed by the  wavefunction, the coherent 
state solutions may be parameterized in terms of a single collective parameter 
$\beta$ that governs the mixture of the $S$ and $D$ pairs defined in 
\eq{coupled1.4} contributing in the ground state.

\item
The collective parameter $\beta$ may also be interpreted physically in terms of 
the pair configurations displayed in \fig{S_D_pairs_brillouin}.

\end{enumerate}

\noindent
The SO(8) coherent states corresponding to the symmetry structure in 
\fig{dynamicalChains_coherentSO5} have been developed previously in Ref.\ 
\cite{zhang88} for nuclear physics applications and will be adapted extensively 
to development of the present formalism.

\subsection{\label{sh:}SO(8) Coherent State Energy Surfaces}

Let us now consider the energy surfaces that may be computed from the coherent 
states, which link the SO(8) solutions to Ginzburg--Landau theory. Within the 
coherent state formalism the ground state energy may be determined through the 
variational requirement
$
\delta \mel{\eta}{H}{\eta} = 0 ,
$
where $\ket \eta$ is the coherent state, $H$ is the SO(8) Hamiltonian,
\begin{equation}
H = H' + G_0 S^\dagger S + \sum_{r=1,2,3} b_r P^r \tightdot P^r,
 \label{coherent1.12}
\end{equation}
and the coefficients $G_0$ and $b_r$ are 
functions of the effective interaction. 
For the dynamical symmetry chains having SO(5) as a subgroup the energies take 
the general form 
\cite{zhang88}
\begin{equation}
E\tsub g(n,\beta) = N\tsub g \left[ A\tsub g\beta^4 + B\tsub g(n)\beta^2 + 
C\tsub g(n) + D\tsub g(n,\beta) \right] ,
\label{generalEnergyFormula}
\end{equation}
where the group-dependent parameters $N\tsub g$, $A\tsub g$, $B\tsub g(n)$, 
$C\tsub g(n)$, and $D\tsub g (n,\beta)$ are given in 
\tableref{stiffnessParameters}.%
{\renewcommand\arraystretch{1.0}
\begin{table*}[tpb]
  \vspace*{\tabup}
  \centering
  \caption{Parameters for the energy surfaces $E\tsub g(n,\beta)$ defined in 
\eq{generalEnergyFormula}}
  \label{tb:stiffnessParameters}
    \begin{centering}
      \setlength{\tabcolsep}{6 pt}
      \vspace{\tabtitlesep}
      \begin{tabular}{cccccc}
      
        \tableLineOne
        
            g &
            $N\tsub g$ &
            $A$ &
            $B(n)$ &
            $C(n)$ &
            $D(n,\beta)^\dagger$

        \tableLineTwo
        
            SU(2) &
            $G_0$ &
            $2\Omega(\Omega-2)$ &
            $-n(\Omega-2)$ &
            $n(\Omega-\tfrac12 n + \frac {n}{\Omega})/4$ &
            $\tfrac12 \Omega^2 F(n,\beta)$

        \\ [\tablerowMore]  
        
            SO(5) &
            $b_3$ &
            $-8\Omega$ &
            $4n$ &
            $n(1-\frac{n}{2\Omega})$ &
            $-2\Omega F(n,\beta)$

        \\  [\tablerowMore]     
        
            SU(4) &
            $b_2$ &
            $-4\Omega(\Omega+3)$ &
            $2n(\Omega+3)$ &
            $\frac{5}{4\Omega}(2\Omega -n)$ &
            $0$

        \\  [\tablerowMore]  
        
            SO(7) &
            $G_2$ &
            $2\Omega(\Omega+4)$ &
            $-n(\Omega+4)$ &
            $-\frac n4 \left(\Omega - \tfrac{n}{2} - \tfrac{2n}{\Omega} 
+4\right)
+\frac \Omega4 (\Omega+10)$ &
            $-\tfrac12 \Omega(\Omega+4) F(n,\beta)$

        \tableLineThree
        
        \noalign{\vskip 5pt}
            \multicolumn{6}{l}{\footnotesize 
            $^\dagger$The function $F(n,\beta)$ is defined by
$F(n,\beta) \equiv \left( \tfrac{n}{2\Omega} - 2\beta^2\right)
\left[
(1-\tfrac{n}{2\Omega})^2 -4\left( \tfrac{n}{2\Omega} -\beta^2\right) \beta^2
\right]^{1/2}$.
} 
            
      \end{tabular}
    \end{centering}
\end{table*}
}%

Our primary interest in this discussion is in the ground state properties of 
graphene in a strong magnetic field.  The ground states in the coherent state 
approximation at fixed $n/2\Omega$ will be given by those values of $\beta 
\equiv \beta_0$ that correspond to minima of the energy surface $E(n,\beta)$. 
 These are determined by the values of $\beta$ satisfying
\begin{equation}
\pardiv{E\tsub g(n, \beta))}{\beta} = 0
\qquad
\frac{\partial^2 E\tsub g(n, \beta))}{\partial\beta^2} > 0.
\label{Eminima}
\end{equation}
Evaluating these constraints for the energy surfaces 
\eqnoeq{generalEnergyFormula}, one finds that the minima $\beta\tsup g_0$ are 
given by \cite{zhang88}
\begin{equation}
\beta^{\sutwo\times\sofive}_0 = 
\beta^{\soseven}_0 = 0
\qquad
\beta^{\sufour}_0 = \pm \sqrt{\frac{n}{4\Omega}} .
\label{betaMinima}
\end{equation}
The coherent state energy surfaces for the $\sofive\times\sutwo$, SO(7), and 
SU(4) symmetry limits computed from \eq{generalEnergyFormula} using the entries 
in \tableref{stiffnessParameters} are shown as functions of $\beta$ for several 
values of the fractional occupation $\shellfill$ in 
\fig{coherentEnergySurfaces_graphene}.
\singlefig
{coherentEnergySurfaces_graphene}  
{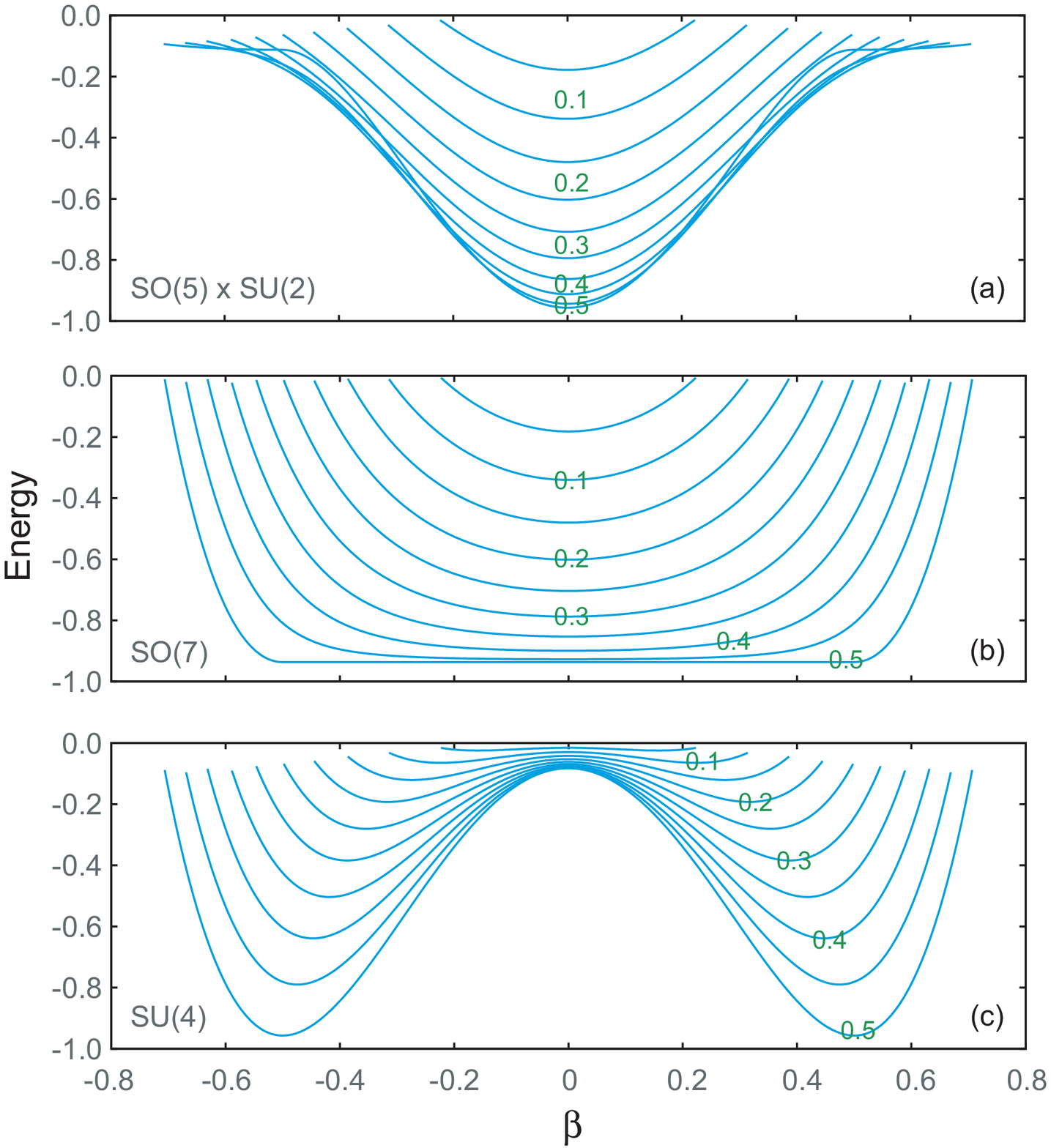}
{\figup}
{0pt}
{0.44}
{Coherent state energy surfaces as a function of the order parameter $\beta$ for 
three of the SO(8) dynamical symmetry limits.  The curves are labeled by the 
fractional occupation $\shellfill=n/2\Omega$ defined in \eq{fillingFraction}, 
where $n$ is the particle number and $2\Omega$ is the maximum number of 
particles that can be accommodated in the Landau level. The formalism is 
particle--hole symmetric so curves for fractional occupations with $\shellfill > 
\tfrac12$ are equivalent to those shown above, but with the fractional 
occupation counted in terms of number of holes. For example, the curves for 
$\shellfill=0.4$ and $\shellfill=0.6$ are equivalent. For the direct product 
group $\sutwo\times\sofive$ the energy surface is the sum of contributions to 
\eq{generalEnergyFormula} from SU(2) and SO(5). As will be demonstrated later, 
the SU(2) contribution typically dominates that of SO(5).}
There one sees that indeed the minima for the $\sofive\times\sutwo$ and SO(7) 
limits are at $\beta_0 = 0$, and the minima for the SU(4) limit are at $ 
\beta_0 = \pm 
\sqrt{n/4\Omega}$. 

Although the minimum energies for both the $\sofive\times\sutwo$ and SO(7) 
limits are consistent with $\beta_0=0$, \fig{coherentEnergySurfaces_graphene} 
shows that these symmetries differ fundamentally in the localization of the 
minimum.  For  $\sofive\times\sutwo$ the energy surface has a deep minimum at 
$\beta_0=0$ but for SO(7)  the energy surface is very flat around $\beta_0 = 0$, 
with a broad range of $\beta$ giving essentially the same ground state energy.  
This highly-degenerate SO(7) ground state has significant physical implications 
that will be discussed further below.

\section{\label{h:conservationLaws}Elementary Conservation Laws}

The SO(8)  generalized coherent state is equivalent to the 
Hartree--Fock--Bogoliubov (HFB) approximation subject to a symmetry constraint. 
Since HFB  is  a BCS-type approximation married to a Hartree--Fock 
mean field, its solutions correspond to symmetry-breaking intrinsic states. In 
particular, the BCS-like state conserves
the physical particle number only on average, and the Hartree--Fock mean field 
may break both translational and rotational invariance.  Let us address these 
issues for the SO(8) coherent state.

\subsection{\label{sh:fluctuationN}Fluctuation in Particle Number}

The fractional uncertainty in electron number $\Delta n$ for the SO(8) coherent 
state is given by \cite{zhang88}
\begin{equation}
(\Delta n)^2 =
\ev{\hat n^2} - \ev{\hat n}
=
2n - \frac{n^2}{\Omega} + 16 \Omega \beta_0^4 - 8n \beta_0^2 ,
 \label{particleNumberFluctuation}
\end{equation}
where $\beta_0$ is the value of $\beta$ at the minimum energy, 
given by \eq{betaMinima} in the symmetry limits.
Expressing \eq{particleNumberFluctuation} in terms of the fractional occupation 
$\shellfill = n/2\Omega$,
in the $\sofive\times\sutwo$ and SU(4) limits, respectively,
one obtains
\begin{equation}
\left[ \frac{\Delta n}{n} \right]_{{\rm SO}_5\times\sutwo}
\hspace{-5pt}
= \sqrt{\frac{1-\shellfill}{\shellfill\Omega}}
\qquad
\left[\frac{\Delta n}{n}\right]_{{\rm SU}_4}
\hspace{-5pt} = 
\sqrt{\frac{1-2\shellfill}{\shellfill\Omega}} ,
\label{particleNumberFluctuation3}%
\end{equation}  
From these results one may notice two important things.

\begin{enumerate}

\item 
The fluctuation in particle number is large at low degeneracy $\Omega$ but 
decreases 
 with increasing  $\Omega$.  

\item
If SU(4) symmetry is realized $\Delta n/n$ decreases with increasing 
$\shellfill$ and {\em vanishes  identically} at $\shellfill = \tfrac12$ for any 
$\Omega$, which corresponds to the 
fractional occupation for the ground state of undoped graphene.

\end{enumerate} 

\noindent
Thus it is expected that the 
current theory applied to graphene has negligible particle number fluctuation 
$\Delta n/n$ in the SU(4) limit. In the $\sofive\times\sutwo$ limit the particle 
number fluctuation $\Delta n/n$ remains finite for all $\shellfill$ but it 
becomes very small as $\Omega$ becomes large, particularly near $\shellfill = 
\tfrac12$. Thus it too may be neglected in the large-$\Omega$ limit.  
Comparison with
Table \ref{tb:degeneracyTable} suggests that graphene quantum Hall experiments 
involve sufficient degeneracy that particle number fluctuation in the 
coherent state solution is not significant.

\subsection{\label{sh:rotation-translation}Translational and Rotational 
Invariance}

The coherent state approximation represents a mean field localized in 
spatial position and orientation, so it violates translational and rotational 
invariance.  However, because the crystal is generally macroscopic, the net 
violation of these symmetries may be expected to be negligible.  One concludes 
that for applications of coherent state methods to graphene, violations of 
particle number conservation, rotational invariance, and translational 
invariance are negligible in realistic systems.

\section{\label{h:wfCS} Coherent-State Wavefunctions and Order Parameters}

The generalized coherent state method has been used above to calculate total 
energy surfaces for quantum Hall states in graphene, but one also  may use the 
coherent state wavefunctions and appropriate operators to calculate matrix 
elements of other relevant observables.  This section addresses the nature 
of the wavefunction and the matrix elements that can serve as order parameters.

\subsection{\label{sh:AForder} Order Parameters}

A significant consequence of the SO(8) dynamical symmetry structure displayed in 
\fig{dynamicalChains_coherentSO5} is that the phases may be distinguished in 
terms of a single order parameter and its fluctuations, which may be taken to 
be 
$\beta$. Let us now characterize in more depth the physical meaning of this 
order parameter. In \S\ref{sh:orderParameters} an antiferromagnetic 
order parameter $\ev{N_z}$ was defined.  In the coherent state approximation 
the onset of 
AF order is signaled by an energy-surface minimum at a finite value of $\beta$. 
 Because $N_z = P^2_0$ [see \eq{ngx9}], the antiferromagnetic order parameter 
$\ev{N_z}$ is related to the coherent state AF order parameter $\beta$  by the 
intrinsic state matrix element of $P^2_0$ \cite{zhang88}
\begin{align}
\ev{N_z} &=
|b_2| \mel{{\rm int}, \beta, \gamma, n}{P^2_0}{{\rm int}, \beta, \gamma, n}
\nonumber
\\
&=
2 \Omega |b_2| \left(\shellfill - \beta^2\right)^{1/2} \ \beta ,
\label{intrinsic1.1}
\end{align}
where $b_2$ is the coupling strength for the $P^2\tightdot P^2$ term in the 
Hamiltonian.
Each value of $\beta$ corresponds to a unique value of $\ev{N_z}$, so $\beta$ 
is a measure of antiferromagnetic order.

The location of the maxima  may be obtained by 
setting the derivative with respect to $\beta$ of \eq{intrinsic1.1} equal to 
zero, which yields that  $\ev{N_z}\tsub {max}$ for a given $n$ occurs at a 
$\beta$ of
\begin{equation}
\beta = \sqrt{\frac{n}{4\Omega}} = \beta^{\sufour}_0,
\label{maxNz}
\end{equation}
where \eq{betaMinima} was used to make the identification in the last step.
Thus, if $\beta \ne 0$ the maximum value of $\ev{N_z}$ maps to a value of 
$\beta$ that corresponds to a minimum of the energy surface (a ground state) in 
the SU(4) limit.  Substituting \eq{betaMinima} for $\beta$ in 
\eq{intrinsic1.1}, for $\beta \ne 0$ ground states the AF order parameter 
$\ev{N_z}$ depends on the electron number $n$ as 
\begin{equation}
\ev{N_z}\tsub {max} = 2\Omega |b_2|\left( \frac{n}{4\Omega}\right)
= \Omega |b_2| \shellfill .
\label{NzVs_n}
\end{equation}
The SO(8) model is particle--hole symmetric so $n$ or $f$ count electrons up 
to half filling and holes for greater than half filling.  Hence the maximum AF 
collectivity occurs for half filling of the single valence Landau level.

\subsection{\label{sh:cswf} Coherent State Wavefunctions}

As was discussed in \S\ref{h:conservationLaws}, the coherent state 
wavefunction corresponding physically to $N=2n$ pairs conserves particle number 
only on average and so is a superposition of terms having different numbers of 
pairs. In Eq.\ (5.27) of Ref.\ \cite{zhang88} the SO(8) coherent state is 
decomposed into terms of definite pair number $p$ according to
\newcommand{\Dcoeff}{\kappa}
\newcommand{\Ccoeff}{C_p}
\begin{equation}
\ket{\beta}
= \sum _{p} \Ccoeff \left( S^\dagger + \Dcoeff D_0^\dagger \right)^p
\ket0,
\label{coherentFN1.1}
\end{equation}
where  $\ket\beta$ denotes 
an intrinsic state with order parameter $\beta$ and closed forms for $C_p$ and 
$\kappa$ are given in Ref.\ \cite{zhang88}.
According to Eq.\ (5.28) of Ref.\ \cite{zhang88}, the values of
$\Dcoeff$ that correspond to the minima of the potential energy surface at 
$\beta=0$ for the $\sofive\times\sutwo$ limit and $\beta = \pm 
\sqrt{n/4\Omega}$ for the 
$\sufour$ limit [see \eq{betaMinima}], respectively, are
\begin{equation}
\Dcoeff_{{\rm SO}_5\times{\rm SU}_2} = 0
\qquad
\Dcoeff_{\sufour} = \pm 1,
\label{coherentFN1.3}
\end{equation}
so the SO(8) coherent state wavefunction \eqnoeq{coherentFN1.1} in the 
$\sofive\times\sutwo$ and SU(4) limits, respectively, becomes
\begin{equation}
\begin{array}{c}
\displaystyle
\ket{{\rm SO}_5\times{\rm SU}_2} = 
\sum _{p} \Ccoeff \left( S^\dagger \right)^p
\ket0,
\\[5pt]
\displaystyle
\ket{{\rm SU}_4} =
\sum _{p} \Ccoeff \left( S^\dagger \pm D_0^\dagger \right)^p \ket0
=2\sum _{p} \Ccoeff \left( Q_\pm^\dagger \right)^p \ket0 ,
\end{array}
\label{coherentFN1.4}%
\end{equation}
where \eq{Qdef} was used.

As discussed in \S\ref{h:conservationLaws},  fluctuations in particle number 
are negligible in the large-$\Omega$ limit for SO(8) coherent states, 
implying that the summations in Eqs.\ \eqnoeq{coherentFN1.4}  become 
dominated by terms with $p \simeq N$.  Thus for large $\Omega$ the coherent 
state wavefunctions are well approximated up to a normalization by
\begin{subequations}
\begin{align}
\ket{{\rm SO}_5\times{\rm SU}_2} &\simeq 
 \left( S^\dagger \right)^N
\ket0
\label{largeOmegWF.a}
\\
\ket{{\rm SU}_4} &\simeq
 \left( Q_\pm^\dagger \right)^N  \ket0 .
\label{largeOmegWF.b}
\end{align}
\label{largeOmegWF}%
\end{subequations}
As seen from \tableref{orderParms}, the SU(4) state of \eq{largeOmegWF.b} is a 
coherent superposition of $Q_-$ or $Q_+$ pairs, each contributing vanishing 
ferromagnetic order  $\ev{\spin_z}$ and charge density wave order 
 $\ev{T_z}$, but non-zero AF order  $\ev{N_z}$.  Conversely, 
the $\sofive\times\sutwo$ state of \eq{largeOmegWF.a} is a coherent 
superposition of $S$  pairs, each with vanishing $\ev{\spin_z}$, $\ev{T_z}$, and 
$\ev{N_z}$.  

Thus the $\sofive\times\sutwo$ and SU(4) limits of the SO(8) symmetry are 
distinguished by the order parameter $\ev{N_z}$, which is zero in the 
$\sofive\times\sutwo$ state and is non-zero in the SU(4) state.  Equivalently, 
the coherent state order parameter $\beta$ vanishes in the pure 
$\sofive\times\sutwo$ limit and is equal to $\pm \sqrt{n/4\Omega}$ in the pure 
SU(4) limit [see \eq{betaMinima} and \fig{coherentEnergySurfaces_graphene}]. 
Equation (\ref{generalEnergyFormula}) depends only on even powers of $\beta$ so 
the sign for the two possible spontaneously broken symmetry solutions does not 
affect the energy.

For undoped graphene the Fermi surface corresponds to 
the $\shellfill = 0.5$ curves of \fig{coherentEnergySurfaces_graphene}.  These 
are shown in \fig{coherentEnergySurfaces_grapheneGS} for the 
$\sofive\times\sutwo$, SO(7), and SU(4) limits.
\singlefig
{coherentEnergySurfaces_grapheneGS}  
{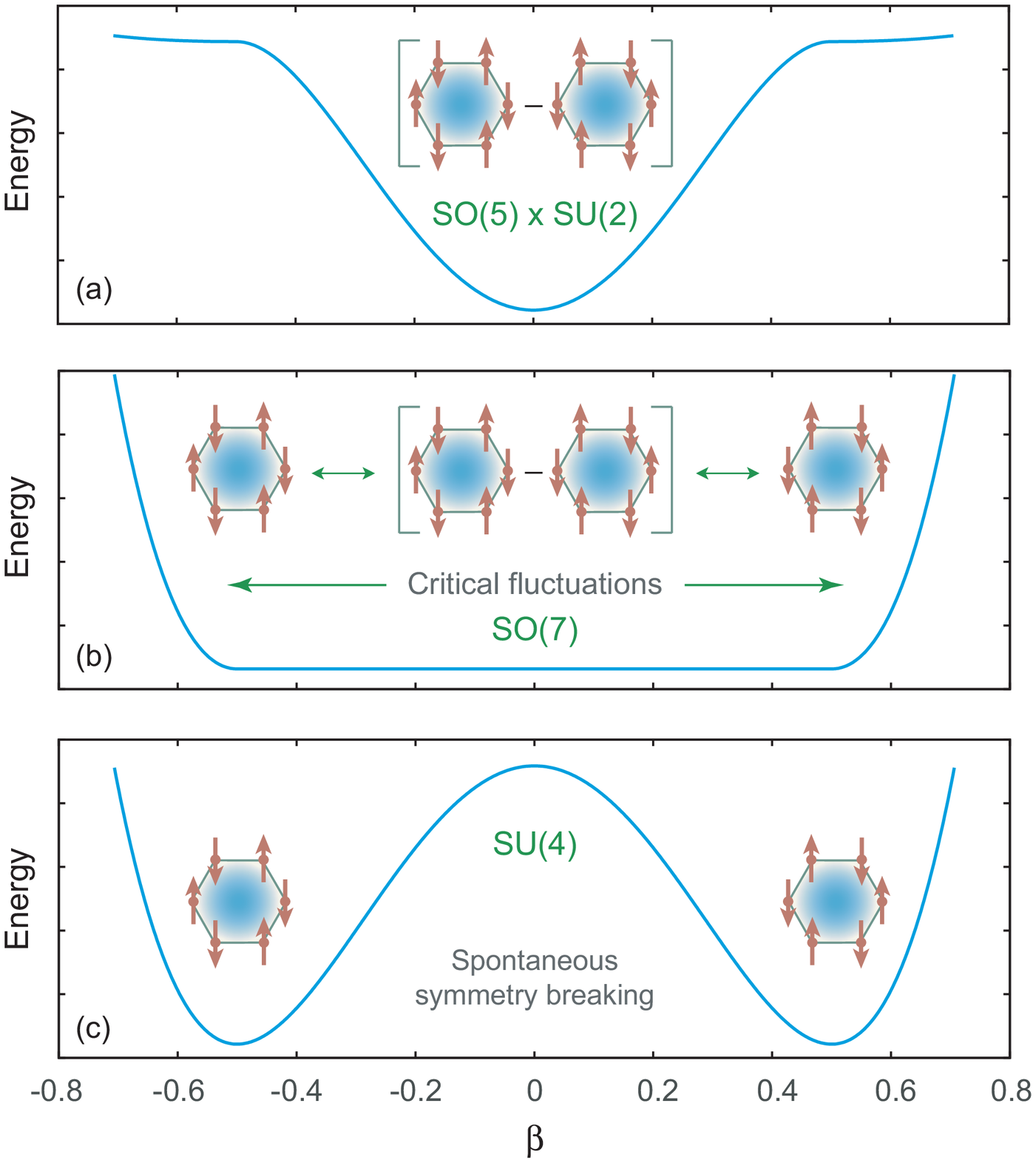}
{\figup}
{0pt}
{0.48}
{Ground-state  energy surfaces in coherent state approximation for three of the 
SO(8) dynamical symmetry limits.  (a)~The $\sofive\times\sutwo$ limit.  (b)~The 
SO(7) limit.  (c)~The SU(4) limit. The diagrams indicate schematically the 
corresponding wavefunctions, as suggested by Eqs.\ 
\eqnoeq{coherentFN1.1}--\eqnoeq{largeOmegWF} and \fig{S_D_pairs_brillouin}.
}
\begin{itemize}

\item
The $\sofive\times\sutwo$ limit in \fig{coherentEnergySurfaces_grapheneGS}(a) 
has its minimum $\beta_0$ at $\beta = 0$.  It corresponds to states of the form 
\eqnoeq{largeOmegWF.a}, with vanishing expectation values of $\spin_z$, $T_z$, 
and $N_z$. 

\item
The SU(4) limit of \fig{coherentEnergySurfaces_grapheneGS}(c) corresponds to 
states of the form \eqnoeq{largeOmegWF.b}, with minima located at
$$
\beta_0 = \pm \sqrt{\frac{n}{4\Omega}} = 
\pm \frac12.
$$
When the symmetry is broken spontaneously by choosing one of these possibilities 
(they are equivalent since the energy depends only on even powers of $\beta$), 
the resulting state has $\ev{\spin_z} = \ev{T_z} = 0$ (no spin or isospin 
order), but $\ev{N_z}\ne0$ (spin density wave or AF order).

\item
The SO(7) limit of \fig{coherentEnergySurfaces_grapheneGS}(b) corresponds to a 
critical dynamical symmetry that interpolates  between the 
$\sofive\times\sutwo$ 
and SU(4) states through critical fluctuations in the antiferromagnetic order.

\end{itemize}

 \noindent
Thus the SO(8) dynamical symmetry limits illustrated in 
\fig{coherentEnergySurfaces_grapheneGS} represent a rich set of collective 
states that can be distinguished by the expectation value and fluctuations 
associated with the order parameter $\beta$.

\section{\label{h:quantumPhaseso8} SO(8) Quantum Phase Transitions}

The SO(8) coherent state solution can be used to study transitions among the 
phases defined in \fig{coherentEnergySurfaces_grapheneGS}. For the $u=0$ space 
(no broken pairs) assumed here, $H'$ yields a constant that is neglected and 
\eq{coherent1.12} may be expressed as
\begin{align}
 H = 
 G_0 S^\dagger S + b_2 P^2\tightdot P^2 + b_3 \casimir\sofive + 
\frac{b_1-b_3}{5} \,\casimir\sutwo,
 \label{coherent1.13a}
\end{align}
The last two terms yield constants when evaluated in a given representation,
and $\casimir\sofive$ is found to contribute negligibly to the total energy 
compared with $\casimir\sufour$. 
Therefore, it will be instructive to set $b_1 = b_3 =0$ and 
study the approximate SO(8) Hamiltonian
\begin{equation}
H = 
G_0 S^\dagger S + b_2 P^2\tightdot P^2 .
 \label{modelHamiltonian}
\end{equation}
From Eqs.\ \eqnoeq{casimir_so5}--\eqnoeq{casimir_so8}, one 
finds that 
$$
\begin{array}{c}
\ev{S^\dagger S} \sim \ev{\casimir{\sutwo}} \qquad
\ev{P^2 \tightdot P^2} \sim \ev{\casimir{\sufour}} 
\\[5pt]
\ev{S^\dagger S} + \ev{ P^2\tightdot P^2} \sim \ev{\casimir{\soseven}}
\end{array}
$$
if constants are neglected. Thus the model Hamiltonian \eqnoeq{modelHamiltonian} 
may be tuned to favor the $\sofive\times\sutwo$, SO(7), or SU(4) phases by 
varying the ratio of the coupling parameters $G_0$ and $b_2$.

\newcommand{\tuner}{q}

\subsection{\label{sh:tuningHam} Tuning Quantum Phase Transitions}

To study the quantum phase transitions of the SO(8) model with the 
approximate Hamiltonian \eqnoeq{modelHamiltonian}, it is convenient to define 
a parameter $q\equiv b_2/G_0$
and to rewrite \eq{modelHamiltonian} as
\begin{equation}
H = G_0(S^\dagger S + \tuner P^2\tightdot P^2) .
 \label{modelHamiltonian2}
\end{equation}
Thus the value of $\tuner$ tunes the Hamiltonian \eqnoeq{modelHamiltonian2} 
between $\sutwo\times\sofive$ and SU(4) phases via an intermediate SO(7) 
phase exhibiting quantum critical behavior.
\begin{enumerate}
\item 
If $\tuner <<1$ the ground-state energy surface is approximated by 
\fig{coherentEnergySurfaces_grapheneGS}(a), with a minimum  at $\beta = 0$, no 
antiferromagnetic order, and $\sutwo\times\sofive$ symmetry.
\item
If $\tuner >> 1$ the ground-state energy surface is approximated by 
\fig{coherentEnergySurfaces_grapheneGS}(c), with an energy minimum at $\beta \ne 
0$ implying $\sufour$ symmetry and antiferromagnetic order.
\item
If $\tuner \sim 1$, the ground-state energy surface is approximated by 
\fig{coherentEnergySurfaces_grapheneGS}(b) and the system exhibits SO(7) 
critical dynamical symmetry, with large fluctuations in the AF order parameter 
$\beta$.
\end{enumerate}
Let us now use the Hamiltonian \eqnoeq{modelHamiltonian2} to study the quantum 
phase transitions and spontaneously broken symmetry of the SO(8) model.

\subsection{\label{sh:tuningSurfaces} Energy Surfaces and Quantum Phase 
Transitions}

If  terms involving $\ev{P^1\tightdot 
P^1}$ and $\ev{P^3\tightdot P^3}$ are ignored (as justified above),
\eq{generalEnergyFormula} with the parameters in \tableref{stiffnessParameters} 
imply that 
$$
E_\sutwo + E_\sufour - E_\sofive 
\simeq G_0 \ev{S^\dagger S} + b_2 \ev{P^2\tightdot P^2} .
$$
Hence the energy surfaces corresponding to the Hamiltonian 
\eqnoeq{modelHamiltonian2} may be expressed as
\begin{align}
E(n,\beta) &= \ev H = G_0 \ev{S^\dagger S} + b_2 \ev{P^2\tightdot P^2}
\nonumber
\\
& \simeq
E_\sutwo(n,\beta) + E_\sufour(n,\beta) - E_\sofive(n,\beta),
\nonumber
\\
& \simeq
E_\sutwo(n,\beta) + E_\sufour(n,\beta) .
 \label{coherent1.13d}
\end{align}
The variation of the energy surface computed from \eq{coherent1.13d} with the 
control parameter $\tuner = G_0/b_2$ for half filling (ground state for undoped 
graphene) is shown in \fig{coherentEnergySurfaces_control2D};%
\singlefig
{coherentEnergySurfaces_control2D}  
{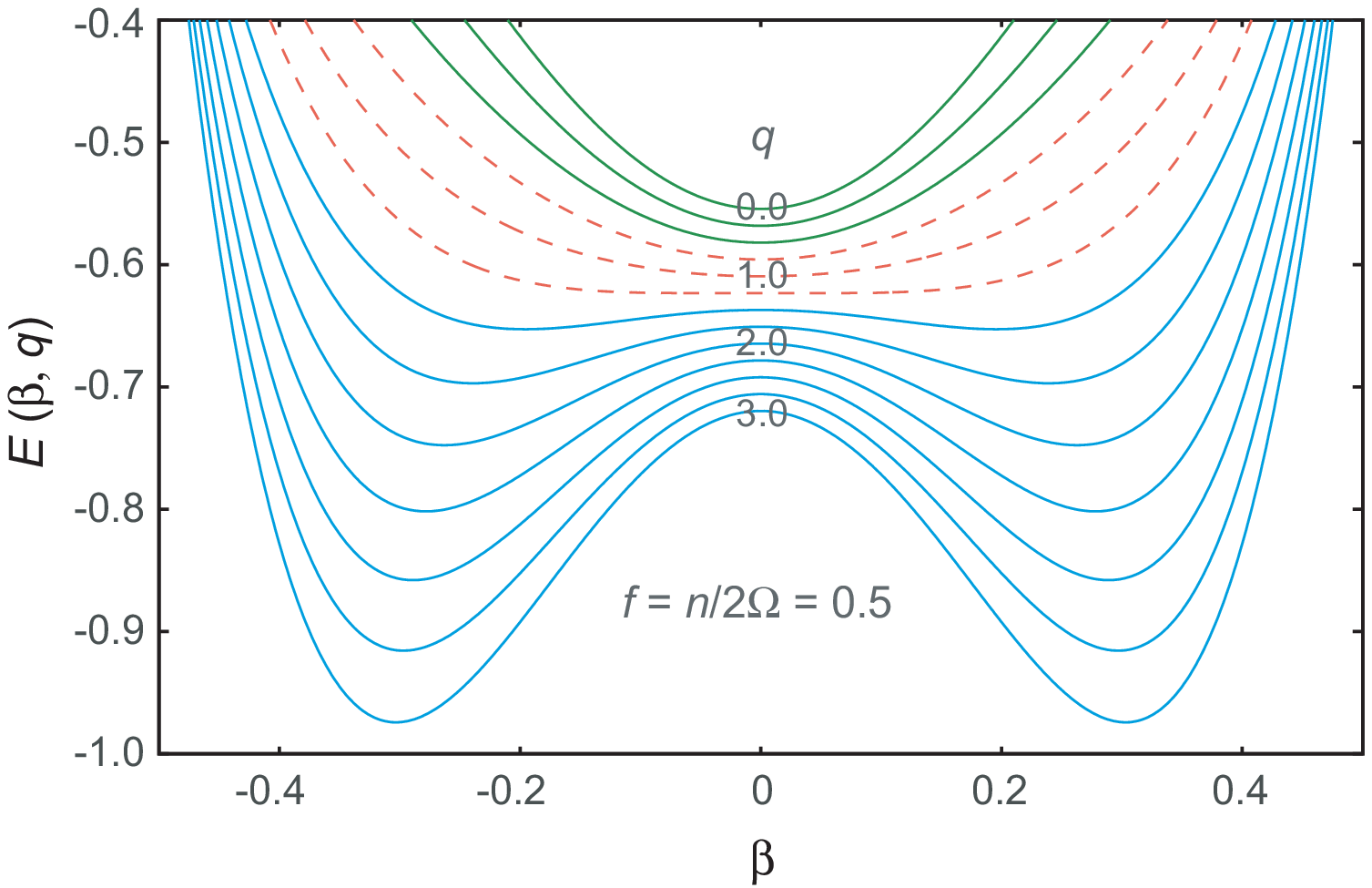}
{\figup}
{0pt}
{0.56}
{Quantum phase transitions with coupling strength as control parameter: 
coherent state energy surface as a function of the control parameter $\tuner 
\equiv G_0/b_2$ for a degeneracy parameter $\Omega=50$ and a fractional 
occupation $\shellfill = n/2\Omega = 0.5$. The solid green curves from $\tuner 
\sim 0-0.5$ correspond to approximate $\sofive\times\sutwo$ symmetry, the solid 
blue curves for $\tuner \ge 1.5$ correspond to approximate SU(4) symmetry.  The 
dashed red curves near $\tuner \sim 1$ correspond to an approximate SO(7) 
symmetry mediating the quantum phase transition from $\sofive\times\sutwo$ to 
SU(4) symmetry. }
By tuning the control parameter from $\tuner=0$ to $\tuner >> 1$, one sees that 
the system undergoes a quantum phase transition near $\tuner = 1$ from an 
approximate $\sofive \times \sutwo$ state with energy minimum at $\beta=0$ [see 
\fig{coherentEnergySurfaces_grapheneGS}(a)] to an approximate SU(4) state having 
energy minima at $\beta=\pm(n/4\Omega)^{1/2}$ [see 
\fig{coherentEnergySurfaces_grapheneGS}(c)  and \eq{betaMinima}].  For 
$\tuner\sim 1$ the system has an approximate SO(7) dynamical symmetry [see 
\fig{coherentEnergySurfaces_grapheneGS}(b)], with no well-defined minimum for 
the energy as a function of $\beta$, but with large fluctuations in $\beta$ 
implied by a highly-degenerate ground state.

For fixed values of the coupling parameters $G_0$ and $b_2$, phase transitions 
may be mediated by changing the particle occupancy. 
\fig{coherentEnergySurfaces_vs_n_2D} illustrates for different values of 
$n/2\Omega$ at fixed $b_2 = 2.5 G_0$.  One sees that as the particle number is 
increased the system makes a transition from approximate $\sofive\times\sutwo$ 
symmetry with $\beta=0$ to SU(4) symmetry with $\beta \ne 0$ through a critical 
SO(7) symmetry for which the energy is highly degenerate in $\beta$.

\section{\label{h:zeemanTerm} Effect of the Zeeman Term}

In the coherent state approximation the dynamical symmetry 
structure of \fig{dynamicalChains_coherentSO5} has been examined
and not the full group structure given in \fig{dynamicalChains_graphene}.  For 
the group chains that contain the SO(5) subgroup, the only physical implication 
is to omit the physical effect of Zeeman splitting from dynamical symmetry 
Hamiltonian [which would break the SO(5) subgroup down into a U(1) subgroup 
generated by the $z$ component of the physical spin].  Our primary concern in 
this discussion is the structure associated with the $n=0$ Landau level, for 
which the effect of the Zeeman term is expected to be small (see 
\S\ref{h:applicationsGraphene}). Thus, one may view the effect of the Zeeman 
term as a perturbation on the results obtained thus far that will act only on 
the spin part of the wavefunction.  As 
Kharitonov \cite{khar2012} has already discussed, the competition of the Zeeman 
term with the valley interactions will convert the antiferromagnetic solution 
into a canted antiferromagnetic solution.

\singlefig
{coherentEnergySurfaces_vs_n_2D}  
{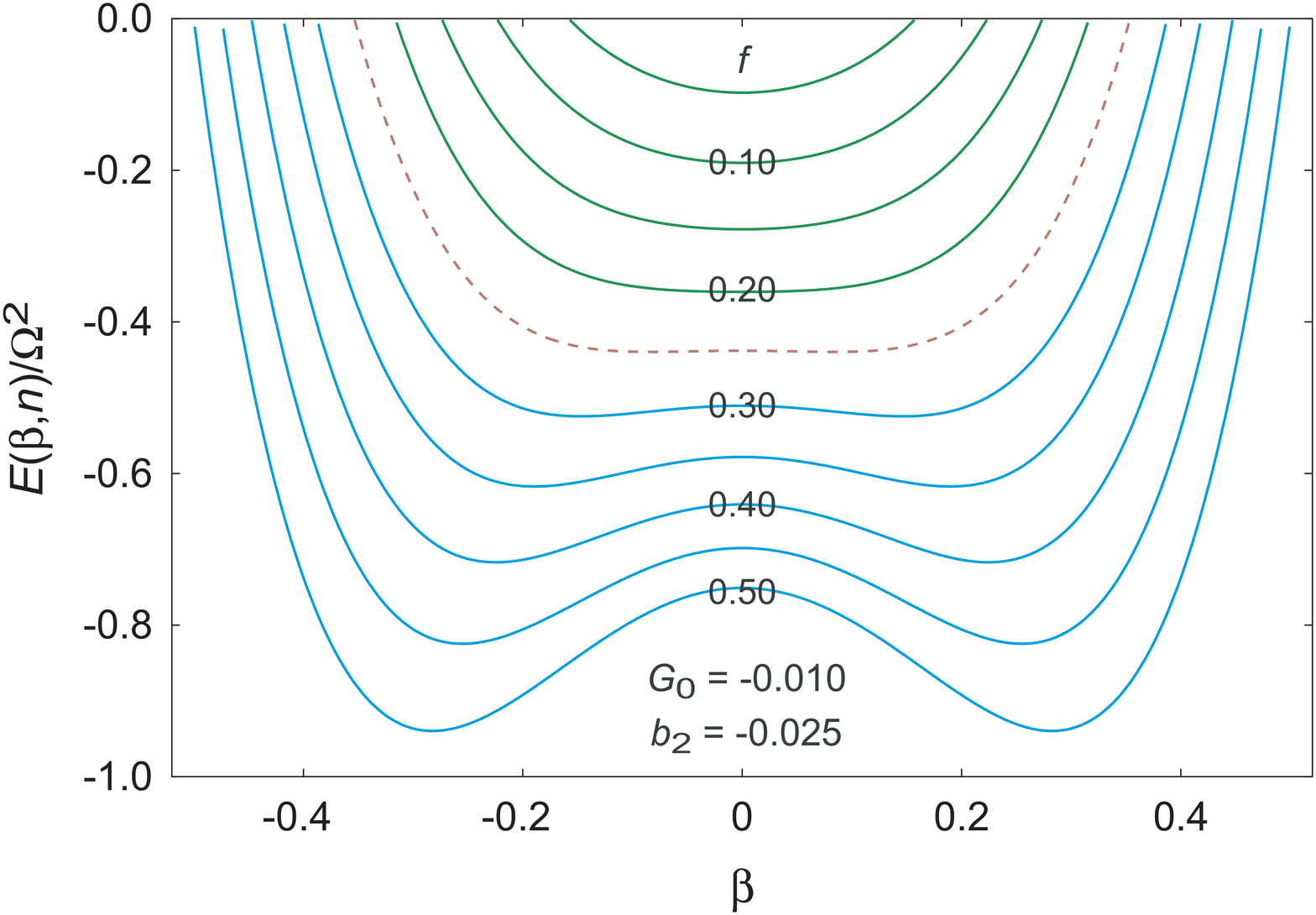}
{\figup}
{0pt}
{0.22}
{Quantum phase transitions with particle number as control parameter: coherent 
state energy surfaces for different filling fraction $f=n/2\Omega$ at 
fixed values for $G_0$ and $b_2$ with $\Omega=10,000$. The solid green curves 
for $n/2\Omega \sim 0-0.2$ correspond to approximate $\sofive\times\sutwo$ 
symmetry, the solid blue curves for $n/2\Omega \sim 0.3-0.5$ correspond to 
approximate SU(4) symmetry.  Curves near $n/2\Omega \sim 0.25$ (dashed red) 
correspond to an approximate SO(7) symmetry mediating the quantum phase 
transition from $\sofive\times\sutwo$ to SU(4) symmetry.
}

\section{\label{h:applicationsGraphene} Physical Graphene 
States}

Placing a strong magnetic field on graphene leads to high level degeneracy at 
energies corresponding to the quantized Landau levels.  In general, interacting 
electronic systems with high level degeneracy near the Fermi surface can produce 
(even for relatively weak interactions) a rich variety of collective states that 
differ qualitatively from the non-interacting ground state.  These states 
correspond to a spontaneous breaking of the symmetry and generally cannot be 
obtained through small perturbations of the weakly-interacting ground state 
since they are emergent in nature.

The SU(4) symmetry of quantum Hall ferromagnetism gives rise to a ground state 
symmetry reflected in the SU(4)-symmetric Hamiltonian \eqnoeq{so5_1.1a} and a 
possible symmetry breaking structure that has been outlined in 
\S\ref{h:symmetriesFQHEgraphene}.  However, the symmetry-breaking patterns 
illustrated in \fig{dynamicalChains_QHF} represents {\em perturbations around 
the symmetric ground state (explicit symmetry breaking).} They cannot capture 
the nature of these possible spontaneously-broken symmetry states, since the 
broken-symmetry states may differ fundamentally from the possible states in 
\fig{dynamicalChains_QHF}:  the states corresponding to the most general linear 
combination \eqnoeq{upairs1.8} represent a complex superposition of many 
SU(4)-symmetric components and generally cannot be classified by pure or any 
simple linear combination of SU(4) irreducible representations.

Since the nature of the broken symmetry states cannot be studied directly 
within the SU(4) framework because they are unlikely to be anywhere near 
eigenstates of an SU(4)-symmetric Hamiltonian, the broken-symmetry states have 
typically been studied numerically, or by effective field theory methods.  
However, as has been shown, the kinds of collective configurations that have 
been 
proposed as candidates for low-lying broken-symmetry states in graphene (see 
\fig{S_D_pairs_brillouin} and \cite{khar2012}) bear strong resemblance to {\em 
eigenstates} of SO(8) dynamical symmetry chains.  Thus, the present SO(8) 
symmetry holds the promise of providing {\em analytical} solutions for possible 
broken-symmetry states in graphene.  This is the most important result of the 
present paper.

At specific filling factors the ground state of graphene will be determined 
by the competition among the SU(4) symmetry breaking terms.  The most obvious 
SU(4)-anisotropic effect is the Zeeman term, which favors a spin-polarized 
state, but the graphene sublattice structure introduces additional interactions 
that favor ground states without spin polarization that are characterized by 
spin density wave or charge density wave order at the lattice scale.  The 
competition between Zeeman-term spin polarization and the lattice-scale 
polarizations can be studied by changing the in-plane component of the magnetic 
field relative to the perpendicular component, since this changes the Zeeman 
energy but not the orbital energies \cite{youn14}. Such studies indicate that 
for the higher-energy Landau levels the Zeeman term is dominant, producing spin 
ferromagnets that have skyrmionic excitations at half filling, but in the $n=0$ 
Landau level the lattice-scale interactions dominate the Zeeman interaction and 
drive the system into a spin-unpolarized state \cite{youn12}.  

The remainder of this discussion will concentrate on these spin-unpolarized 
collective states that are candidates for the ground state in the $n=0$ Landau 
level for charge-neutral graphene, with the Zeeman interaction viewed as a 
perturbation on a collective structure that is dominated by lattice-scale 
interactions. Consideration in this paper will be restricted further to those 
states that can arise from the dynamical symmetries of 
\fig{dynamicalChains_graphene} that contain the SO(5) subgroup (those displayed 
in \fig{dynamicalChains_coherentSO5}).

It has been shown that SO(8) describes analytically a number of 
spontaneously-broken-SU(4)  candidates for the states observed in modern 
experiments such as those described in 
Refs.\ \cite{youn12,feld2012,benj2013,youn14}.  These solutions provide a 
spectrum of excited states as well as ground states.  The excited states will 
not be discussed here, except to note that all ground state solutions have a gap 
to electronic and collective excitations and so correspond to incompressible 
states. The general  theory to be discussed in forthcoming papers can 
accommodate FM, CDW, and AF states, but  for the dynamical symmetries containing 
 SO(5) that were the focus here, all solutions may be classified by a {\em 
single parameter} $\beta$ measuring AF order:  SU(4) states have finite AF order 
but no CDW or FM order, $\sofive\times\sutwo$ states have  no AF, CDW, or FM 
order, and SO(7) states correspond to a critical dynamical symmetry 
interpolating between SU(4) and $\sofive\times\sutwo$ with large AF fluctuations 
but no static AF order, and with no CDW or FM order. 

In a strong magnetic field the zero-energy state of graphene has four-fold spin 
and valley degeneracy per Landau level, and (neglecting the lattice-scale 
interactions) near the sample boundary one might expect the zero Landau level to 
split into one positively-dispersing (electron-like) and one negatively 
dispersing (hole-like) mode for each spin projection.  This would suggest a 
ground state having a bulk energy gap at charge neutrality but with 
electron-like and hole-like states of opposite spin polarization crossing at the 
edge of the sample (with the edge-state structure being analogous to the quantum 
spin Hall effect) \cite{aban2006,fert2006,youn14}.  However, experiments 
indicate that the ground state of charge-neutral graphene becomes {\em strongly 
insulating} at high magnetic fields \cite{chec2008}.  The detailed nature of 
this state remains uncertain, but it is generally expected to correspond to a 
spontaneously broken symmetry caused by the strong Coulomb interactions among 
the electrons in the zero Landau level.

Transport properties are not manifest in the algebraic solutions presented here 
but the coherent state approximation is equivalent to symmetry-constrained 
Hartree--Fock--Bogoliubov (HFB) theory \cite{zhan90,weimin1989}, suggesting 
that SO(8) theory can be mapped onto Hartree--Fock (HF) transport calculations. 
HF calculations for armchair nanoribbons found that AF and CDW states similar 
to ours have no edge currents \cite{jung2009}. It may be speculated that our AF 
states also are insulating and thus strong candidates for the high-field ground 
state, but confirmation requires more work.

Solutions depend on $G_0$ and $b_2$  in  \eq{modelHamiltonian}, which define 
effective interactions in the truncated space [highly renormalized relative to  
parameters in \eq{so5_1.1a}]. They may be fixed by systematic comparison with 
data, enabling a robust prediction for the nature of the  ground and other 
low-energy states. One expects modest impurity levels to modify the effective 
interaction parameters but to leave dynamical symmetries intact.

\section{\label{h:analogy_su4}Analogy with High-Temperature Superconductors}

A unified model of conventional superconductivity and high-temperature 
superconductors having an $\soeight$ highest symmetry has been 
proposed 
\cite{guid01,lawu03,su4review,guid99,guid04,sun05,sun06,sun07,guid09,guid09b,
guid10,guid11}.
For sufficiently strong on-site Coulomb repulsion (true for the cuprates and 
approximately true for the Fe-based compounds), the most general SO(8) symmetry 
is reduced to its SU(4) subgroup \cite{guid04}, which forms the basis of an 
SU(4) dynamical symmetry model of high-temperature superconductivity. This SU(4) 
model has three dynamical symmetry subchains ending in an $\sutwo\times \uone$ 
subgroup representing conservation of spin and charge. Physically, these 
SU(4) subgroup chains represent

\begin{enumerate}
 \item 
An SU(2) pseudospin subgroup chain that describes a $d$-wave singlet 
superconductor (SC). 
\item
An SO(4) subgroup chain that describes an antiferromagnetic Mott 
insulator (AF).
\item
An SO(5) subgroup chain representing a critical dynamical symmetry that is soft 
with respect to AF and SC fluctuations and interpolates between the AF and SC 
collective modes.
\end{enumerate}
Thus, in the SU(4) model of high-$T\tsub c$ superconductivity the SO(5) subgroup 
chain plays a similar physical role as the SO(7) subgroup chain of the graphene 
SO(8) dynamical symmetry model and the SO(7) subgroup chain of the nuclear SO(8) 
model. In all three cases the subgroup chains represent the generalization of a 
quantum critical point to an entire quantum critical phase that exhibits 
large fluctuations (in order-parameter space) connecting collective modes 
defined by other 
dynamical symmetries of the problem. 

\doublefig
{nuc_SC_graphene_chains}  
{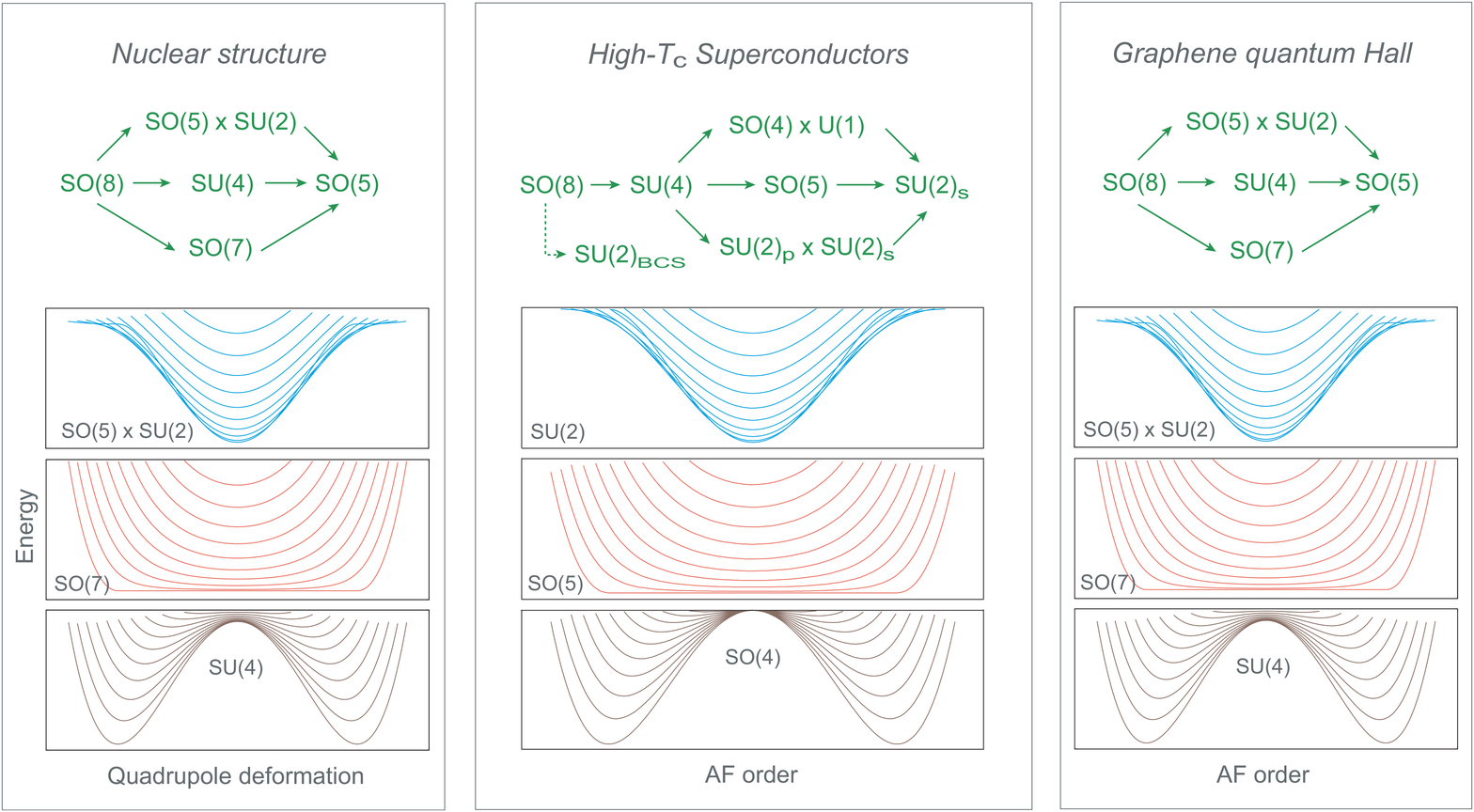}
{\figup} 
{0pt}
{0.24}
{Striking similarity in the dynamical symmetry chains and the coherent 
state energy surfaces for dynamical symmetry in nuclear structure physics, 
high-temperature superconductors, and graphene in a strong magnetic field.  }

In the high-$T\tsub c$ case the modes connected by the $\sufour \supset \sofive$ 
critical dynamical symmetry are antiferromagnetism and superconductivity, in 
graphene the critical $\soeight \supset \soseven \supset \sofive$ dynamical 
symmetry connects states with and without to N\'eel (AF) order, and in the 
nuclear structure case the critical $\soeight \supset \soseven \supset \sofive$ 
dynamical symmetry connects collective states that  differ in the relative 
contributions of nucleon pairs carrying total angular momentum 0 and total 
angular momentum 2.  Thus physically these three applications of dynamical 
symmetry have little in common, but one sees that mathematically they have a 
deep similarity. Although the microscopic physics differs fundamentally, at the 
level of the observed emergent collective modes in the system one sees that 
graphene quantum Hall states, high-temperature superconductors, and broad 
classes of nuclear collective states have a unified description in terms of 
dynamical symmetries associated with very similar compact Lie algebras. This 
remarkable similarity is illustrated  in \fig{nuc_SC_graphene_chains}

\section{\label{h:summary} Summary}

The well-known quantum Hall ferromagnetic SU(4) symmetry of 
graphene in strong magnetic fields has been extended by adding to the 
particle--hole operators 
that generate SU(4) a set of six creation and six annihilation operators that 
create or destroy fermion pairs in either a total valley isospin triplet, total 
spin singlet state, or a total valley isospin singlet, total spin triplet state 
(the only possibilities allowed by the Pauli principle).  This extended set of 
operators is shown to close an SO(8) algebra under commutation, which is 
formally analogous to the SO(8) algebra of the Fermion Dynamical Symmetry Model 
of nuclear structure physics.  This permits immediate adaptation of mathematical 
tools developed in nuclear physics to the graphene problem.

The previously-known SU(4) quantum Hall ferromagnetism symmetry is recovered as 
one subgroup, but one finds a richer set of low-energy collective modes 
associated with the full subgroup structure of SO(8).  By exploiting the 
established methodology of fermion dynamical symmetries, it was possible to 
decouple a collective-pair subspace from the full Hilbert space of the problem, 
permitting exact, analytical, many-body solutions to be obtained in several 
physically-interesting limits. In addition to exact solutions in specific 
dynamical symmetry limits, a generalized SO(8) coherent state 
approximation has been introduced that permits a broad range of solutions to be 
obtained even when not in the dynamical symmetry limits.

The pairs spanning the collective subspace are shown to be analogous to pairs 
that have already been discussed at a qualitative level in the graphene 
literature as defining the possible broken symmetry ground states in the 
presence of strong electron--electron and electron--phonon correlations in the 
$n=0$ Landau level.  The development here places these pairs on a firm, unified 
mathematical footing and permits analytical solutions to be developed that 
explore the possible collective states that previously required numerical 
simulation for their quantitative description.

Finally, it has been shown that there are uncanny dynamical symmetry analogies 
among broken symmetry states for graphene in a strong magnetic field, 
high temperature superconductors, and strongly collective states in atomic 
nuclei. On the one hand this has the practical utility of allowing technology 
already developed in one field to be adapted easily to another.  On the other 
hand, it implies a deep and intriguing mathematical affinity among physical 
problems that are not usually viewed as having more than a superficial 
connection.

{\it Acknowledgments.}{\bf--}
This work is supported by the Basque Government (Grant No.~IT472-10), 
the Spanish MICINN (Project No.~FIS2012-36673-C03-03), and the Basque Country 
University UFI (Project No.~11/55-01-2013).  Partial support was provided by 
LightCone Interactive LLC.

\begin{appendix}

\section{\label{h:appendixA} Extension of SU(2N) to SO(4N)}

The formalism discussed in this paper was introduced by postulating a set of 
physical operators that were shown explicitly to generate an SO(8) Lie algebra 
under commutation.  It is illuminating to consider a somewhat different 
perspective on the motivation for introducing an SO(8) symmetry for the 
graphene problem.

\subsection{\label{sh:addingPairs}Adding Pair Operators to a Unitary Algebra}

It is well known that in a fermionic space having $2N$ degrees of freedom the 
most general set of bilinear products $c^\dagger_i c_j$ of 
creation--annihilation operators and their hermitian conjugates generates an 
SU($2N$) Lie algebra under commutation \cite{unitaryAlgebra,wyb74}.  
(Physically, the restriction to bilinear products corresponds 
to limiting consideration to two-body interactions.)  Likewise, it is well 
known that adding to this particle--hole operator set the  most general pair 
creation and annihilation operators $c_i^\dagger c^\dagger_j$ and $c_ic_j$ 
extends the SU($2N$) algebra to  SO($4N$).

This extension is useful because sometimes more is less.  The advantage of 
expanding the space from SU($2N$), with $4N^2-1$ generators, to SO($4N$), with 
$8N^2-2N$ generators, is that the added pair operators permit the definition of 
a (collective) subspace of the full Hilbert space spanned by the products of 
pair creation operators acting on the pair vacuum.  If an effective Hamiltonian 
is then constructed by writing the most general polynomial in the Casimir 
invariants of all groups in the subgroup chains of the highest symmetry SO($4N$) 
(restricted to quadratic Casimir invariants if one considers only two-body 
interactions), it will correspond to the most general Hamiltonian that can be 
written in the collective subspace, and will be diagonal in the collective 
subspace basis for specific dynamical symmetry subgroup chains.  Thus, the 
Schr\"odinger equation can be solved analytically in the symmetry limits defined 
by each dynamical symmetry subgroup chain, and even away from the symmetry 
limits it can be solved analytically in coherent-state approximation.

\subsection{\label{sh:inadequacySU2N} Inadequacy of SU(2N) Alone}

The SU($2N$) particle--hole algebra alone can be used to construct a 
Hamiltonian that commutes with its generators, and the corresponding 
Schr\"odinger equation can be solved analytically.  But when one considers the 
realistic case of adding symmetry-breaking terms to the Hamiltonian that do not 
commute with the generators of SU($2N$), the best that can be done analytically 
is to assume that these terms are small and that the physical solutions can be 
treated as small perturbations around the symmetric solution.  

For the non-perturbative case where the added terms lead to spontaneously broken 
symmetry and new possible ground states that are not connected perturbatively to 
the symmetric ground state, one has no systematic way to construct the new 
ground state from the symmetric one except to guess it.  But since the 
SU($2N$)-symmetric solution is not connected analytically to non-perturbative 
broken symmetry solutions, the symmetries of the unperturbed ground state could 
be a poor guide to guessing the nature of the broken-symmetry states and one 
must rely on numerical solutions or other approximations and not symmetry to 
determine their properties.

\subsection{\label{sh:applicationGraphene}Application to Graphene}

Let us now apply this general discussion to quantum Hall magnetism in graphene. 
As discussed in \S\ref{h:symmetriesFQHEgraphene}, it is widely accepted that an 
approximate  SU($2N$) particle--hole symmetry with $N=2$ is relevant in the 
$n=0$ Landau level because the dominant long-range Coulomb interaction is 
SU(4)-symmetric.  However, the short-range terms that break this symmetry in the 
effective Hamiltonian prevent the SU(4) symmetry from providing a solution for 
the broken symmetry ground state, unless it is assumed that those terms only 
perturb the SU(4)-symmetric solution (small explicit symmetry breaking).  But 
experimental evidence suggests that the true ground state of graphene in a 
strong magnetic field at low temperature breaks SU(4) symmetry {\em 
spontaneously, not explicitly} (see the Introduction), and therefore is produced 
by a non-perturbative effect that cannot be explained in terms of small 
fluctuations around the SU(4)-symmetric solution.

The possible (spontaneously) broken-symmetry states for the $n=0$ Landau level 
have been described in terms of the most general sets of electron pairs 
occupying the two valley isospin and two spin degrees of freedom 
\cite{khar2012},
\begin{equation}
\Psi = \left[
\prod_m \left(
\sum_{\lambda \sigma, \lambda' \sigma'}
\Phi^*_{\lambda\sigma, \lambda'\sigma'}
c^\dagger_{0m\lambda\sigma}  c^\dagger_{0m\lambda'\sigma'}
\right)
\right] \ket 0 ,
\label{basis1.0}
\end{equation}
where the vacuum state $\ket 0$ corresponds to completely filled Landau levels 
for $n<0$ and completely empty Landau levels for $n\ge 0$.  Each 
factor in the product $\prod_m$ creates a pair of electrons in the state
$\Phi = \{\Phi_{\lambda\sigma, \lambda'\sigma'} \}$ at orbital $m$ of the 
$n=0$ LL, with $\lambda,\lambda'$ equal to sublattice A or B, $\sigma,\sigma'$ 
equal to spin up or down, and with the valley isospin and sublattice pseudospin 
identified: $K \leftrightarrow A$ and $K' \leftrightarrow B$. 

But the pair creation operators $c^\dagger c^\dagger$ in \eq{basis1.0} and their 
hermitian conjugates are not generators of SU(4) and  do not commute with the 
SU(4)-symmetric Hamiltonian, and the collective states of the form 
\eqnoeq{basis1.0} that are of interest in the present context are unlikely to 
represent small fluctuations around the SU(4)-symmetric solution. Thus the 
nature of these collective states is not determined by the SU(4) symmetry and 
had to be investigated by numerical calculations using a small basis in prior 
work \cite{khar2012}.  

On the other hand, the SO(8) pair generators introduced in \eq{coupled1.4} are 
included in the most general collective pairs generated by the $c^\dagger 
c^\dagger$ operators in \eq{basis1.0}, and include the collective degrees of 
freedom discussed in Ref.\ \cite{khar2012} (see \S\ref{sh:otherStates} and 
\fig{S_D_pairs_brillouin}).  Thus, the SO(8) fermion dynamical symmetry permits 
the nature of possible spontaneously broken symmetries to be investigated in 
terms of {\em symmetry properties} that permit analytical solutions for the 
broken-symmetry states.

\subsection{\label{sh:alternativeMotivation} An Alternative Motivation}

Hence, the formalism described in this paper may also be introduced by the 
following logic.  The SU(4) particle--hole symmetry generated by the operators 
$B_{ab}$ defined in \eq{algebra1.2} is known to provide a good starting point 
for graphene quantum Hall states dominated by the long-range Coulomb 
interaction, but does not describe quantitatively the broken-symmetry modes 
discussed by Kharitonov \cite{khar2012} in terms of collective pairs resulting 
from short-range correlations.  Motivated by the preceding discussion in this 
Appendix,  the SU(4) generator set may be extended to include the possible pair 
creation and pair annihilation operators operating in the space corresponding to 
the indices in the SU(4) basis.  By the general SU($2N$) $\rightarrow$ SO($4N$) 
extension discussed above, this gives the SO(8) Lie algebra of \eq{algebra1.3}.  
Hence, expanding the algebra from SU(4) to SO(8) introduces the capability 
to explore {\em analytically} the possible collective states following from 
perturbation of the SU(4) quantum Hall ferromagnet by short-range interactions 
that break SU(4) both explicitly and spontaneously.

\clearpage

\section{\label{h:appendixB}Transformations between Bases}
 
This Appendix collects some useful transformations among the several bases 
that have been employed in this paper. For brevity, in the following 
$
\{P^1, P^2, P^3, S_0, S, S^\dagger, D\phantomdagger_{\mu}, 
D^\dagger_\mu \}
$
will be termed  the nuclear SO(8) basis and 
$
\{ \spin_\alpha,  \,T_\alpha,\, N_\alpha, \,\Piop \alpha x, \,
\Piop \alpha y,  S_0, S, 
S^\dagger, D\phantomdagger_{\mu}, D^\dagger_\mu \}
$
will be termed the graphene SO(8) basis.

In transforming from the nuclear SO(8) basis to the graphene SO(8) basis  
the particle number (charge) operator $n$ or $S_0$ and the 12 pairing operators 
$\{D_\mu, \, D^\dagger_\mu,\, S,\, S^\dagger\}$ are retained, but 
the 15 SU(4) 
generators $\{ P^1,\, P^2, \,P^3\}$ in the nuclear representation are 
replaced with the 15 
SU(4) generators $\{\spin_\alpha,\,  T_\alpha,\, N_\alpha, \,\Piop \alpha x, 
\,\Piop \alpha y\}$ defined in the graphene representation of \eq{algebra1.4}.
The explicit transformation from the  $\{ P^1,\, P^2,\, P^3\}$ generators to the 
$\{\spin_\alpha,  \,T_\alpha,\, N_\alpha, \,\Piop \alpha x, \, \Piop \alpha y\}$ 
generators is given by
\begin{subequations}
\begin{align}
\spin_x &=
\sqrt{\tfrac{6}{5}}\left(P^1_{-1} - P^1_1 \right) + 
\tfrac{2}{\sqrt5}\left(P^3_{-1} - P^3_1 \right)
\label{ngx1}
\\
\spin_y &=
i\left(\sqrt{\tfrac{6}{5}}\left(P^1_1 + P^1_{-1}\right) + 
\tfrac{2}{\sqrt5}\left(P^3_1 + P^3_{-1}\right)\right)
\label{ngx2}
\\
\spin_z &=
\tfrac{2}{\sqrt5} P^1_0 + \tfrac{4}{\sqrt5} P_0^3 = n_1 -n_2 + n_3 -n_4
\label{ngx3}
\\
T_x &=
-\sqrt2 \left( P^2_2 + P^2_{-2}\right)
\label{ngx4}
\\
T_y &=
i\sqrt2 \left( P^2_2 - P^2_{-2}\right)
\label{ngx5}
\\
T_z &=
\tfrac{4}{\sqrt 5} P^1_0 - \tfrac{2}{\sqrt5} P^3_0 = n_1 + n_2 -n_3 -n_4
\label{ngx6}
\\
N_x &=
\tfrac{1}{\sqrt2}\left(P^2_{-1} - P^2_1\right)
\label{ngx7}
\\
N_y &=
\tfrac{i}{\sqrt2}\left(P^2_{-1} + P^2_1\right)
\label{ngx8}
\\
N_z &=
P^2_0 = n_1 - n_2 + n_4 - n_3
\label{ngx9}
\\
\Piop xx &=
\tfrac{1}{2}\left[ \vphantom{\sqrt{\tfrac{2}{5}}} P^3_{-3} - P^3_3  \right.
\nonumber
\\
&\left. +\sqrt{\tfrac{2}{5}}\left(P^1_{-1} - 
P^1_1\right)  + \sqrt{\tfrac{3}{5}}\left(P^3_1 - P^3_{-1}\right)\right]
\label{ngx10}
\\
\Piop yx &=
\tfrac{i}{2}\left[ \sqrt{\tfrac{2}{5}} P^3_{-3} + P^3_3 \right.
\nonumber
\\
&\left. + \sqrt{\tfrac{2}{5}}\left(P^1_{-1} + 
P^1_1\right)  - \sqrt{\tfrac{3}{5}}\left(P^3_1 + P^3_{-1}\right)\right]
\label{ngx11}
\\
\Piop zx &=
-\tfrac{1}{\sqrt2} \left( P^3_2 + P^3_{-2}\right)
\label{ngx12}
\\
\Piop xy &=
\tfrac{i}{2}\left[\sqrt{\tfrac{2}{5}} P^3_{-3} + P^3_3 \right.
\nonumber
\\
&\left.- \sqrt{\tfrac{2}{5}}\left(P^1_{-1} + 
P^1_1\right) + \sqrt{\tfrac{3}{5}}\left(P^3_1 + P^3_{-1}\right)\right]
\label{ngx13}
\\
\Piop yy &=
\tfrac{1}{2}\left[ \vphantom{\sqrt{\tfrac{2}{5}}} -P^3_{-3} + P^3_3 \right.
\nonumber
\\
&\left.- 
\sqrt{\tfrac{2}{5}}\left(P^1_1 - P^1_{-1}\right) + 
\sqrt{\tfrac{3}{5}}\left(P^3_1 - P^3_{-1}\right)\right]
\label{ngx14}
\\
\Piop zy &=
-\tfrac{i}{\sqrt2} \left( P^3_2 - P^3_{-2}\right)
\label{ngx15}
\end{align}
\label{ngx}
\end{subequations}

In Eqs.\ \eqnoeq{algebra1.8}--\eqnoeq{algebra1.11} the graphene 
basis $\{\spin_\alpha,\,  T_\alpha,\, N_\alpha, \,\Piop \alpha x, \,\Piop \alpha 
y\}$ has been expressed in terms of the generators $B_{ab}$ defined in 
\eq{algebra1.2}.  The 
inverse transformations giving the $B_{ab}$ in terms of the $\{\spin_\alpha,\,  
T_\alpha,\, N_\alpha, \,\Piop \alpha x, \,\Piop \alpha y\}$ are
\begin{subequations}
\begin{align}
B_{12} &=
\tfrac{1}{2}N_{x}+\tfrac{1}{2}iN_{y}+\tfrac{1}{4}\spin_{x}+\tfrac{1}{4}i\spin_{y
}
\label{grapheneToB1}
\\
B_{13} &=
\tfrac{1}{4}T_{x}+\tfrac{1}{4}iT_{y}+\tfrac{1}{2}\Pi_{zx}-\tfrac{1}{2}i\Pi_{zy}
\label{grapheneToB2}
\\
B_{14} &=
\tfrac{1}{2}\Pi _{xx}-\tfrac{1}{2}i\Pi _{yx}-\tfrac{1}{2}i\Pi _{xy}-
\tfrac{1}{2}\Pi _{yy}
\label{grapheneToB3}
\\
B_{23} &=
\tfrac{1}{2}\Pi _{xx}+\tfrac{1}{2}i\Pi _{yx}-\tfrac{1}{2}i\Pi _{xy}+
\tfrac{1}{2}\Pi _{yy}
\label{grapheneToB4}
\\
B_{24} &=
\tfrac{1}{4}T_{x}+\tfrac{1}{4}iT_{y}-\tfrac{1}{2}\Pi _{zx}+\tfrac{1}{2}
i\Pi _{zy}
\label{grapheneToB5}
\\
B_{34} &=
\tfrac{1}{4}\spin_{x}-\tfrac{1}{2}iN_{y}-\tfrac{1}{2}N_{x}+\tfrac{1}{4}i\spin_{y
}
\label{grapheneToB6}
\\
B_{11} &=
\tfrac14\spin_z +\tfrac14 T_z + \tfrac12 N_z + \tfrac14 (n-\Omega)
\label{grapheneToB7}
\\
B_{22} &=
-\tfrac14\spin_z +\tfrac14 T_z - \tfrac12 N_z + \tfrac14 (n-\Omega)
\label{grapheneToB8}
\\
B_{33} &=
\tfrac14\spin_z -\tfrac14 T_z - \tfrac12 N_z + \tfrac14 (n-\Omega)
\label{grapheneToB9}
\\
B_{44} &=
-\tfrac14\spin_z -\tfrac14 T_z + \tfrac12 N_z + \tfrac14 (n-\Omega)
\label{grapheneToB10}
\end{align}
\label{grapheneToB}%
\end{subequations}
where the unlisted operators may be obtained from $B_{ba} = B_{ab}^\dagger$
and the diagonal operators have been assumed to obey the U(4) constraint
\begin{equation}
 B_{11} + B_{22} + B_{33} + B_{44} =   n-\Omega,
 \label{u4Constraint}
\end{equation}
with $n = n_1 + n_2 + n_3 + n_4$ the total particle number and $\Omega$ the 
total pair degeneracy given by \eq{degen1.2}.

Since from \eq{orderp1.1} the order parameters for the quantum Hall ground 
states are functions of the expectation values for the number operators $n_a$ 
specifying the population of the four basis states in 
\tableref{quantumNumberMapping} and \fig{grapheneBasis} labeled by the index 
$a$, it is useful to have explicit expressions for them in terms of the 
$P^r_\mu$ operators.  These are
\begin{subequations}
\begin{align}
n_1 &=
\tfrac14 n + \tfrac{3\sqrt5}{10}P^1_0 + \tfrac12 P^2_0 + \tfrac{\sqrt5}{10}P^3_0
\label{Ptoi1}
\\[2pt]
n_2 &=
\tfrac14 n + \tfrac{\sqrt5}{10}P^1_0 - \tfrac12 P^2_0 - \tfrac{3\sqrt5}{10}P^3_0
\label{Ptoi2}
\\[2pt]
n_3 &=
\tfrac14 n - \tfrac{\sqrt5}{10}P^1_0 -\tfrac12 P^2_0 + \tfrac{3\sqrt5}{10}P^3_0
\label{Ptoi3}
\\[2pt]
n_4 &=
\tfrac14 n - \tfrac{3\sqrt5}{10}P^1_0 +\tfrac12 P^2_0 -\tfrac{\sqrt5}{10}P^3_0
\end{align}
\label{Ptoi}%
\end{subequations}
where the total number operator $n$ is
\begin{equation}
 n = \tfrac12 N = n_1 + n_2 + n_3 + n_4 = 2\left( P^0_0 + 
\tfrac12 \Omega\right).
\end{equation}

\end{appendix}

\clearpage

\bibliographystyle{}

\begin{thebibliography}{99}

\bibitem{vonk1980}K. von Klitzing, G. Dorda, M. Pepper, Phys.\ Rev.\ Lett.\ 
{\bf 45}, 494 (1980).

\bibitem{laug1981}R. B. Laughlin, Phys.\ Rev.\ {\bf B23}, 5632 (1981).

\bibitem{tsui1982}D. C. Tsui, H. L. Stormer, and A. C. Gossard, Phys.\ Rev.\ 
Lett.\ {\bf 48}, 1599 (1982).

\bibitem{laug1983}R. B. Laughlin, Phys.\ Rev.\ Lett.\ {\bf 50}, 1395 (1983).

\bibitem{novo2005}K. S. Novoselov et al, Nature {\bf 
438}, 197 (2005).

\bibitem{zhan2005} Y. Zhang et al, Nature {\bf 438}, 201 (2005).

\bibitem{bolo2009}K. I. Bolotin, F. Ghahari, M. D. Shulman, H. L. Stormer, and 
P. Kim, Nature {\bf 462}, 196 (2009).

\bibitem{du2009}X. Du, I Skachko, F. Duerr, A. Luican, and E. Y. Andrei,
Nature {\bf 462}, 192 (2009).

\bibitem{feld2012} B. E. Feldman, B. Krauss, J. H. Smet, and A. Yacoby, 
Science {\bf 337}, 1196 (2012).

\bibitem{benj2013}B. E. Feldman, A. J. Levin, B. Krauss, D. 
A. Abanin, B. I. Halperin, J. H. Smet, and A. Yacoby, Phys.\ Rev.\ 
Lett.\ {\bf 111} 076802 (2013).


\bibitem{chec2008} J. G. Checkelsky, L. Li, and N. P. Ong, Phys.\ Rev.\ Lett.\ 
{\bf 100}, 206801 (2008).

\bibitem{jung2009} J. Jung and A. H. MacDonald, Phys.\ Rev.\ {\bf B80}, 235417 
(2009).

\bibitem{herb07a}I. F. Herbut, Phys.\ Rev.\ {\bf B75}, 165411 (2007).

\bibitem{herb07b}I. F. Herbut, Phys.\ Rev.\ {\bf B76}, 085432 (2007).

\bibitem{roy14}B. Roy, M. P. Kennett, and S. Das Sarma, Phys.\ Rev.\ {\bf B90}, 
201409(R) (2014)

\bibitem{clwu86} C.-L. Wu, D. H. Feng, X.-G. Chen, J.-Q. Chen, and M. W. Guidry,
Phys.\ Lett {\bf B168}, 313 (1986).

\bibitem{clwu87} C.-L. Wu, D. H. Feng, X.-G. Cheng, J.-Q. Chen, and M. W. 
Guidry, Phys.\ Rev.\ {\bf C36}, 1157 (1987).

\bibitem{FDSM} C.-L. Wu, D. H. Feng and M. W. Guidry,
{\em Adv.\ in Nucl.\ Phys} {\bf 21}, 227 (1994).

\bibitem{guid01} M. W. Guidry, L.-A. Wu, Y. Sun, and C.-L. Wu, Phys.\ Rev.\ {\bf
B63}, 134516 (2001);

\bibitem{lawu03} L.-A. Wu,  M. W. Guidry, Y. Sun, and C.-L. Wu Phys.\ Rev.\ {\bf
B67}, 014515 (2003).

\bibitem{su4review}M. W. Guidry, Y. Sun, and C.-L. Wu, ``Fermion Dynamical 
Symmetry and Strongly-Correlated Electrons: A
Comprehensive Model of High-Temperature Superconductivity'', to be published.

\bibitem{wu2016}L.-A. Wu and M. W. Guidry, Sci.\ Rep.\ {\bf 6}, 22423 (2016).

\bibitem{cast2009}A. H. Castro Neto, F. Guinea, N. M. R. Peres, K. S. 
Novoselov, and A. K. Geim, Rev.\ Mod.\ Phys.\ {\bf 81}, 109 (2009).

\bibitem{goer2011}M. O. Goerbig, Rev.\ Mod.\ Phys.\ {\bf 83}, 1193 (2011).

\bibitem{ando1998} T. Ando, T. Nakaishi, and R. Saito, J. Phys.\ Soc.\ Jpn.\ 
{\bf 67}, 2857 (1998).

\bibitem{miki1999}G. P. Jikitik and Y. V. Sharai, Phys.\ Rev.\ Lett.\ {\bf 82}, 
2147 (1999).

\bibitem{jain1989} J. K Jain, Phys.\ Rev.\ Lett.\ {\bf 63}, 199 (1989).


\bibitem{khar2012}M. Kharitonov, Phys.\ Rev.\ {\bf B85}, 155439 (2012).

\bibitem{wufe2014}F. Wu, I. Sodemann, Y. Araki, A. H. MacDonald, and T. 
Jolicoeur, Phys.\ Rev.\ {\bf B90}, 235432 (2014) [arXiv:1406.2330].

\bibitem{chen86} J.-Q. Chen, D. H. Feng, and C.-L. Wu, Phys.\ Rev.\ {\bf C34} 
2269 (1986).

\bibitem{wyb74}B. G. Wybourne, {\em Classical Groups for Physicists,} Wiley 
Interscience
(1974).

\bibitem{angularMomentum} For example, see {\em Angular Momentum,} D. M. Brink 
and G. R. Satchler, Clarendon Press (1968) or {\em Angular Momentum in Quantum 
Mechanics,} A. R. Edmonds, Princeton University Press (1959).

\bibitem{weimin1989}W.-M. Zhang, D. H. Feng, C.-L. Wu, H. 
Wu, and J. N. 
Ginocchio, Nucl.\ Phys.\ {\bf A505}, 7 (1989).

\bibitem{gino80} J.N. Ginocchio, Ann.\ Phys., {\bf 126}, 234 (1980).

\bibitem{IBM} F. Iachello and A. Arima, {\it The Interacting Boson Model}
(Cambridge University Press, Cambridge, 1987).

\bibitem{wmzha87}W.-M. Zhang, D. H. Feng, and J. N. Ginocchio,
Phys.\ Rev.\ Lett.\ {\bf 59}, 2032 (1987)

\bibitem{zhang88}W.-M. Zhang, D. H. Feng, and J. N. Ginocchio,
Phys.\ Rev.\ {\bf C37}, 1281 (1988)

\bibitem{arec1972} F. T. Arecchi, E Courtens, R. Gilmore, and H. Thomas, Phys.\ 
Rev.\ {\bf A6}, 2211 (1972).
\bibitem{gilm1972} R. Gilmore, Ann.\ Phys.\ {\bf 74}, 391 (1972).
\bibitem{pere1972} A. M. Perelomov, Commun.\ Math.\ Phys.\ {\bf 26}, 222 (1972).
\bibitem{gilm1974} R. Gilmore, Rev.\ Mex.\ de Fisica {\bf 23}, 143 (1974).

\bibitem{zhan90}W.-M. Zhang, D. H. Feng, and R. Gilmore, {\em Rev.\ Mod.\
Phys.} {\bf 62}, 867 (1990).

\bibitem{youn12} A. F. Young, C. R. Dean, L. Wang, H. Ren, P. Cadden-Zimansky,	
K. Watanabe, T. Taniguchi, J. Hone, K. L. Shepard	
and P. Kim Nature Physics {\bf 8}, 550--556 (2012).


\bibitem{youn14} A. F. Young, J. D. Sanchez-Yamagishi, B. Hunt, S. H.
Choi, K. Watanabe, T. Taniguchi, R. C. Ashoori, and
P. Jarillo-Herrero, Nature {\bf 505}, 528 (2014).

\bibitem{aban2006} D. A. Abanin, P. A. Lee, and L. S. Levitov, Phys.\ Rev.\ 
Lett.\ {\bf 96}, 176803 (2006).

\bibitem{fert2006} H. A. Fertig and L. Brey, Phys.\ Rev.\ Lett.\ {\bf 97}, 
116805 (2006).

\bibitem{guid99} M. W. Guidry, Rev.\ Mex.\ F\'is.\ {\bf 45 S2}, 132 (1999).

\bibitem{guid04} M. W. Guidry, Y. Sun, and C.-L. Wu, Phys.\ Rev.\ {\bf B70},
184501, (2004).

\bibitem{sun05}
Y. Sun, M. W. Guidry, and C.-L. Wu, Phys.\ Rev.\ {\bf B73}, 134519 (2006).

\bibitem{sun06}
Y. Sun, M. W. Guidry, and C.-L. Wu, Phys.\ Rev.\ {\bf B75}, 134511 (2007).

\bibitem{sun07}
Y. Sun, M. W. Guidry, and C.-L. Wu, Phys.\ Rev.\ {\bf B78}, 174524 (2008).

\bibitem{guid09}M. W. Guidry, Y. Sun, and C.-L. Wu, Front. Phys.
China 4, 233 (2009).

\bibitem{guid09b} M. W. Guidry, Y. Sun, and C.-L. Wu, New J. Phys. 11, 123023
(2009).

\bibitem{guid10}  Mike Guidry, Yang Sun, and Cheng-Li Wu, Front. Phys.
China, 5(2), 171-175 (2010).

\bibitem{guid11}M. W. Guidry, Y. Sun, and C.-L. Wu, Chinese Science Bulletin,
56, 367-371 (2011).

\bibitem{unitaryAlgebra} B. R. Judd, {\em Operator Techniques in Atomic
Spectroscopy,} McGraw--Hill (1963).


\end{thebibliography}

\end{document}